\documentclass[traditabstract]{aa}
\usepackage{graphicx,float,psfrag}
\usepackage{tabularx}
\usepackage{latexsym,amsmath,amssymb}
\usepackage{natbib}
\usepackage{verbatim}
\usepackage{rjwmath}
\usepackage{multirow}

\bibpunct{(}{)}{;}{a}{}{,}

\def \george    {G. Younes}
\def \delphine  {D. Porquet}
\def \bassem    {B. Sabra}
\def \jreeves   {J.~N. Reeves}

\def \strasbourg {Observatoire Astronomique de Strasbourg, Universit\'e de Strasbourg, CNRS, UMR7550, 11 rue de
l'Universit\'e, F-67000 Strasbourg, France}

\def \ndu {Department of Physics \&\ Astronomy,  Notre Dame University-Louaize, P.O.Box 72, Zouk Mikael, Lebanon}
\def \keele {Astrophysics Group, School of Physical \&\ Geographical Sciences, Keele University, Keele, Staffordshire ST5 5BG}

\newcommand {\rosat} {{\it ROSAT}}

\newcommand {\xmm} {\textsl{XMM-Newton}}
\newcommand {\chandra} {\textsl{Chandra}}

\def \rsun {\ifmmode$R$_{\odot}\else R$_{\odot}$}

\def \hcm {\hbox {\ifmmode $ atoms cm$^{-2}\else atoms cm$^{-2}$\fi}}

\def\approxgt{\mathrel{\hbox{\rlap{\lower.55ex \hbox {$\sim$}}
        \kern-.3em \raise.4ex \hbox{$>$}}}}
\def\approxlt{\mathrel{\hbox{\rlap{\lower.55ex \hbox {$\sim$}}
        \kern-.3em \raise.4ex \hbox{$<$}}}}

\def\arcmin{\hbox{$^\prime$}}
\def\arcsec{\hbox{$^{\prime\prime}$}}

\newcommand {\chisq} {$\chi ^{2}$}

\def \eddratio {$L_{2-10~keV}/L_{Edd}$}

\begin{document}

\title{Study of  LINER sources with broad  H$\alpha$ emission.\\ X-ray
  properties and comparison to luminous AGN and X-ray binaries.}

\author{\george\inst{1} \and \delphine\inst{1} \and \bassem\inst{2} \and \jreeves\inst{3}}

\institute{\strasbourg \and \ndu \and \keele}

%\date{Received; Accepted:}
\date{Received / Accepted}
%\date{Received / Accepted}

\authorrunning{G. Younes et al.}

\titlerunning{X--ray properties of a sample of LINERs showing broad H$\alpha$ emission}

\abstract{An  important  number  of  multiwavelength  studies  of  low
  ionization nuclear emission-line  regions (LINERs) were dedicated to
  investigate  the excitation mechanism  responsible for  the detected
  emission  lines.  Radiative  emission  from accretion  into a  super
  massive black  hole (SMBH) is now  believed to be,  in an increasing
  number of LINERs, the  source of excitation.  However, the accretion
  mode is not yet firmly  understood, and could be explained in either
  a thin  accretion disk or  a radiatively inefficient  accretion flow
  (RIAF).   Our purpose  is to  study  the X-ray  properties of  LINER
  sources with  definite detection of a broad  H$\alpha$ emission line
  in their  optical spectra, LINER~1s  from Ho et al.   sample.  These
  objects preferentially harbor a low luminosity active nucleus at the
  center    and    show    small    or   no    intrinsic    absorption
  ($\le$10$^{22}$cm$^{-1}$).   We compare  their  X-ray properties  to
  both  X-ray binaries  and luminous  AGN. We  analyzed  all available
  X-ray  archived  \xmm\ and  \chandra\  observations  of 13  LINER~1s
  satisfying the above criterion  in a systematic homogeneous way.  We
  looked  for any correlations  between the  X-ray properties  and the
  intrinsic  parameters  of  our   sample  of  LINER~1s.  An  absorbed
  power-law gave a good fit to the  spectra of 9 out of the 13 sources
  in our sample.  A combination of a thermal component and an absorbed
  power-law  were required  in the  remaining 4  sources.  We  found a
  photon  index for  our sample  between $1.3\pm0.2$  for  the hardest
  source and $2.4^{+0.2}_{-0.3}$ for the softest one with a mean value
  of $1.9\pm0.2$ and a dispersion $\sigma=0.3$.  The thermal component
  had a mean temperature kT$\approx0.6$~keV.  Significant short (hours
  to days) time-scale variability is  not common in the present sample
  and was observed  in only 2 sources (NGC~3226  and NGC~4278).  Three
  other sources  indicate a possible  variability with a low  K-S test
  probability (2\%--4\%)  that the nuclear emission  originates from a
  constant source.   On the other  hand, significant variability  on a
  longer time-scale  (months to years) is  detected in 7 out  of the 9
  sources  observed  more  than  once.   No  significant  Fe~K$\alpha$
  emission line at 6.4~keV was  detected and upper limits were derived
  for the  4 sources with a  high enough signal to  noise ratio around
  6~keV.  Finally, we established, for  the first time for a sample of
  LINER~1s,   that  the   photon  index   $\Gamma$   is  significantly
  anticorrelated    to     $L_{2-10~keV}/L_{Edd}$.     Whereas    this
  anticorrelation is in contrast to the positive correlation found for
  type 1 AGN, it is similar to  the one seen in XRBs in their low/hard
  state where  a radiatively inefficient accretion flow  is thought to
  be responsible for the X-ray emitted energy.}


\maketitle

\section{Introduction}
\label{sec:intro}

It is  widely known  that a large  fraction of local  galaxies contain
emission   line  nuclei   that   are  the   result   of  low   nuclear
activity. \citet{ho95apjs:optspec,ho97apjs:specparam,ho97apjs:broadHal}
showed   that   more  than   40\%   of   486   nearby  galaxies   with
$B_{T}\le12.5$~mag could be considered  as active with optical spectra
classified  as Seyfert  nuclei,  low-ionization nuclear  emission-line
regions   \citep[LINERs,][]{heckman80aap},   or   transition   objects
(objects having intermediate spectra  between LINERs and H~II nuclei).
Pure  LINER sources would  be the  most common  component representing
20\% of  all 486 galaxies.   The ionization mechanism  responsible for
the excitation of emission lines in LINER sources is an ongoing matter
of  debate  and could  be  explained in  terms  of:  shock heated  gas
\citep{dopita95apj:shocliner},            starburst           activity
\citep{alonso-herrero00ApJ:starburstinliners,terlevich95mnras:starburstliner},
or   a   low  luminosity   active   Galactic   nucleus  (AGN).    Many
multiwavelength studies were attributed to this subject, looking for a
radio, sometimes  variable, core \citep{4278nagar05aap}  or a variable
UV   core  \citep{maoz05apj:linervarUV}   in  nearby   LINER  sources.
Nevertheless, the most used tool to  search for an active nucleus in a
LINER is to look for a hard 2-10~keV unresolved core that could not be
due to  diffuse emission from shock  heated gas or  from unusually hot
stars
\citep{terashima00ApJ:liners,ho01apjl,dudik05apj:hardcoreliner,flohic06apj,gonzalezmartin06aa,gonzalezmartin09aa,zhang09apj:llagnxray}.
How do  LINERs harboring  a low luminosity  active nucleus  compare to
luminous Seyfert galaxies and quasars?

\citet{maoz07MNRAS},  using  high  angular resolution  multiwavelength
observations  of 13  LINER sources,  demonstrated that  the luminosity
ratios  in different wavebands,  mainly UV  to X-ray  and radio  to UV
luminosities, follow the same  trend as luminous Seyfert galaxies. The
authors  did  not  find  any  sharp  change  in  the  spectral  energy
distribution  (SED) of  their sample  of  13 LINERs  compared to  more
luminous Seyfert  and quasar nuclei, suggesting that  a thin accretion
disk    may    persist   at    low    accretion   rates.     Moreover,
\citet{pianmnras10} detected up to 30\%\ flux variations on half a day
time-scale in 2  (NGC~3998 and M81) out of 4  LINER and low luminosity
AGN sources observed in X-ray  with the XRT onboard {\sl Swift}.  They
combined their  X-ray fluxes with  simultaneous UV fluxes  coming from
the UVOT instrument  and showed that the SED and the  UV to X-ray flux
ratios of  their 4  sources sample are  consistent with those  of more
luminous  sources  and that  LINERs  may  have  similar accretion  and
radiative  processes  at their  center  compared  to luminous  Seyfert
nuclei.

On the other hand, the faintness of LINER sources compared to luminous
Seyfert  galaxies  and quasars  has  been  attributed  to a  different
accretion mechanism  owing to some observational  contrast between the
two  classes.   No broad  nor  narrow  Fe~K$\alpha$  emission line  at
6.4~keV have  been detected in the  spectra of the  LINER sources with
the         highest        signal        to         noise        ratio
\citep{ptak04apj:ngc3998,binder09apj:ngc3226},  X-ray short time-scale
variability   has    been   detected    in   only   a    few   sources
\citep{ptak98apj:variance,awaki01pasj:varliner}, and the disappearance
of  the big  blue bump  in the  UV band  in the  SED of  LINER sources
\citep{ho99sed,ho08aa:review}, all these signatures could indicate the
disappearance of the thin accretion  disk at low luminosities and that
a  different accretion mechanism  is responsible  for the  emission in
LINER sources.   It has  been suggested that  when the  mass accretion
rate falls below a critical value $\dot{M}_{crit}$, the density of the
disk could become too low  for radiative cooling to be effective.  The
trapped   heat  will   expand   the  thin   accretion   disk  into   a
pressure-supported  thick disk  with a  very low  radiative efficiency
\citep[see  ][for  reviews]{quataert01aspc:riaf,narayan08:riafreview}.
Such radiatively inefficient accretion flow (RIAF) models successfully
explained the spectral energy distribution  of a large number of LINER
sources
\citep{ptak04apj:ngc3998,nemmen06apj:ngc1097,nemmen10:lineradaf}.

Another way to assess the geometry  of the accretion mode in AGN is to
compare  them  to  their  less massive  counterparts,  X-ray  binaries
(XRBs).   \citet{shemmer06apj:rqagn} showed  that  the X-ray  spectral
slope, $\Gamma$,  of Seyfert~1 galaxies and quasars  and the Eddington
ratio, $L_{bol}/L_{Edd}$,  are positively correlated,  similar to XRBs
in   their   high/soft   state   \citep[see  also   ][and   references
  therein]{shemmer08apj:gamvsedd}.  Such a behavior could be explained
in  an   AGN  accretion  mode  consistent  with   an  optically  thick
geometrically     thin     accretion     disk     \citep{shakura73aa}.
\citet{gu09mnras:gamVSeddllagn} performed  a similar study  on a broad
sample of  LINERs and  low luminosity Seyfert  galaxies. They  found a
significant anticorrelation  between $\Gamma$ and  the Eddington ratio
for the  local Seyfert galaxies in  their sample analogous  to XRBs in
the   low/hard  state   where   a  RIAF   mode   of  accretion   takes
place. However, no strong  correlation was found when considering only
the LINER sources in their sample, owing, as suggested by the authors,
to heterogeneous fitting models as they have collected their data from
different     studies.     In     a     separate    study,     \citet{
  constantin09ApJ:liners} analyzed  the X-ray emission of  a sample of
107  nearby  galaxies   including  low  luminosity  Seyferts,  LINERs,
transitions (nuclei  with spectra  between Seyferts and  LINERs), H~II
regions,   and  passive   galaxies   (lacking  optical   emission-line
activity), {\sl  none of which  show broad-line components}.   Using a
Spearman-rank  correlation, the authors  found an  anticorrelation for
their   sample  between  $\Gamma$   and  the   $L_{bol}/L_{Edd}$.   By
considering each  class separately,  a spearman-rank test  showed that
the anticorrelation persists for the different objects, except for the
low luminosity Seyfert galaxies.

Finally, broad  optical emission  lines, a characteristic  property of
classical Seyferts and quasars, are also found in nuclei of much lower
luminosities.  Thirty  three sources out of the  221 nuclei classified
as Seyfert,  LINER, or transition  objects in \citet{ho95apjs:optspec}
sample of nearby galaxies show definite detection of a broad H$\alpha$
emission, 16  of those ($\sim$17\%\  of the total pure  LINER sources)
are     LINERs    \citep[noted     as    LINER~1.9     in    ][LINER~1s
  hereinafter]{ho97apjs:broadHal}.  In  this paper, we  are aiming for
the study of the X-ray  characteristics of these LINER~1s observed with
the current generation of  X-ray telescopes, \xmm\ and \chandra.  Such
a sample insures  the responsibility of accretion into  a SMBH for the
formation of the  broad emission lines (given the  early type class of
this sample  where outflows form  massive stars and/or  supernovae are
not expected  to be relevant),  guarantees the non-existence  of large
obscuration, and enables X-ray comparison of this class with both XRBs
and type 1 AGN.  We introduce  our sample in \S~2, \S~3 represents the
observations and  the data  reduction.  Temporal and  spectral results
are given in  \S~4.  In \S~5 we discuss the results  in the context of
LINER~1s-Seyfert-XRB connections, and a conclusion summarizing the main
results is given in \S~6. We  report, in appendix A, some notes on the
individual sources and  in Appendix B we give  spectral results to the
surrounding  sources  around the  centers  of  galaxies observed  with
\chandra.  In the remainder of this paper, luminosities are calculated
using the  distances given in Table~\ref{galaxy-param}  derived with a
Hubble constant $H_{0}=75$~km~s$^{-1}$~Mpc$^{-1}$.

\section{The sample}
\label{sec:sample}

We  selected  objects classified  as  LINER~1.9  (LINER~1s) sources  in
\citet{ho97apjs:broadHal}  showing  a definite  detection  of a  broad
H$\alpha$ emission  line.  This implies  the definite existence  of an
AGN at  the center  of all  of the sixteen  selected galaxies  and its
responsibility  for the  excitation of  the detected  optical emission
lines.

%-----------
% Table  1
%-----------
\begin{table*}[!th]
\caption{Properties of the 13 LINER~1s showing definite detection of a broad H$\alpha$ emission (taken from \citealt{ho97apjs:broadHal} sample).}
\label{galaxy-param}
\newcommand\T{\rule{0pt}{2.6ex}}
\newcommand\B{\rule[-1.2ex]{0pt}{0pt}}
\begin{center}{
\begin{tabular}{c c c c c c}
\hline
\hline
Galaxy Name\T\B & R.A. & Dec. & Hubble Type & Distance$^{a}$ & log($M_{BH}$)$^{b}$  \\
          \T    &      &      &             & (Mpc)          & ($M_{\odot}$)        \\
\hline
\object{NGC-266}  \T& 00 49 47.8 & +32 16 40& SB(rs)ab       & 62.4 & 8.44   \\
\object{NGC-315}  \T& 00 57 48.9 & +30 21 09& E+:            & 65.8 & 9.06   \\
\object{NGC-2681} \T& 08 53 32.7 & +51 18 49& (R')SAB(rs)0/a & 17.2 & 6.78   \\
\object{NGC-2787} \T& 09 19 18.5 & +69 12 12& SB(r)0+        & 7.48 & 8.15   \\
\object{NGC-3226} \T& 10 23 27.0 & +19 53 55& E2:pec         & 23.6 & 8.06   \\
\object{NGC-3718} \T& 11 32 34.8 & +53 04 05& SB(s)a pec     & 17.0 & 7.61   \\
\object{NGC 3998} \T& 11 57 56.1 & +55 27 13& SA(r)0?        & 14.1 & 9.07   \\
\object{NGC 4143} \T& 12 09 36.0 & +42 32 03& SAB(s)0        & 15.9 & 8.18   \\
\object{NGC-4203} \T& 12 15 05.0 & +33 11 50& SAB0-:         & 15.1 & 7.73   \\
\object{NGC-4278} \T& 12 20 06.8 & +29 16 51& E1+            & 16.1 & 8.72   \\
\object{NGC-4750} \T& 12 50 07.2 & +72 52 28& (R)SA(rs)ab    & 26.1 & 7.27   \\
\object{NGC-4772} \T& 12 53 29.1 & +02 10 06& SA(s)a         & 16.3 & 7.46   \\
\object{NGC-5005} \T& 13 10 56.2 & +37 03 33& SAB(rs)bc      & 21.3 & 7.79   \\
\hline
\end{tabular}}
\end{center}
\begin{list}{}{}
\item[{\bf Notes.}]$^{a}$Distances adapted from \citet{tonry01apj:dist}, otherwise  from \citet{tully88agn:dist}. $^{b}$Black hole mass calculated using \citet{graham10:bhmass} updated M-$\sigma$ relation of \citet{termaine02ApJ:Mbh} with stellar velocity dispersion taken from \citet{ho09apjs:veldisp}.
\end{list}
\end{table*}
%-----------
% Table  1
%-----------

We  excluded three  sources from  the sample:  NGC~3642,  NGC~4636 and
NGC~1052.    NGC~3642   did   not   have   any   archived   \xmm\   or
\chandra\ observations.   As for NGC~4636,  all of the  X-ray archived
observations       were      studied      in       extreme      detail
\citep{jones02apj:ngc4636,xu02apj:ngc4636,ohto03pasj:ngc4636,osullivan05apj:ngc4636,baldi09apj:ngc4636,xu10raa:ngc4636}
and show  a complicated  spectrum that requires,  for a  good spectral
parameters  measurement,  detailed  imaging  analysis  and  should  be
modeled including: sophisticated shock models, temperature and density
gradients, and  last but not  least, steep abundance gradients  in the
core.   Finally, the  broad H$\alpha$  emission line  detected  in the
spectrum  of NGC~1052 is  attributed to  polarization due  to electron
scattering   within   the  opening   cone   of   an  obscuring   torus
\citep{barth99apj:ngc1052}.   NGC~1052 was  classified as  an obscured
AGN  showing  a  large  intrinsic  absorption in  the  X-ray  spectrum
\citep[$N_{H}\approx10^{23}$~cm$^{-2}$,][]{guainazzi99mnras:ngc1052}.

Table~\ref{galaxy-param} shows  the list of galaxies  along with their
corresponding right  ascension and declination,  Hubble type, distance
(taken     from      \citealt{tonry01apj:dist},     otherwise     from
\citealt{tully88agn:dist}), the  mass of  the black hole  derived from
the   M-$\sigma$   relation  \citep{termaine02ApJ:Mbh,graham10:bhmass}
where the velocity  dispersion is taken from \citet{ho09apjs:veldisp}.
Multiple snapshot observations (exposure time $\le5$~ks) were excluded
from  the analysis  either  because of  high background  contamination
(NGC~4143,  obs.ID:   0150010201),  low  number   of  counts  detected
(NGC~2787, obs.   ID: 388), or severe pile-up  (NGC~4203 and NGC~4278,
obs.  IDs:397 and 398, respectively).  The final sample consists of 13
LINER~1s   with   a   total    of   31   observations   summarized   in
Table~\ref{obs-param}.

%-----------
% Table  2
%-----------
\begin{table*}[!th]
\caption{Log of the \chandra\ and \xmm\ X-ray observations.}
\label{obs-param}
\newcommand\T{\rule{0pt}{2.6ex}}
\newcommand\B{\rule[-1.2ex]{0pt}{0pt}}
\begin{center}{
\resizebox{\textwidth}{!}{
\begin{tabular}{c c c c c c}
\hline
\hline
\textbf{Source Name} & \textbf{Satellite} \T \B & \textbf{Instrument} & \textbf{Start Date} & \textbf{Obs. ID} & \textbf{Net Exposure-time}\\
                     &                    \T \B &                     &                     &                  & (ks)\\
\hline
NGC-266  & \chandra \T    & ACIS-S & 2001 June 01      & 1610                 & 2.0  \\
NGC-315  & \chandra \T    & ACIS-S & 2000 October 08   & 855                  & 4.7  \\
         & \chandra \T    & ACIS-S & 2003 February 22  & 4156                 & 55.0 \\
         & \xmm     \T    & EPIC   & 2005 July 02      & 0305290201           & 13.2/23.2/23.2$^b$ \\
NGC-2681 & \chandra \T    & ACIS-S & 2001 January 30   & 2060                 & 80.9 \\
         & \chandra \T    & ACIS-S & 2001 May 02       & 2061$^{a}$           & 79.0 \\
NGC-2787 & \chandra \T    & ACIS-S & 2004 May 18       & 4689                 & 30.8 \\
         & \xmm     \T    & EPIC   & 2004 October 10   & 0200250101           & 14.2/28.5/28.0$^b$ \\
NGC-3226 & \chandra \T    & ACIS-S & 1999 December 30  & 860                  & 46.6 \\
         & \chandra \T    & ACIS-S & 2001 march 23     & 1616                 & 2.2 \\
         & \xmm     \T    & EPIC   & 2002 March 29     & 0101040301           & 30.5/36.9/36.9 \\
         & \xmm     \T    & EPIC   & 2008 January 09   & 0400270101           & 94.3$^b$ \\
NGC-3718 & \chandra \T    & ACIS-S & 2003 February 08  & 3993                 & 4.9 \\
         & \xmm     \T    & EPIC   & 2004 May 02       & 0200430501$^{a}$     & 9.6\\
         & \xmm     \T    & EPIC   & 2004 November 04  & 0200431301$^{a}$     & 8.8\\
NGC 3998 & \chandra \T    & ACIS-S & 2006 July 01      & 6781                 & 13.6 \\
         & \xmm     \T    & EPIC   & 2001 May 09       & 0090020101           & 8.9/12.5/12.5 \\
NGC 4143 & \chandra \T    & ACIS-S & 2001 March 26     & 1617                 & 2.5\\
         & \xmm     \T    & EPIC   & 2003 November 22  & 0150010601$^{a}$     & 9.3/11.9/11.9 \\
NGC-4203 & \chandra \T    & ACIS-S & 2009 March 10     & 10535$^{a}$          & 41.6 \\
NGC-4278 & \chandra \T    & ACIS-S & 2005 February 02  & 4741                 & 37.5 \\
         & \chandra \T    & ACIS-S & 2006 March 16     & 7077                 & 110.3 \\
         & \chandra \T    & ACIS-S & 2006 July 25      & 7078                 & 51.4 \\
         & \chandra \T    & ACIS-S & 2006 October 24   & 7079                 & 105.1 \\
         & \chandra \T    & ACIS-S & 2007 February 20  & 7081                 & 110.7\\
         & \chandra \T    & ACIS-S & 2007 April 20     & 7080                 & 55.8 \\
         & \xmm     \T    & EPIC   & 2004 May 23       & 205010101            & 30.3/35.2/35.2\\
NGC-4750 & \chandra \T    & ACIS-S & 2003 August 27    & 4020                 & 4.9 \\
NGC-4772 & \chandra \T    & ACIS-S & 2003 February 14  & 3999$^{a}$           & 4.7 \\
NGC-5005 & \chandra \T    & ACIS-S & 2003 August 19    & 4021                 & 4.9 \\
         & \xmm     \T    & EPIC   & 2002 December 12  & 0110930501           & 8.7/13.1/13.1 \\

\hline
\end{tabular}}}
\end{center}
\begin{list}{}{}
\item[{\bf Notes.}]$^{a}$Observations reported for the first time for the LINER~1 nucleus study. $^{b}$Exposure time corrected for solar flare intervals.
\end{list}
\end{table*}
%-----------
% Table  2
%-----------

\section{X-ray observations and data reduction}
\label{sec:reduction}

\subsection{\chandra\ observations}
\label{chan-obs}

All   of   the   LINER~1s   in   our   sample   have   at   least   one
\chandra\  observation.  Snapshot observations  with an  exposure time
less  than 5~ks were  performed for  eight sources  (NGC~266, NGC~315,
NGC~3226, NGC~3718, NGC~4143, NGC~4750, NGC~4772, and NGC~5005). Seven
sources  have  observations with  a  sufficient  exposure  time for  a
detailed  temporal and  spectral study  (NGC~315,  NGC~2681, NGC~2787,
NGC~3226,   NGC~3998,   NGC~4203,   and   NGC~4278).    All   of   the
\chandra\  observations  were obtained  with  the spectroscopic  array
\citep[ACIS-S;][]{weisskopf02PASP} where the nucleus was placed on the
aim point, except for  NGC~3226, of the ACIS-S3 back-illuminated chip.
They were taken  in either Faint or Very Faint  mode to increase their
sensitivity.   All of  the  observations are  \chandra\ archival  data
obtained                                                           from
chaser\footnote{\label{chaser}http://cda.harvard.edu/chaser/Descriptions.}.
The log of the \chandra\ observations are listed in Table~2.

All \chandra\ observations were  reduced and analyzed in a systematic,
homogeneous  way  \citep[as in][hereinafter  Y10]{younes10aa:ngc4278}
using  the CIAO  software package  version 4.2,  \chandra\ Calibration
Database,  CALDB, version 4.3.1,  and the  ACIS Extract  (AE) software
package version 3.175  \footnote{\label{AEnote} The {\em ACIS Extract}
software    package    and    User's    Guide   are    available    at
http://www.astro.psu.edu/xray/acis/acis\_analysis.html.}
\citep{broos2010AE}.   We started  by  using the  level  1 event  file
produced by the \chandra\ X-ray  Center (CXC) to suppress the position
randomization  applied  by  the  CXC  Standard  Data  Processing  when
creating a level  2 event file.  We also corrected  for the effects of
charge-transfer  inefficiency  on   event  energies  and  grades.   We
filtered for bad event  grades (only ASCA grades 0, 2, 3,  4 and 6 are
accepted) and hot columns to  take account of several potential issues
such as cosmic  rays and bad pixels. Good  time intervals, supplied by
the pipeline, were applied to the final products.

The  LINER  nucleus source  position  is  determined  after running  a
wavelet  transform detection  algorithm, the  {\sl  wavdetect} program
within the CIAO  data analysis system \citep{wavdetect:freeman02apjs}.
This  position is  then  given to  the  AE software  that refines  it,
extract source  photons, construct local  backgrounds, extract source,
and background spectra, compute redistribution matrix files (RMFs) and
auxiliary response files (ARFs), by  spawning the {\sl mkarf} and {\sl
  mkacisrmf}  routines  of CIAO,  and  perform  spectral grouping  and
fitting.

Source  events are  extracted  around the  source  centroid, inside  a
polygonal     shape    of    the     local    PSF,     generated    by
MARX\footnote{http://space.mit.edu/ASC/MARX/} at  the energy 1.497 keV
using  the {\sl  ae\_make\_psf}  tool implemented  in AE.   Background
region is  defined following the AE procedure.   The background region
is an annular  region centered on the source  position where the inner
radius delimit 1.1~$\times$~99\% encircled energy radius and the outer
radius is set such that the background includes between 100 counts and
200  counts  (depending  on  the  brightness  of  the  source).   This
background is  obtained from a  special image where all  events within
the $\sim1.1\times99$\%  PSF circles of  all the sources in  the field
were excluded ({\sl swiss  cheese image}).  Background was modeled for
snapshot observations.

Piled-up observations were accounted for  by excluding the core of the
PSF  (see Y10 for  more details).   We use  the tool  {\sl dmextract},
called by  the AE  software, to create  spectra over the  energy range
0.5--8~keV.  We used the tool {\sl ae\_group\_spectrum} implemented in
AE to group  the spectra.  Channels between 0.5 and  8 keV are grouped
to  have  a three  sigma  ($3\sigma$)  signal  to noise  ratio,  which
corresponds to  a minimum of 20 counts  per bin, to enable  the use of
the  $\chi^{2}$  statistics  in   the  spectral  analysis.   The  cash
statistic (C-stat) is used  to derive spectral parameters for snapshot
\chandra\ observations with the background being modeled with the {\sl
  cplinear}  background model  developed by  \citet{broos2010AE}.  The
background  model  is  arbitrarily  chosen to  consist  of  continuous
piecewise-linear ({\sl cplinear}) functions with 2 to 10 vertexes. The
model  has 2 to  10 parameters  representing the  X-ray fluxes  at the
different vertexes. These  vertexes are placed on the  energy scale so
that they  divide the energy  range into intervals  with approximately
equal numbers of observed counts in the background spectrum (0.1 to 10
keV).  Vertex  energies are  chosen to coincide  with the  energies of
actual events in the background which helps to prevent the vertex flux
from being driven to the hard limit of zero during the fitting process
\citep[see \S~7.5 of ][]{broos2010AE}.

\subsection{\xmm\ observations}
\label{xmmobs}

The log of the \xmm\ observations is listed in Table~2.  Eight sources
were observed  at least once with \xmm\  (NGC~315, NGC~2787, NGC~3226,
NGC~3718,  NGC~3998, NGC~4143,  NGC~4278, and  NGC~5005) and  two have
multiple  observations  (NGC~3226  and   NGC~3718).   In  all  of  the
observations,     the    EPIC-pn    \citep{struder01aa}     and    MOS
\citep{turner01aa} cameras were operated  in Imaging, Prime Full Frame
or Large Window Mode (except  for the long NGC~3226 observation, where
MOS cameras  were operating in Small  Window Mode\footnote{NGC~3226 is
  off axis during the long  observation and hence is not detected with
  the MOS  cameras when operating in  a Small Window  Mode}) using the
thin or medium  filter.  The Reflection Grating Spectra  show only few
counts for all the different  observations and therefore they were not
included in our analysis.  We did  not make use of the optical/UV data
taken with the optical/UV monitor (OM) instrument \citep{mason01aa:om}
since  this paper concentrates  on the  X-ray characteristics  of this
sample.  A  multiwavelength study of our  sample will be  treated in a
forthcoming  paper.    All  data  products  were   obtained  from  the
XMM-Newton                       Science                       Archive
(XSA)\footnote{http://xmm.esac.esa.int/xsa/index.shtml}   and  reduced
using  the  Science  Analysis  System  (SAS) version  9.0.   Data  are
selected  using  event  patterns  0--4  and  0--12  for  pn  and  MOS,
respectively, during  only good X-ray events  (``FLAG$=$0'').  None of
the  EPIC  observations  were  affected by  pile-up,  although  severe
intervals of enhanced solar  activity, where the background count rate
even  exceeds  the source  count  rate,  were  present during  several
observations (NGC~315, NGC2787, NGC~3226).  In these cases, we reduced
the  background to $5\%$  by excluding  the high  background intervals
which  reduces  the observation  time  usable  for spectral  analysis,
sometimes to less than 30$\%$ of the raw exposure time (NGC~315).

\xmm\ source events of all of the LINER~1s in our sample were extracted
from a  circle centered  at the nucleus  using two different  radii of
10\arcsec\ and 25\arcsec\footnote{The lower 10\arcsec\ limit was taken
  so to encircle  at least 50\%\ of the EPIC  \xmm\ PSF.}. We compared
light curves and spectra of both extraction regions to check if any of
the  sources  (jet   emission,  diffuse  emission,  and/or  unresolved
point-like   sources)  detected   in  the   \chandra\   image  between
10\arcsec\   and  25\arcsec\  around   the  nucleus   contaminate  the
\xmm\  nucleus emission  (see  Appendix~B and  online Fig.~1--4).   No
change is seen in the light  curves, and the fit parameters of the two
extracted spectra were consistent,  within the error bars.  Therefore,
and to achieve better  statistics and better constrain fit parameters,
source  events of  all  of the  LINER~1s  in our  sample observed  with
\xmm\  were taken  from the  25\arcsec-radius circle  centered  on the
nucleus.  We added the  spectral contribution of the different sources
(jet   emission,  diffuse   emission,  and/or   unresolved  point-like
sources),  derived from the  \chandra\ observation  and detected  in a
25\arcsec-radius circle  around the  nucleus (see Appendix~B  for more
details), to the  spectral model used to fit  the \xmm\ spectrum.  The
particular case of NGC~4278 is discussed in detail in Y10.  Background
events are extracted from a source--free circle with a radius twice of
the source on the same  CCD.  We generated response matrix files using
the  SAS  task  {\sl  rmfgen},  while ancillary  response  files  were
generated  using the  SAS task  {\sl arfgen}.   The EPIC  spectra were
created  in   the  energy  range   0.5--10~keV  to  enable   flux  and
model--parameter comparison with \chandra\ spectra.  They were grouped
to have a signal  to noise ratio of 3 with a  minimum of 20 counts per
bin to allow the use of the $\chi^2$ statistic.

\onlfig{1}{
\begin{figure}[]
\centerline{\includegraphics[angle=0,width=0.47\textwidth]{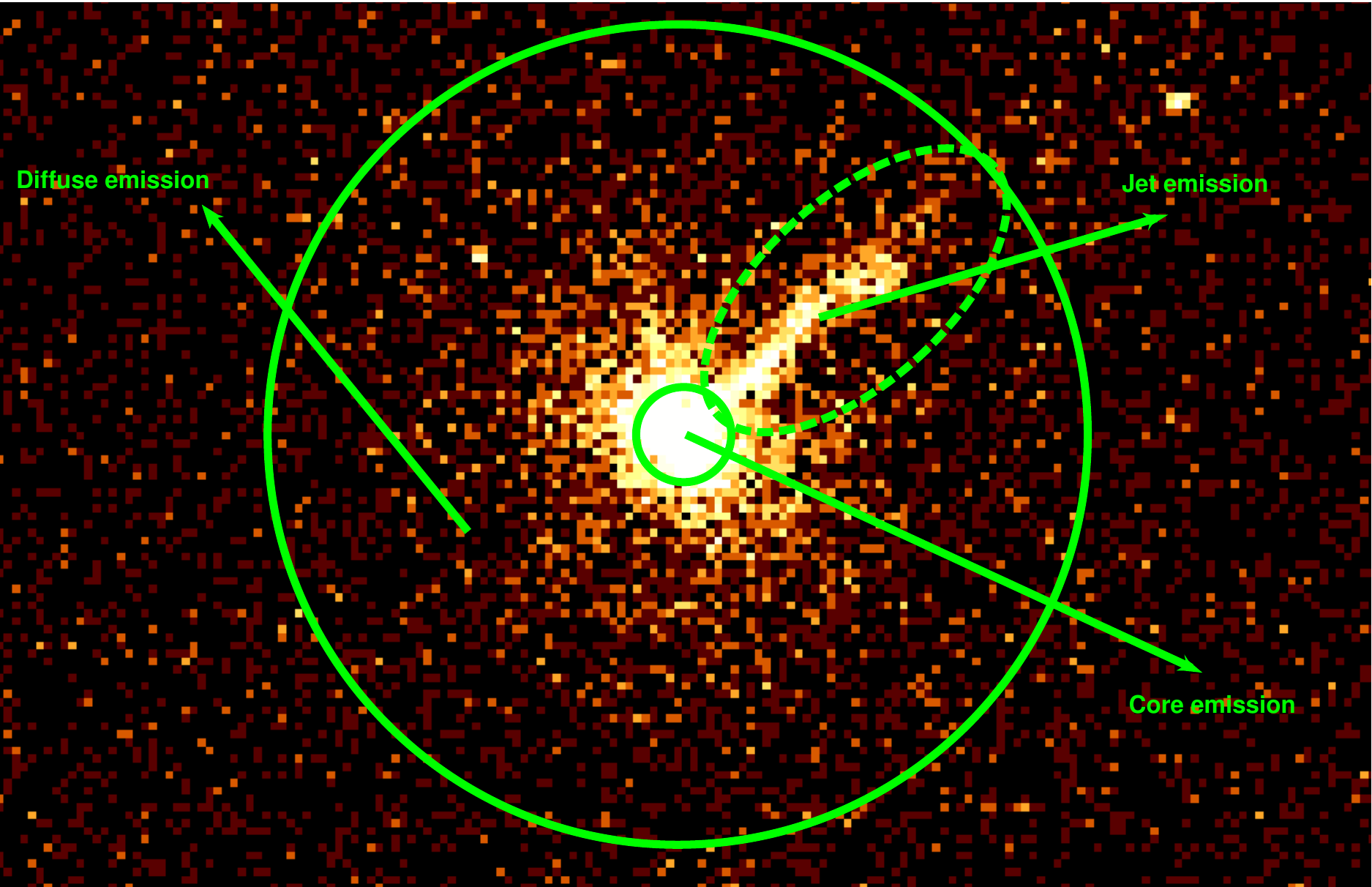}}
\caption{\chandra\ image of the central 25\arcsec\ of NGC~315. Jet spectrum is extracted from an ellipse with a semi-major axis of 11.3\arcsec\ and a semi-minor axis of 5.6 \arcsec. Core emission is extracted from a circle centered on the source with a radius comprising 99\%\ of the PSF ($\sim$2.7\arcsec). The rest of the medium inside a 25\arcsec\ circle is considered diffuse emission. See Appendix~B for more details.}
\label{chanim1}
\end{figure}
}

\onlfig{2}{
\begin{figure}[]
\centerline{\includegraphics[angle=0,width=0.47\textwidth]{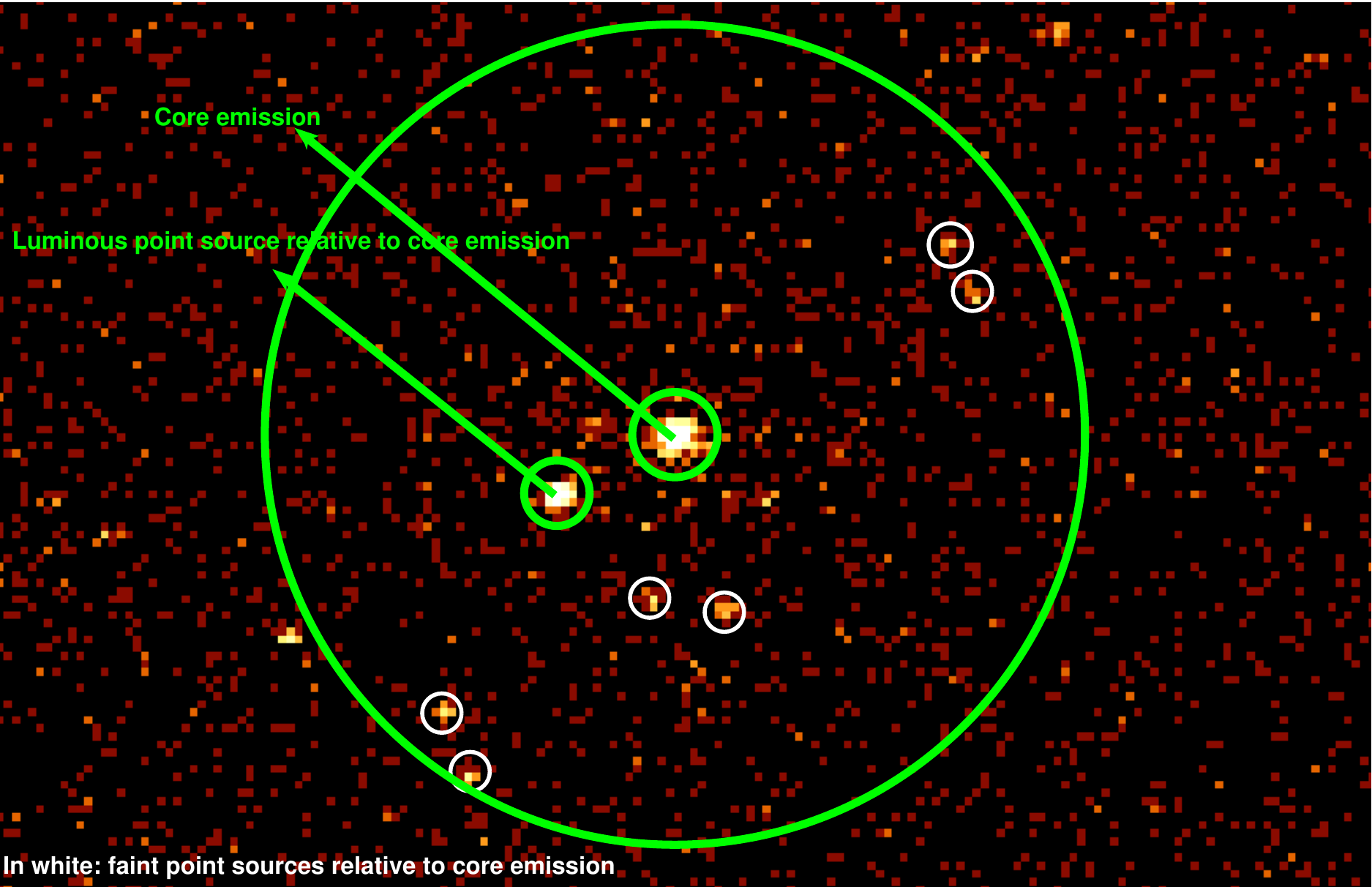}}
\caption{\chandra\ image of the central 25\arcsec\ of NGC~2787. A point-like source south-east of the central LINER is present with a luminosity comparable to the core luminosity. Another six point-like sources, marked in white, are present in the field. See Appendix~B for more details.}
\label{chanim2}
\end{figure}
}

\onlfig{3}{
\begin{figure}[]
\centerline{\includegraphics[angle=0,width=0.47\textwidth]{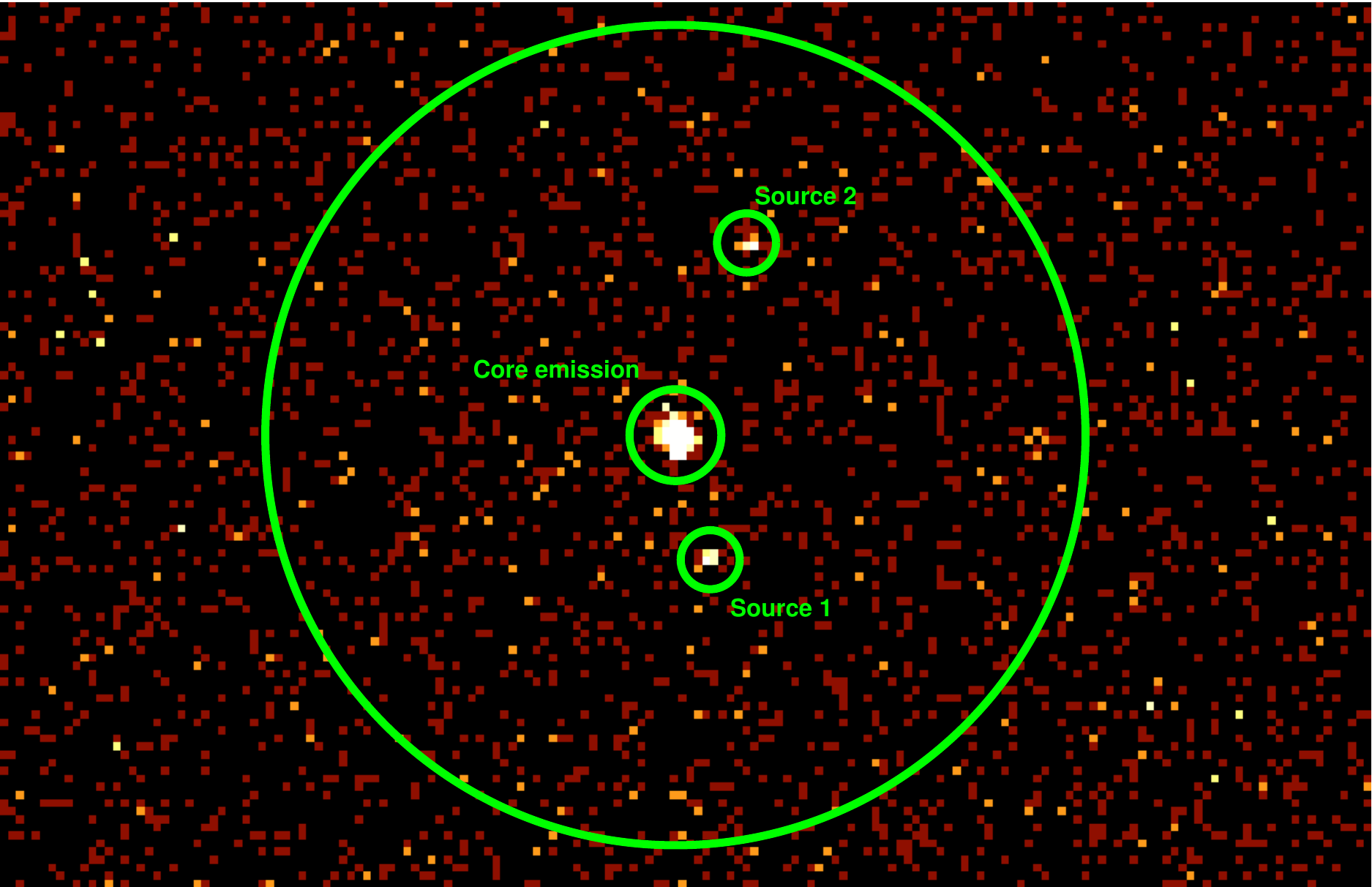}}
\caption{\chandra\ image of the central 25\arcsec\ of NGC~3226. Two sources are present in the field marked source~1 and source~2. See Appendix~B for more details.}
\label{chanim3}
\end{figure}
}

\onlfig{4}{
\begin{figure}[]
\centerline{\includegraphics[angle=0,width=0.47\textwidth]{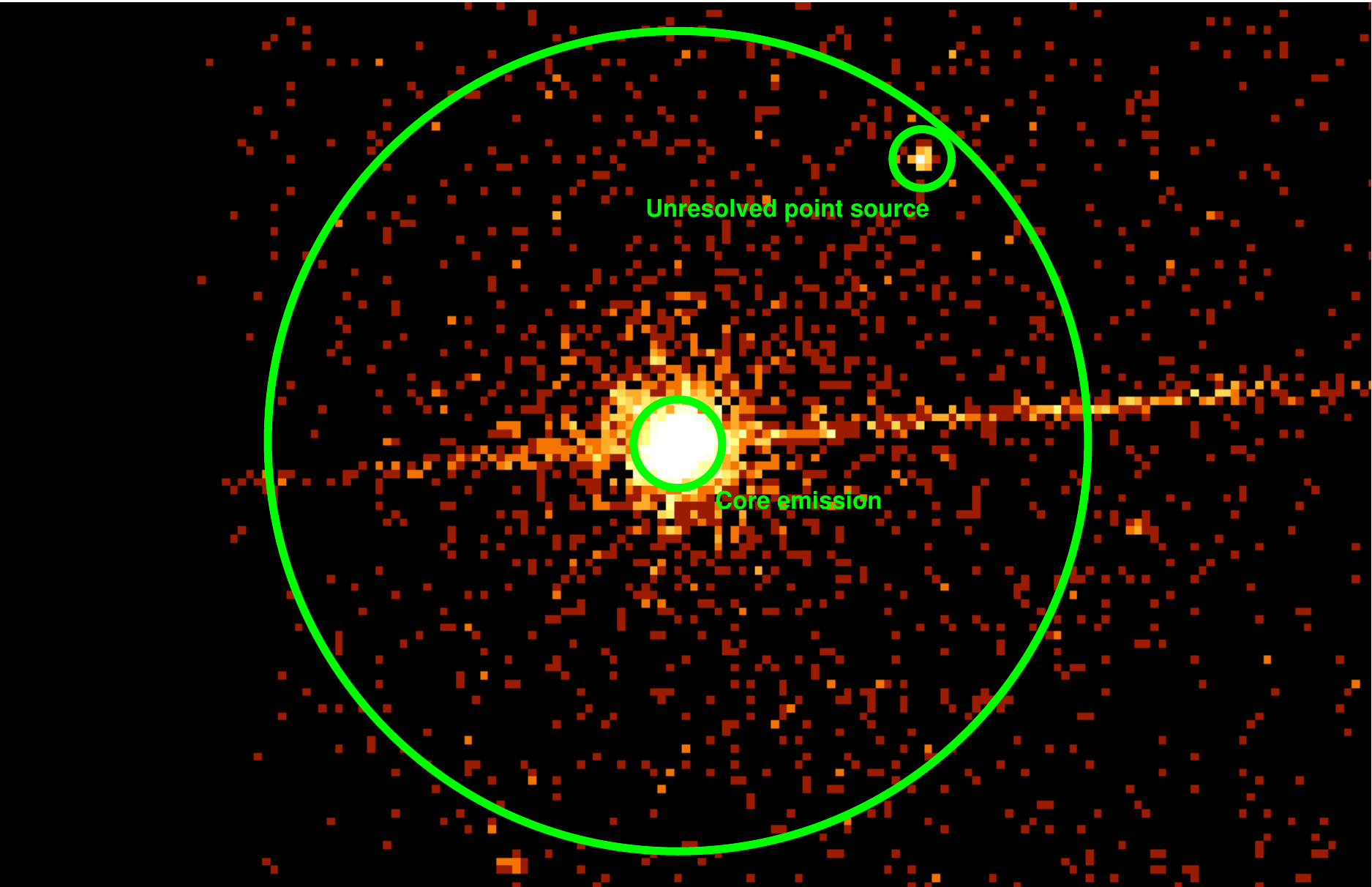}}
\caption{\chandra\ image of the central 25\arcsec\ of NGC~3998. Only one source, much fainter than the core, is present in the field. See Appendix~B for more details. The horizontal bright line corresponds to the readout streak events.}
\label{chanim4}
\end{figure}
}

\section{Results}
\label{resultsec}

\subsection{Light curves and hardness ratios}
\label{lightcurvesec}

Temporal analysis  was only done  for long exposure  observations, not
including   snapshot  \chandra\   observations.    Light  curves   and
corresponding hardness ratios,  defined as $HR=(H-S)/(H+S)$, where $S$
is the count rate in the soft 0.5-2~keV band and $H$ is the count rate
in  the  hard  2-10~keV band,  were  extracted  for  all of  the  long
observations.  We corrected the net count rate of the piled-up sources
for  the excluded  fraction of  the  PSF.  \chandra\  and \xmm\  light
curves were all binned with a time bin size of 1~ks for a reliable rms
variability analysis.

\begin{figure}[]
\includegraphics[angle=0,width=0.5\textwidth]{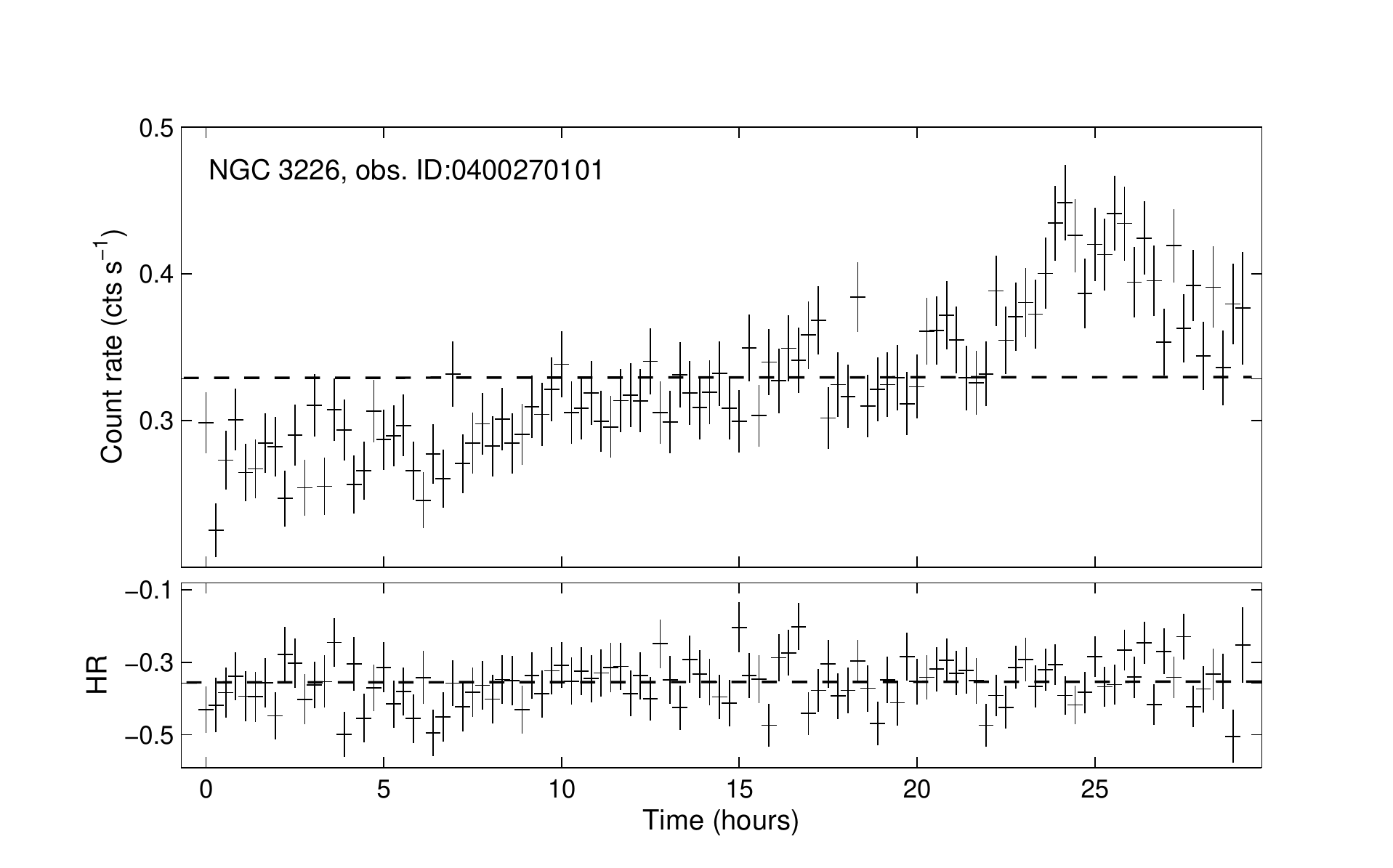}
\caption{Light curve ({\sl upper panel}) and hardness ratio ({\sl lower panel}) of the $\sim$100~ks \xmm\ observation of NGC~3226 binned to have a 1~ks resolution. The dashed lines show the averages on the count rate and hardness ratio.}
\label{light-hardness}
\end{figure}

We  first conducted  a Kolmogorov-Smirnov,  K-S, test  to  examine any
potential variability within each observation.  Based on this test, we
do  not find  short time-scale  (hours  to days)  variability in  5/10
sources (NGC~315,  NGC~2681, NGC~3718, NGC~3998, and  NGC~5005) with a
K-S test  probability $>$10$\%$  that the nuclear  emission originates
from  a  constant  source.   Three sources  (NGC~2787,  NGC~4143,  and
NGC~4203) indicate a possible  variability with a K-S test probability
between  4\%\ and  2\%\ that  the nuclear  emission originates  from a
constant  source.  Two  \xmm\  observations of  two different  sources
exhibit  significant  short   time-scale  variability,  both  already
reported in  the literature, NGC~4278  (obs.  ID: 205010101,  Y10) and
NGC~3226  \citep[obs.ID:  0400270101, ][]{binder09apj:ngc3226},  where
the  K-S  test gives  a  probability less  than  $1\%$  that the  core
emission  is originating from  a constant  source.  NGC~4278  shows an
increase at the beginning of the observation of 10$\%$ on a time-scale
of $\sim$1.5~hour, the  emission remains constant for the  rest of the
observation  following that  hint  of variability.   As for  NGC~3226,
variability  is clear  through  the whole  observation  where a  total
increase of $\sim$60$\%$ is detected between the beginning and the end
of    the   $\sim$100~ks    observation   \citep{binder09apj:ngc3226}.
\xmm\ light  curve of  NGC~3226 is shown  in Fig.~\ref{light-hardness}
and  the \xmm\ and  \chandra\ light  curves of  the other  sources are
given       in        online       Fig.~\ref{LCxmmallLINERs}       and
Fig.~\ref{LCchandraallLINERs}.

\onlfig{6}{
\begin{figure*}[]
\begin{center}
\includegraphics[height=.19\textheight,angle=0,width=.33\textheight]{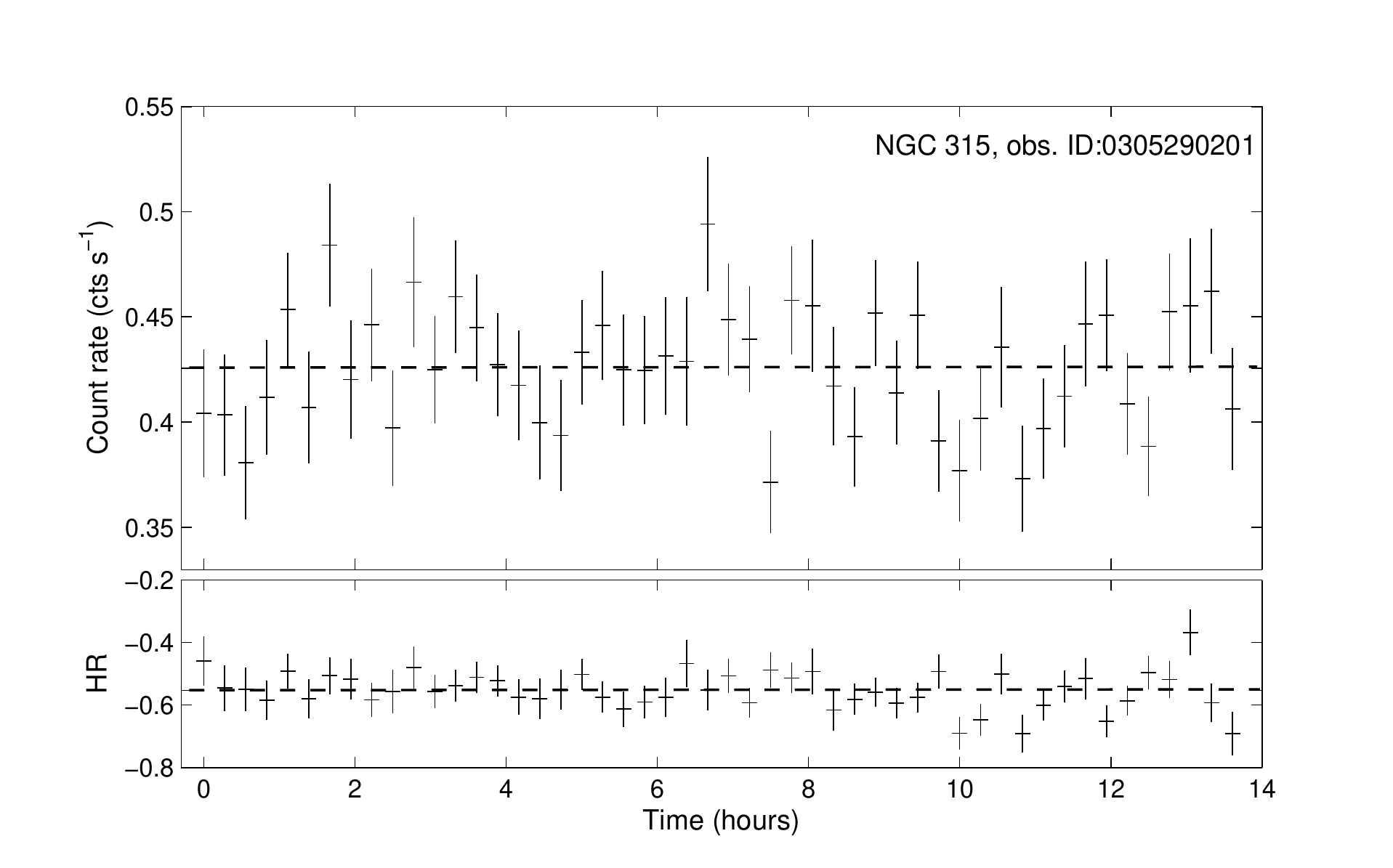}
\includegraphics[height=.19\textheight,angle=0,width=.33\textheight]{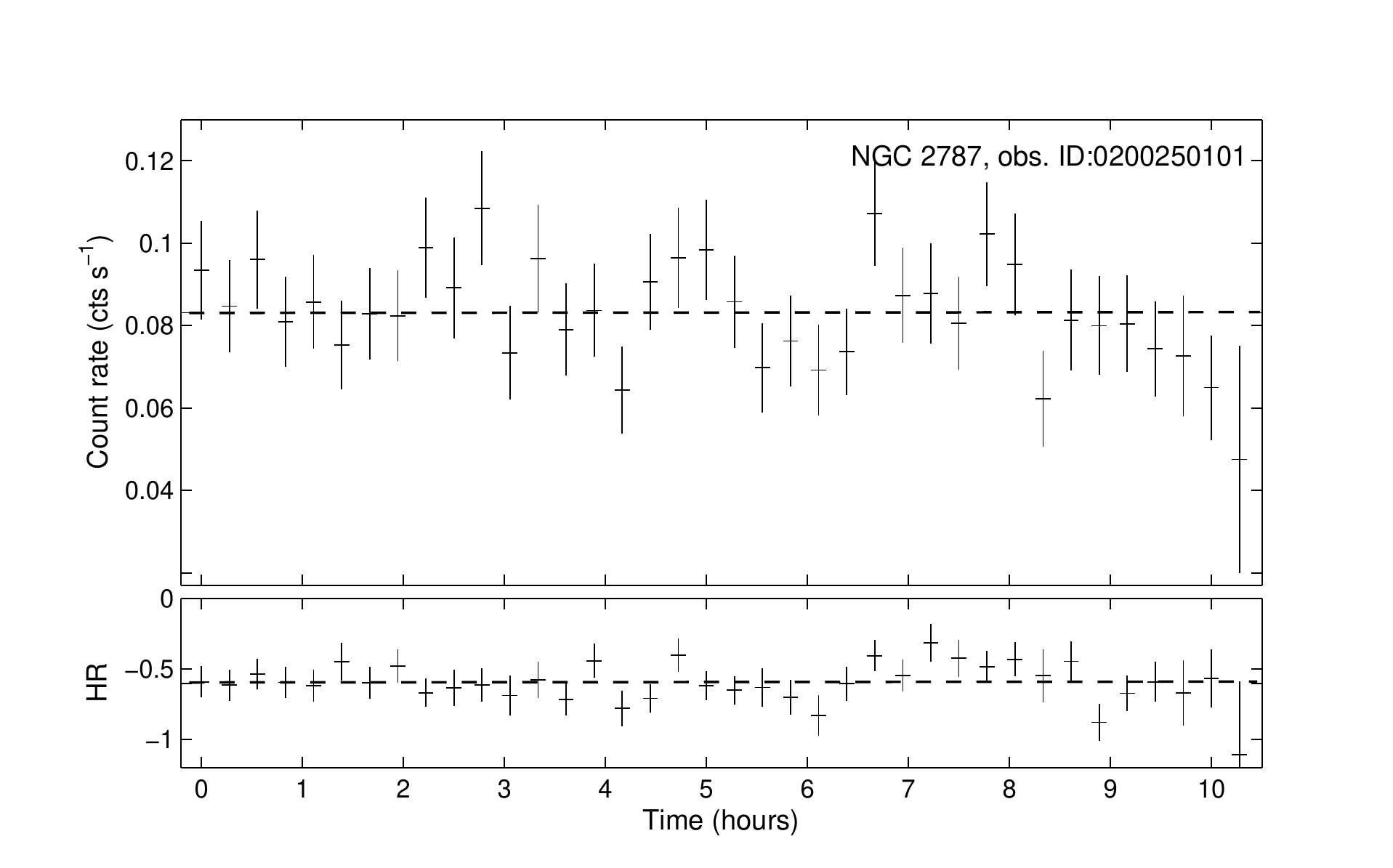}\\
\includegraphics[height=.19\textheight,angle=0,width=.33\textheight]{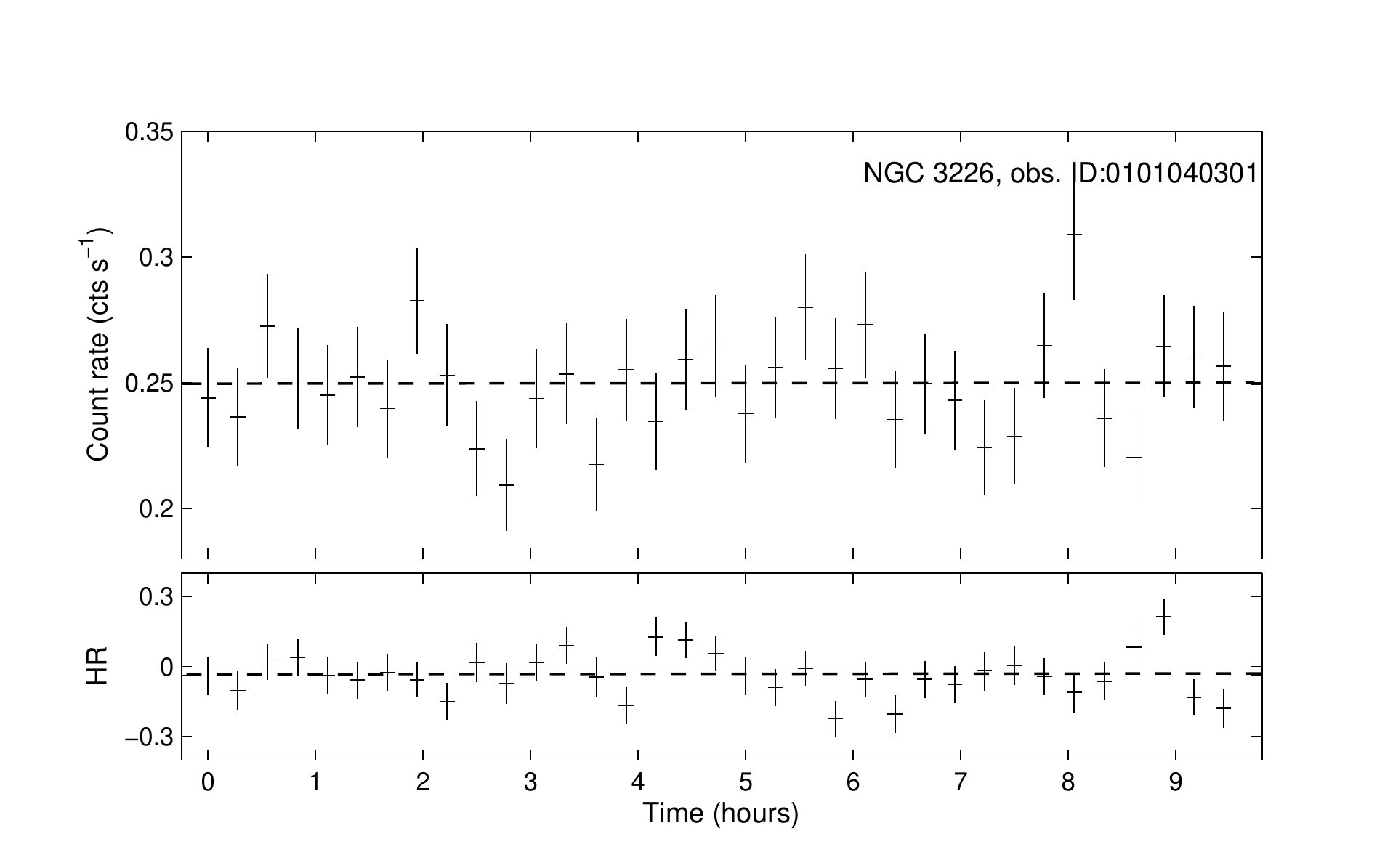}
\includegraphics[height=.19\textheight,angle=0,width=.33\textheight]{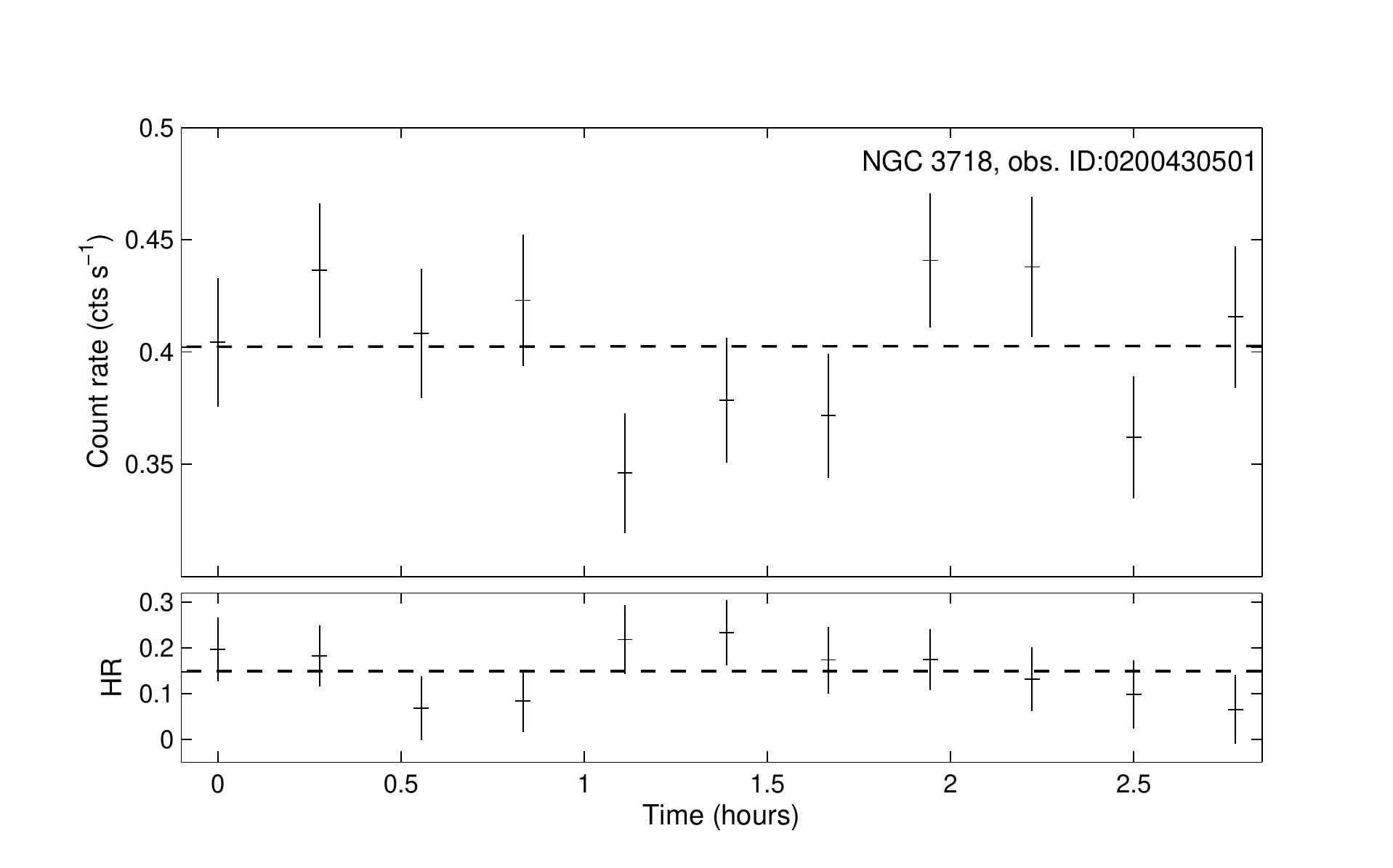}\\
\includegraphics[height=.19\textheight,angle=0,width=.33\textheight]{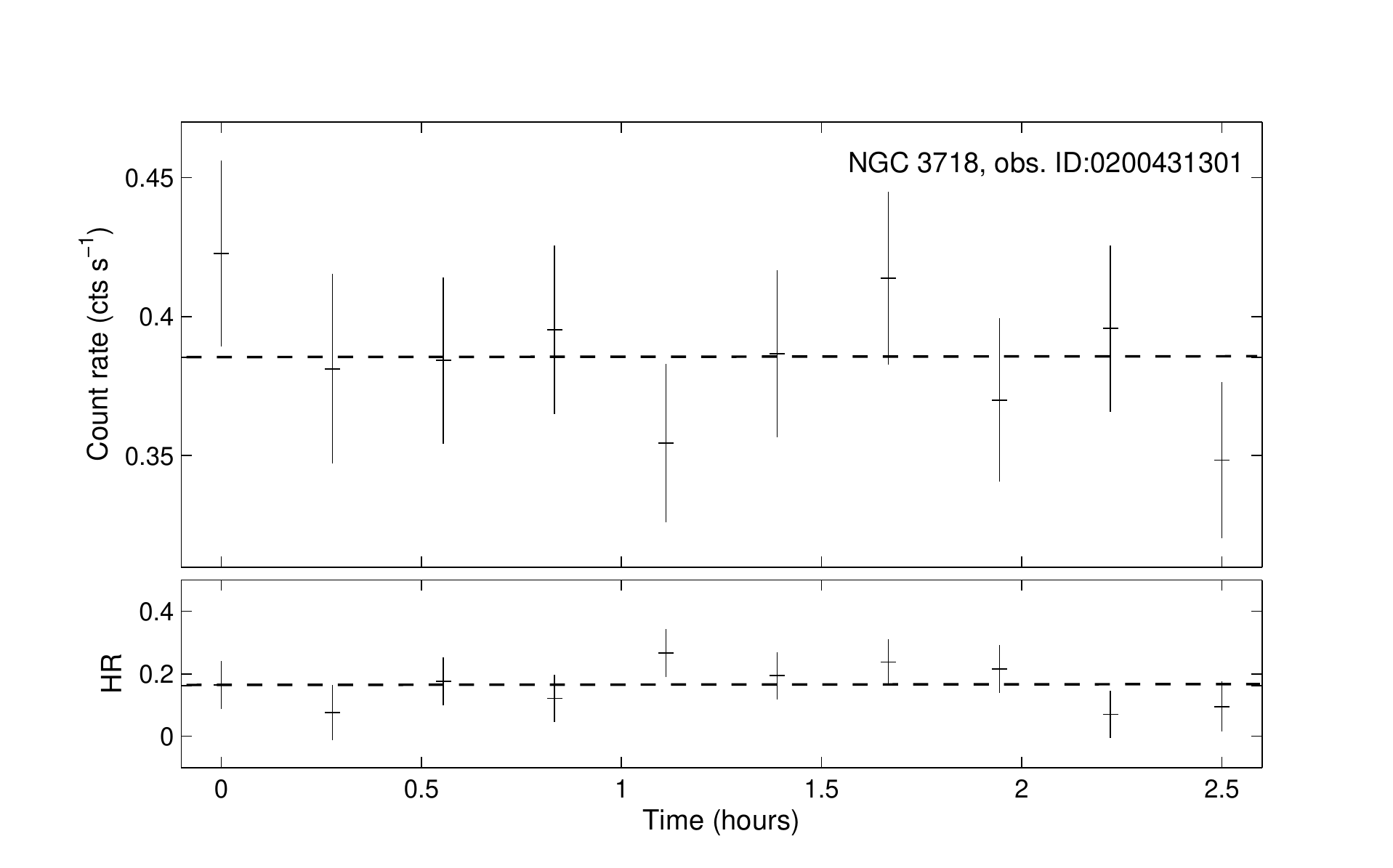}
\includegraphics[height=.19\textheight,angle=0,width=.33\textheight]{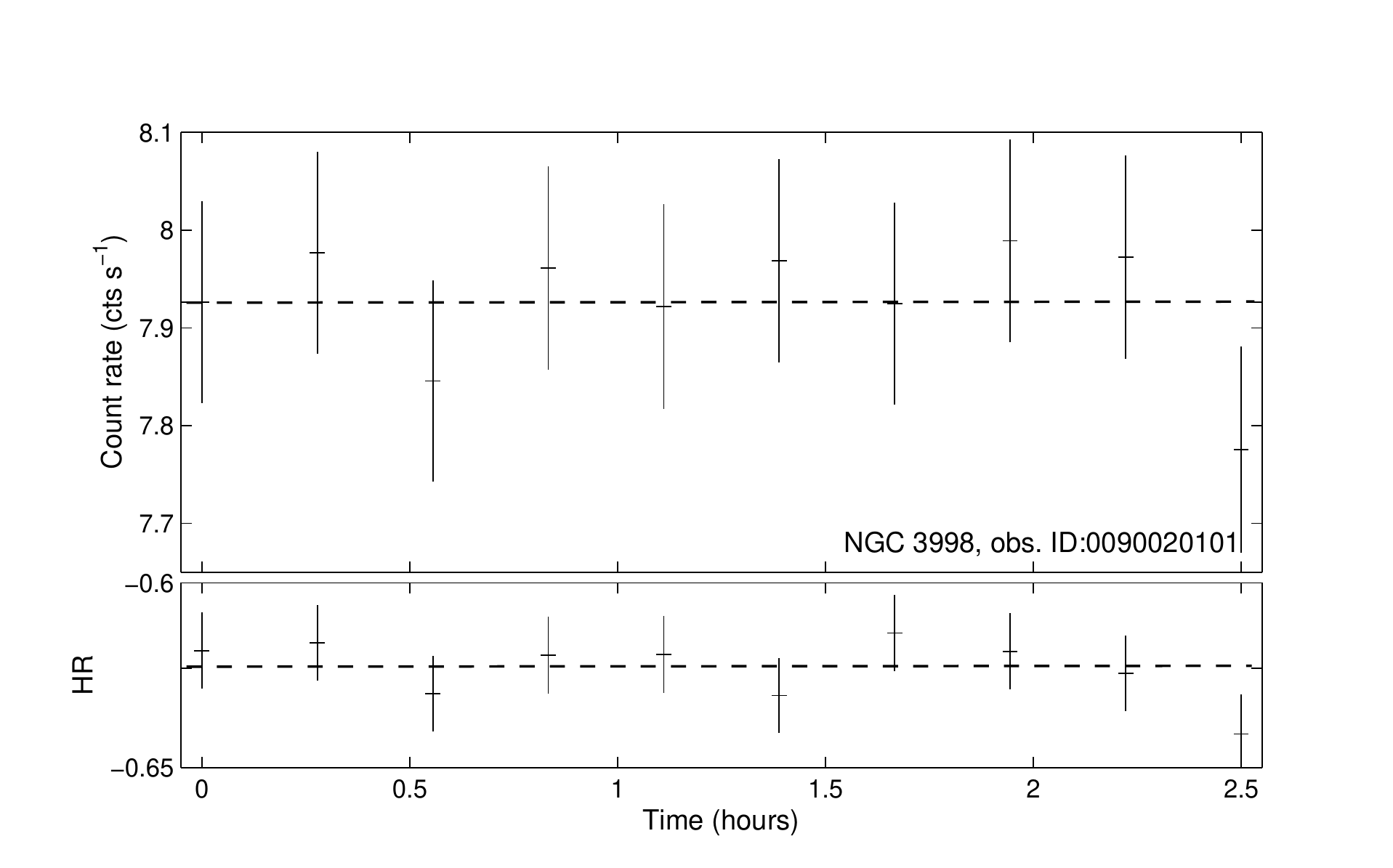}\\
\includegraphics[height=.19\textheight,angle=0,width=.33\textheight]{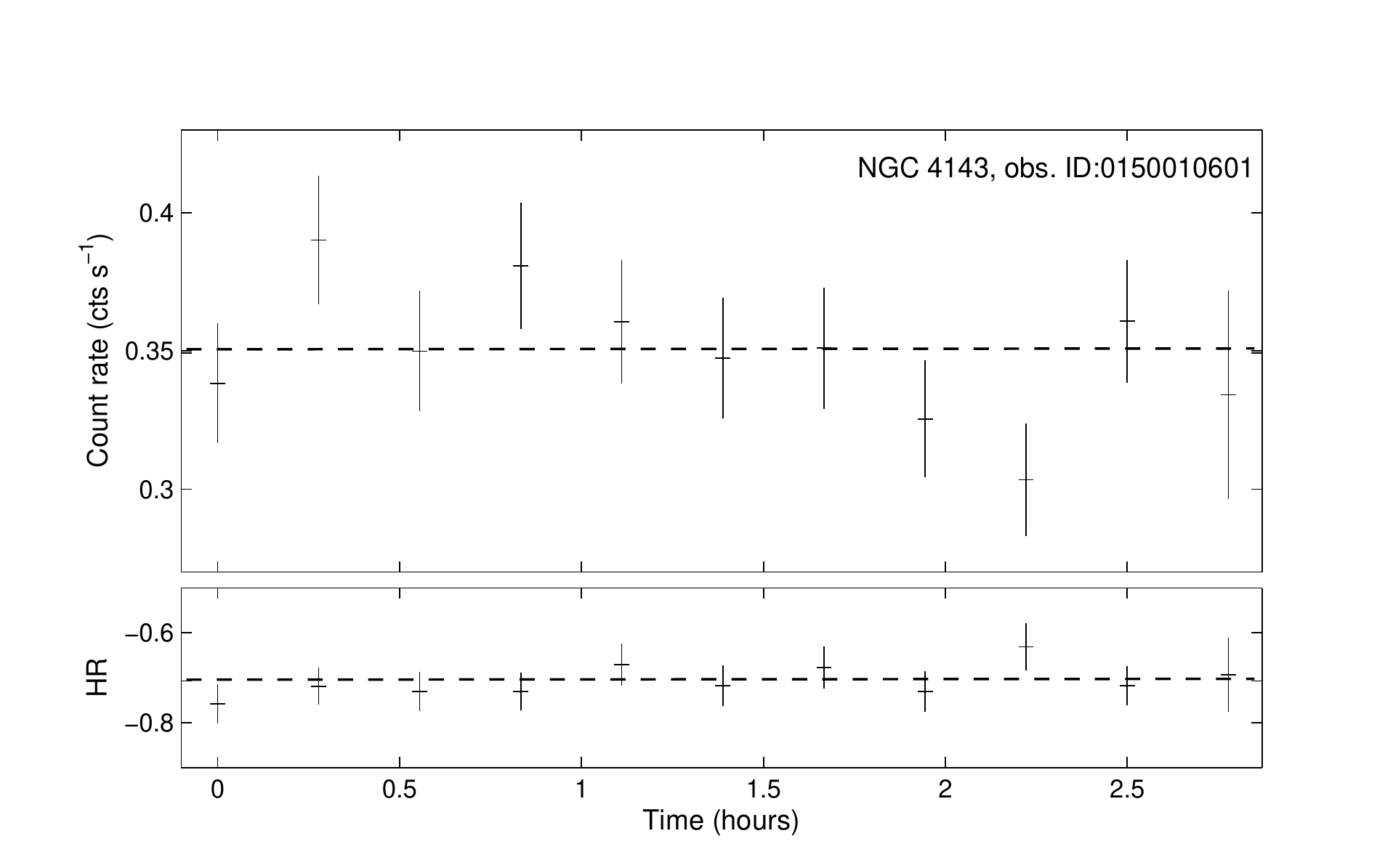}
\includegraphics[height=.19\textheight,angle=0,width=.33\textheight]{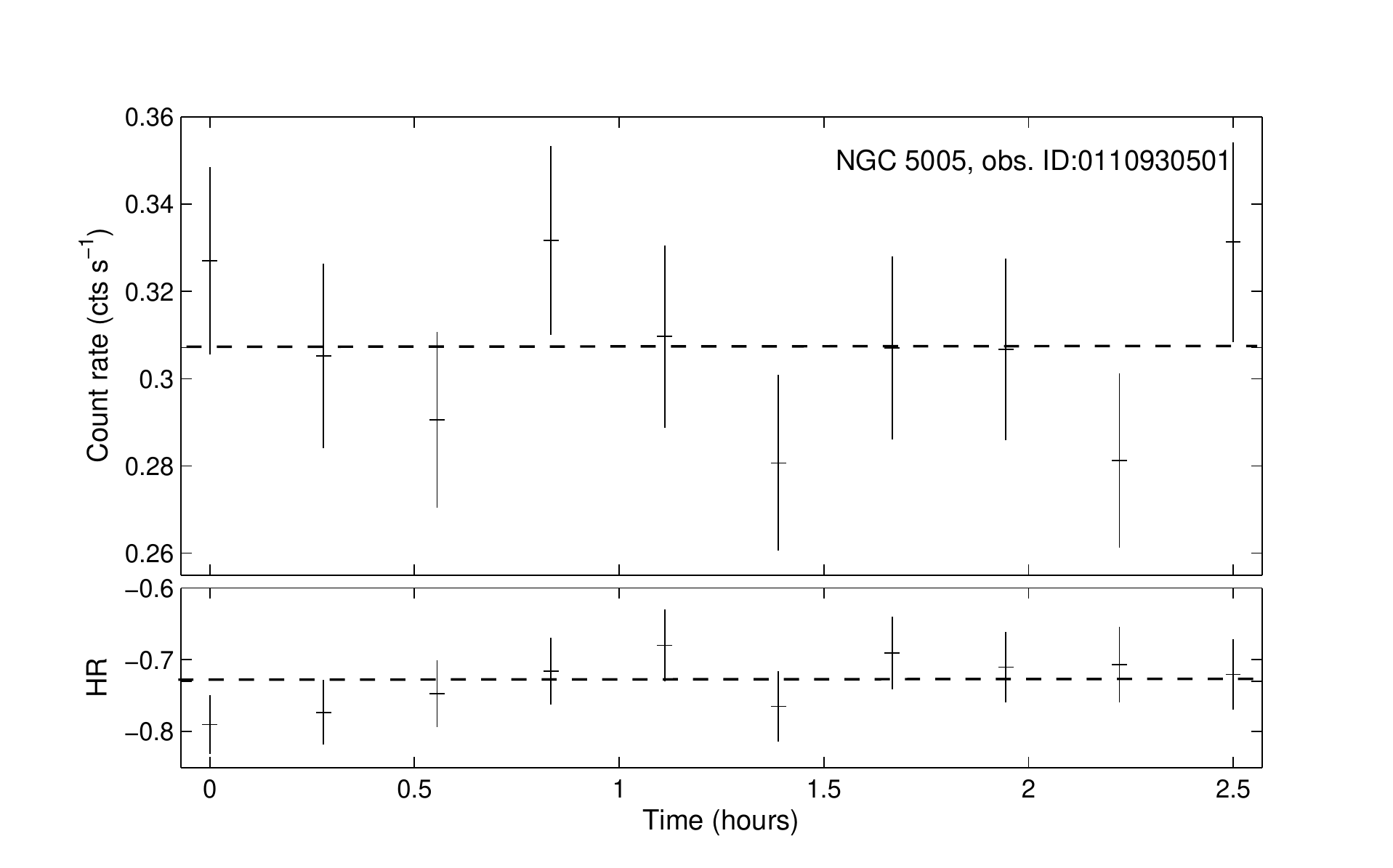}
\caption{Light curves and hardness ratios of the LINER~1s observed with \xmm, all binned with a 1~ks time bin-size.}
\label{LCxmmallLINERs}
\end{center}
\end{figure*}
}

\onlfig{7}{
\begin{figure*}[]
\begin{center}
\includegraphics[height=.19\textheight,angle=0,width=.33\textheight]{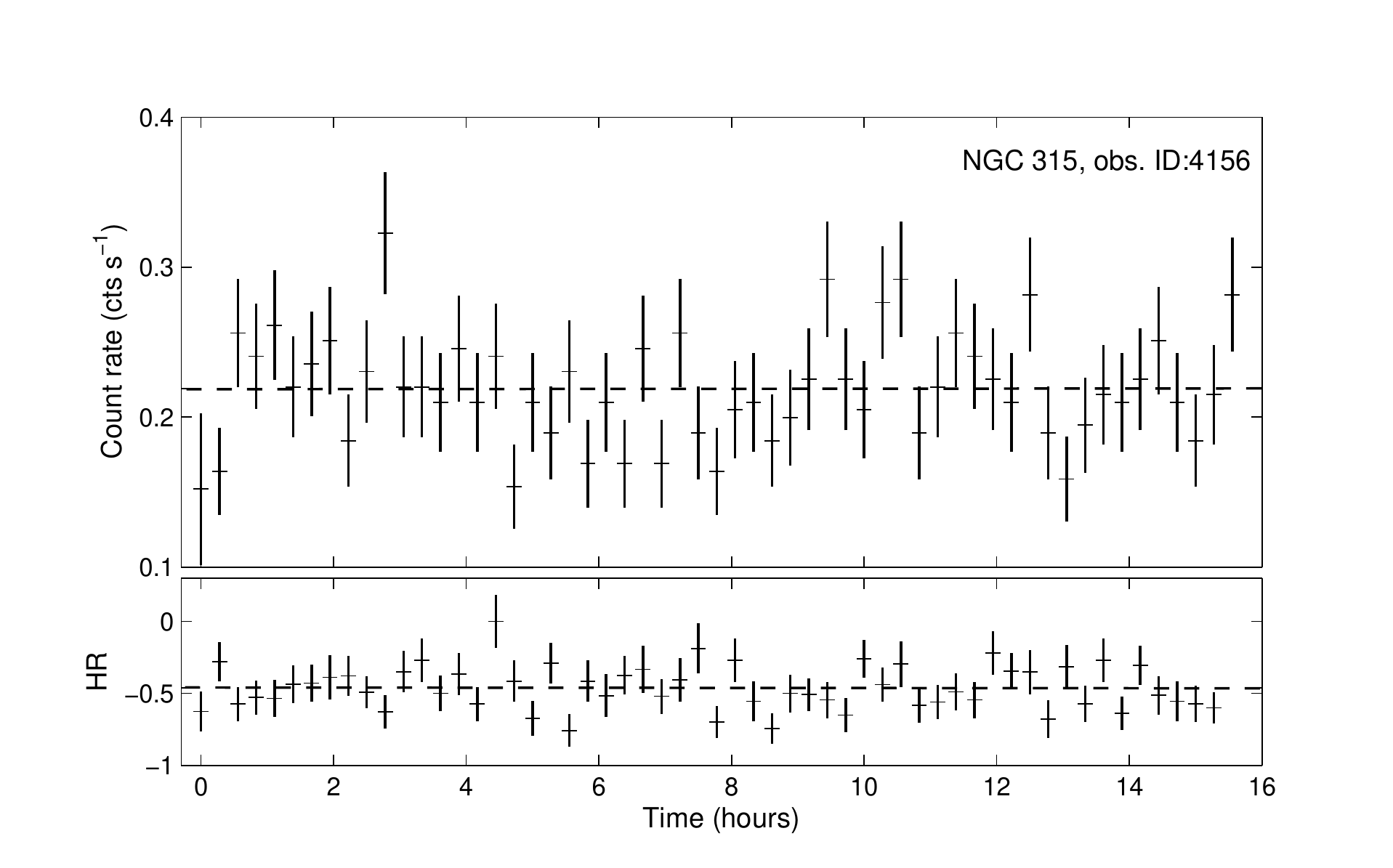}
\includegraphics[height=.19\textheight,angle=0,width=.33\textheight]{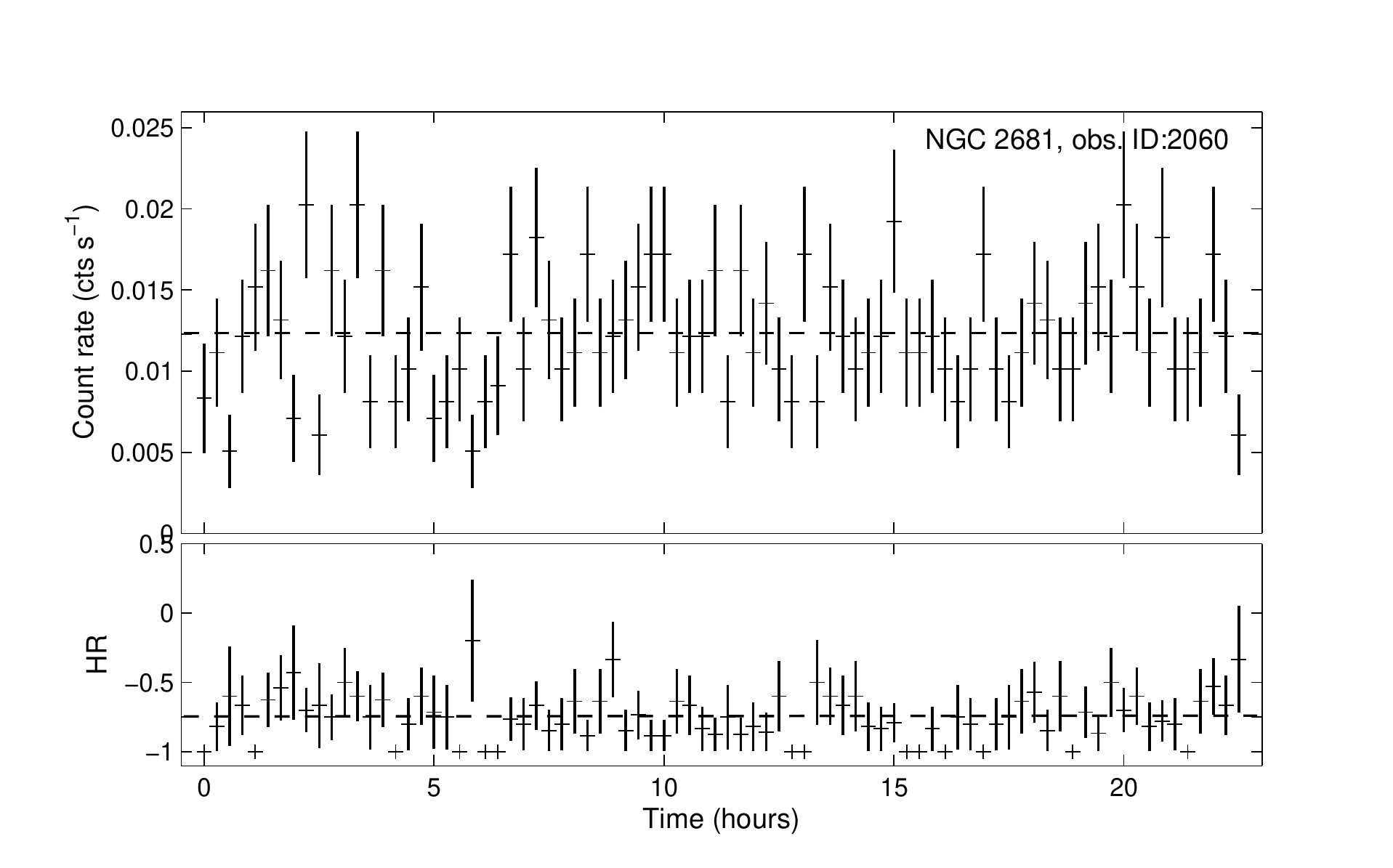}\\
\includegraphics[height=.19\textheight,angle=0,width=.33\textheight]{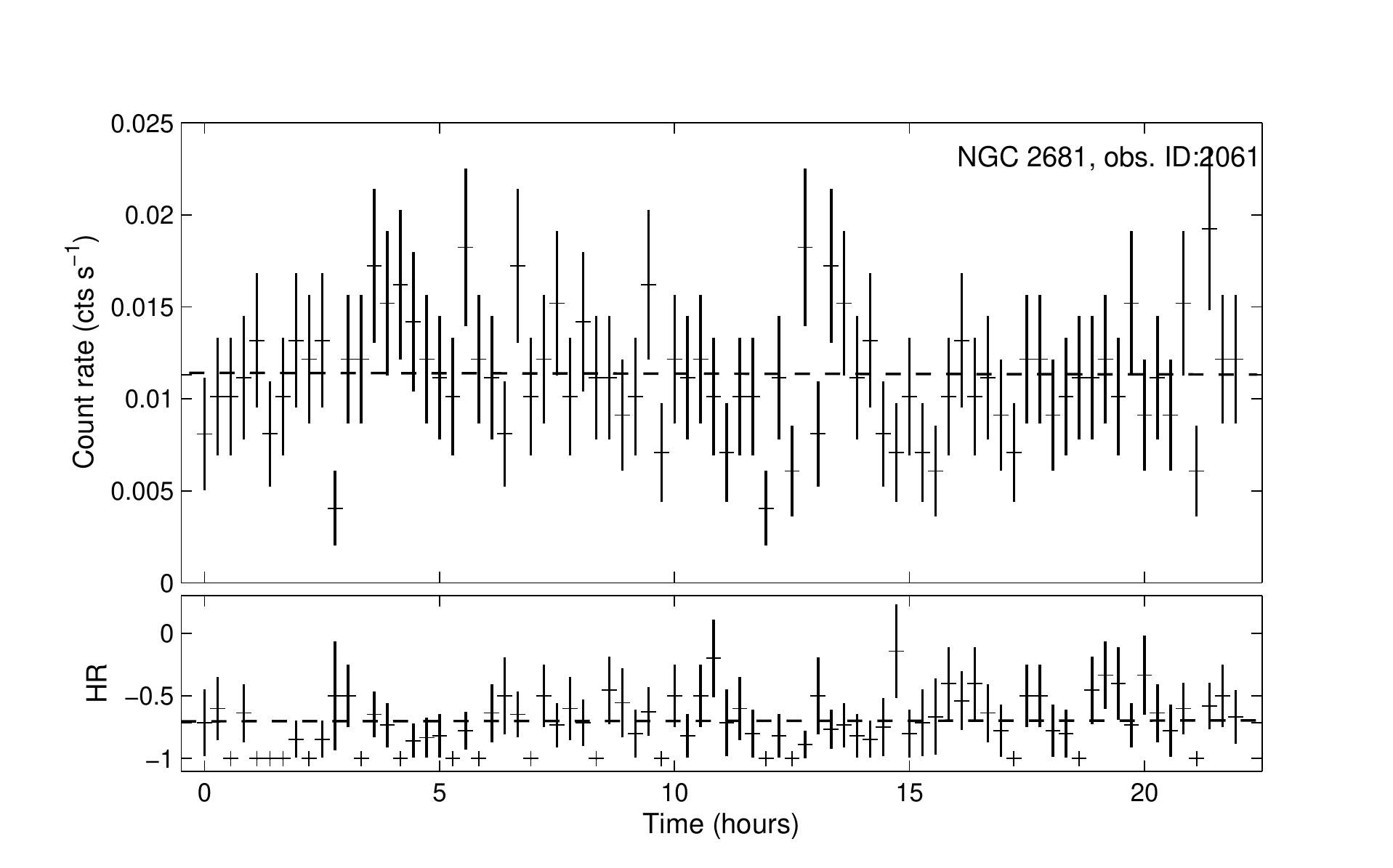}
\includegraphics[height=.19\textheight,angle=0,width=.33\textheight]{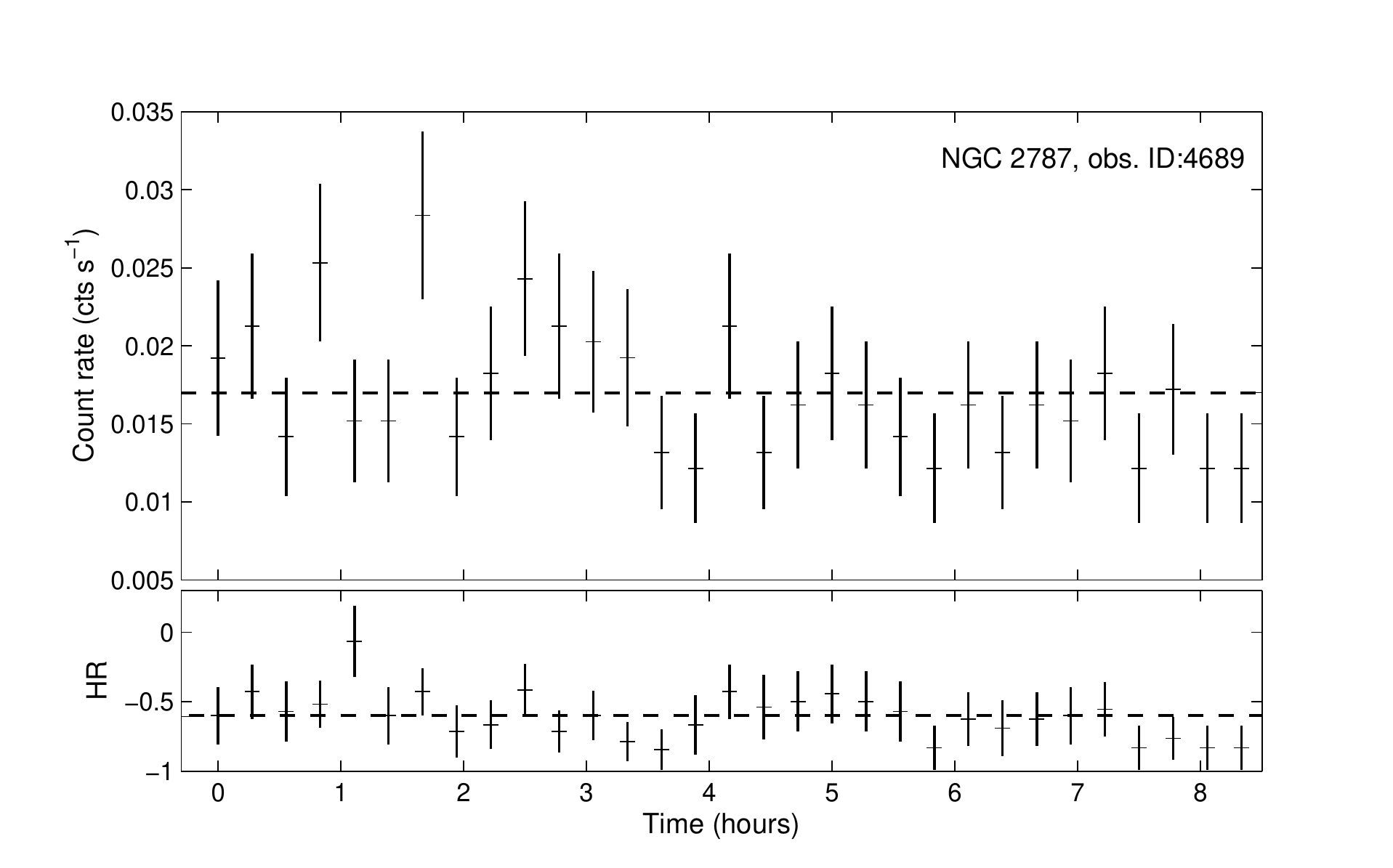}\\
\includegraphics[height=.19\textheight,angle=0,width=.33\textheight]{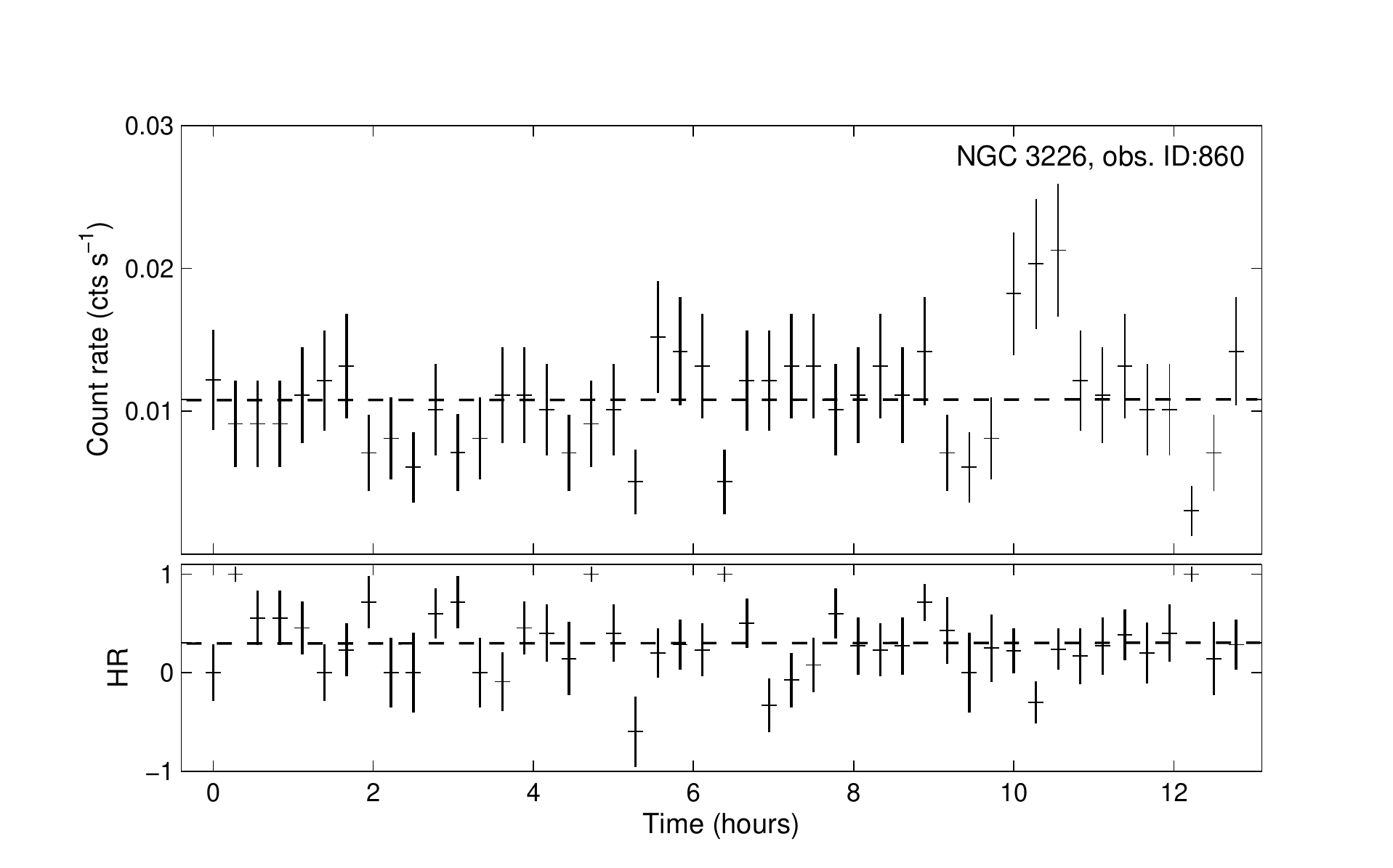}
\includegraphics[height=.19\textheight,angle=0,width=.33\textheight]{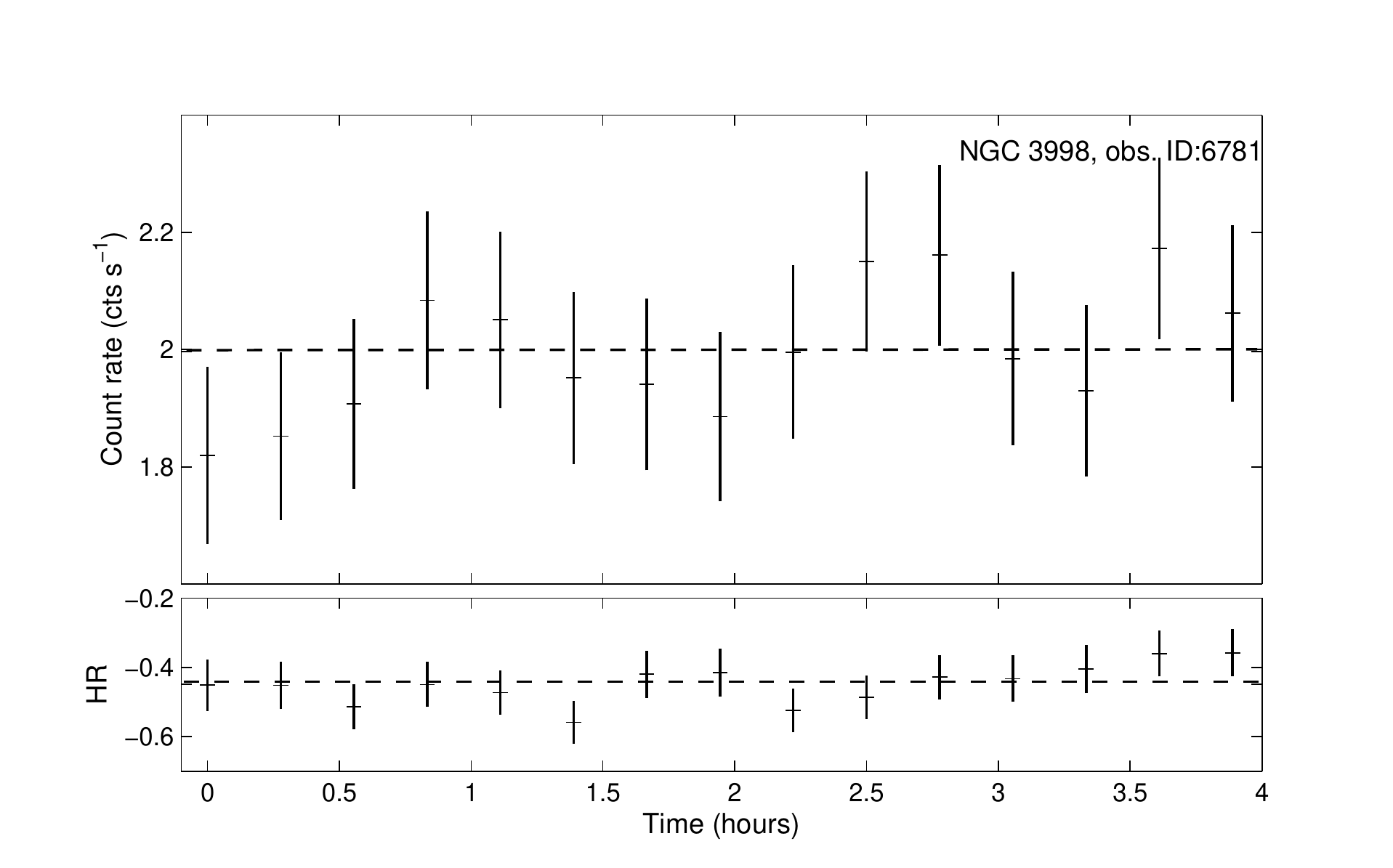}\\
\includegraphics[height=.19\textheight,angle=0,width=.33\textheight]{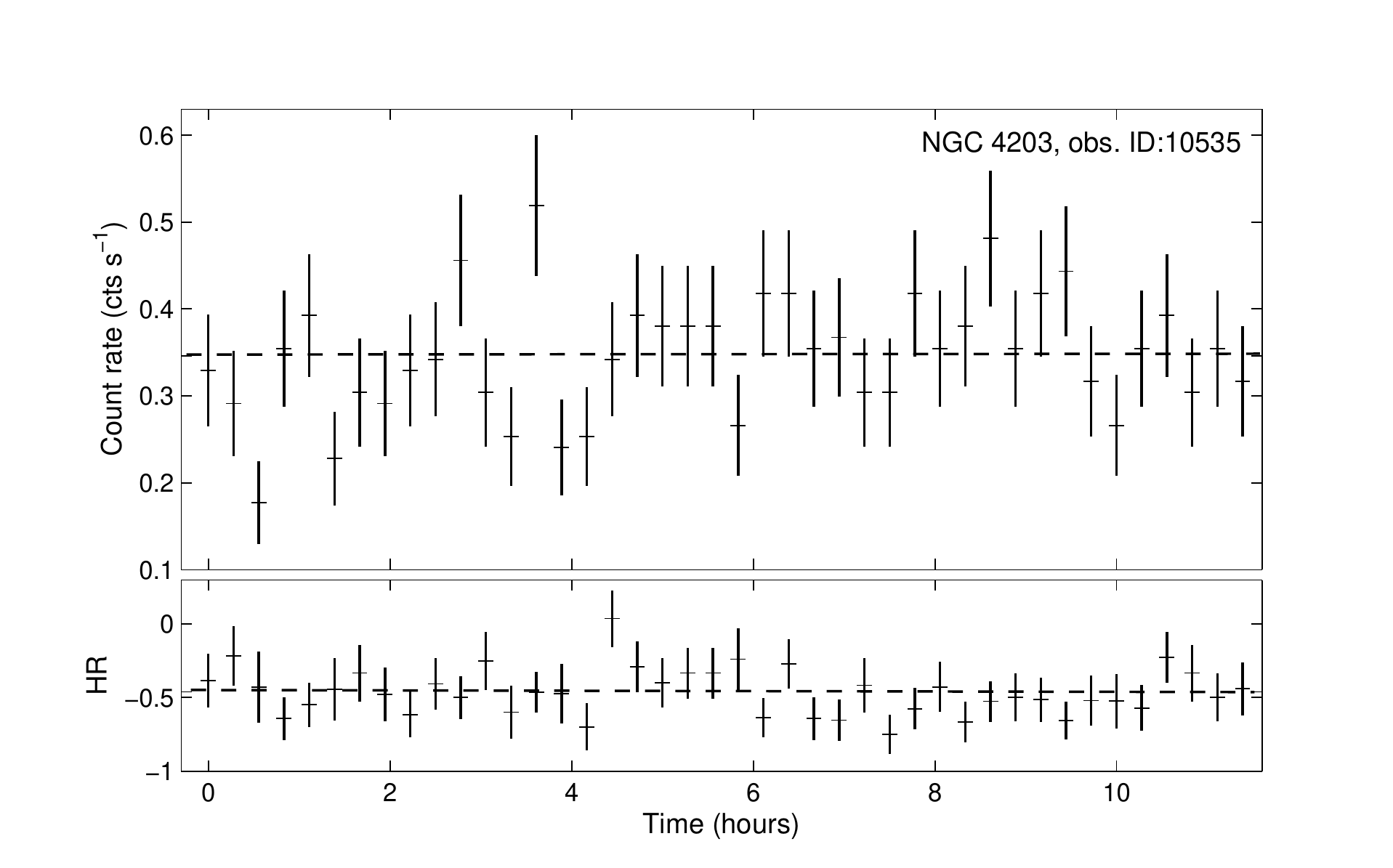}
\end{center}
\caption{Light curves and hardness ratios of the LINER~1s observed with \chandra\ with a long exposure time, all binned with a 1~ks time bin-size.}
\label{LCchandraallLINERs}
\end{figure*}
}

%-----------
% Table  3
%-----------
\begin{table}[!th]
\newcommand\T{\rule{0pt}{2.6ex}}
\begin{center}{
\caption{Normalized excess variance for LINER~1s with a relatively long exposure time.}
\label{varsigma}
\resizebox{0.48\textwidth}{!}{
\begin{tabular}{c c c c c}
\hline
\hline
Galaxy Name \T  & $\sigma_{NXS}^2$& Exposure Time$^a$ & $\sigma_{NXS}^2$ &  Exposure time$^a$ \\
            \T  &      \multicolumn{2}{c}{\xmm}      &       \multicolumn{2}{c}{\chandra}     \\
\hline
NGC-315  \T  & $<0.0027$            & 49.7  & $<0.0160$ & 55.0  \\
NGC-2681 \T  &                      &       & $<0.0400$ & 159.9 \\
NGC-2787 \T  & $<0.0170$            & 37.1  & $<0.0453$ & 30.8  \\
NGC-3226 \T  & $(0.02\pm0.002)^{b}$ & 94.3  & $<0.0700$ & 46.6   \\
NGC-3718 \T  & $<0.0078$            & 18.4  &           & \\
NGC 3998 \T  & $<0.0001$            & 8.9   & $<0.0016$ & 13.6   \\
NGC 4143 \T  & $<0.0060$            & 9.3   &           & \\
NGC-4203 \T  &                      &       & $<0.0026$ & 41.6  \\
NGC-4278 \T  & $<0.0012$            & 30.3  & $<0.0202$ & 470.8 \\
NGC-5005 \T  & $<0.0042$            & 8.7   &           & \\ 
\hline
\end{tabular}}}
\begin{list}{}{}
\item[{\bf Notes.}]$^a$The exposure times in ks used to calculate the value of $\sigma_{NXS}^2$. $^{b}$Value corresponding to only the longest observation of NGC~3226. Including the other \xmm\ observation would result in a $\sigma_{NXS}^2$ upper limit of 0.013.
\end{list}
\end{center}
\end{table}
%-----------
% Table  3
%-----------

To check more accurately any intrinsic variability amplitude from the
different  sources,  we  calculated  the  normalized  excess  variance
\citep{nandra97apj:variance} for all of the long observations with the
following expression:

\begin{equation}
\sigma^2_{NXS}=\frac{1}{N\mu^2}\sum\limits_{i=1}^{N}[(X_i-\mu)^2-\sigma_{i}^2]
\end{equation}

where $N$  is the number of bins  in a given light  curve, $X_{i}$ and
$\sigma_{i}$  are  the  count   rate  and  uncertainty  of  each  bin,
respectively, and $\mu$ is the  arithmetic mean of the counting rates.
To  enable  $\sigma^2_{NXS}$ comparison  between  the different  light
curves, the  bin size and the  light curve segment  duration should be
taken equally.  For that purpose,  we first decided to use light curve
segments  of 20~ks,  as usually  done for  luminous  Seyfert galaxies,
splitting any  longer observations  into multiple ones.   That limited
our sample  to 6 sources observed  with \chandra\ and  5 observed with
\xmm, not enough  to draw any safe conclusions.   Therefore, and owing
to the  heterogeneous sampling  of the observations  for this  type of
study, we decided  to use the whole corrected exposure  time of all of
the long observations.  The mean  of the $\sigma^2_{NXS}$ is taken for
every source with multiple  \chandra\ or \xmm\ observations.  The time
bin size choice of 1~ks for  all of the observations was taken to have
a good signal to  noise ratio with at least 20 counts  in each bin and
an acceptable number of bins in each light curve.

Estimating  the  error  on  the  $\sigma^2_{NXS}$ could  be  a  tricky
task. The variability  in an AGN light curve depends,  on one hand, on
the measurement errors  of the data (e.g.  Poisson  noise) and, on the
other hand,  on the  stochastic nature of  the process  underlying AGN
variability \citep[e.g.  red noise,  see][for a detailed discussion on
this   issue]{vaughan03mnras:varagn};  even   if  a   source   is  not
intrinsically  variable the mean  and the  variance of  different light
curves based on observations performed  at different times will not be
identical.   We  estimated  error  due  to  Poisson  noise  using  the
\citet{vaughan03mnras:varagn} equation

\begin{equation}
\label{errsig}
\resizebox{.43\textwidth}{!}{$err(\sigma^2_{NXS})=\sqrt{\left(\sqrt{\frac{2}{N}}\frac{\overline{\sigma_{err}^{2}}}{\overline{x}^2}\right)^2+\left(\sqrt{\frac{\overline{\sigma^{2}_{err}}}{N}}\frac{2\sigma_{NXS}}{\overline{x}}\right)^2}$}.
\end{equation}

\begin{figure}[!t]
\centerline{\includegraphics[angle=0,width=0.5\textwidth]{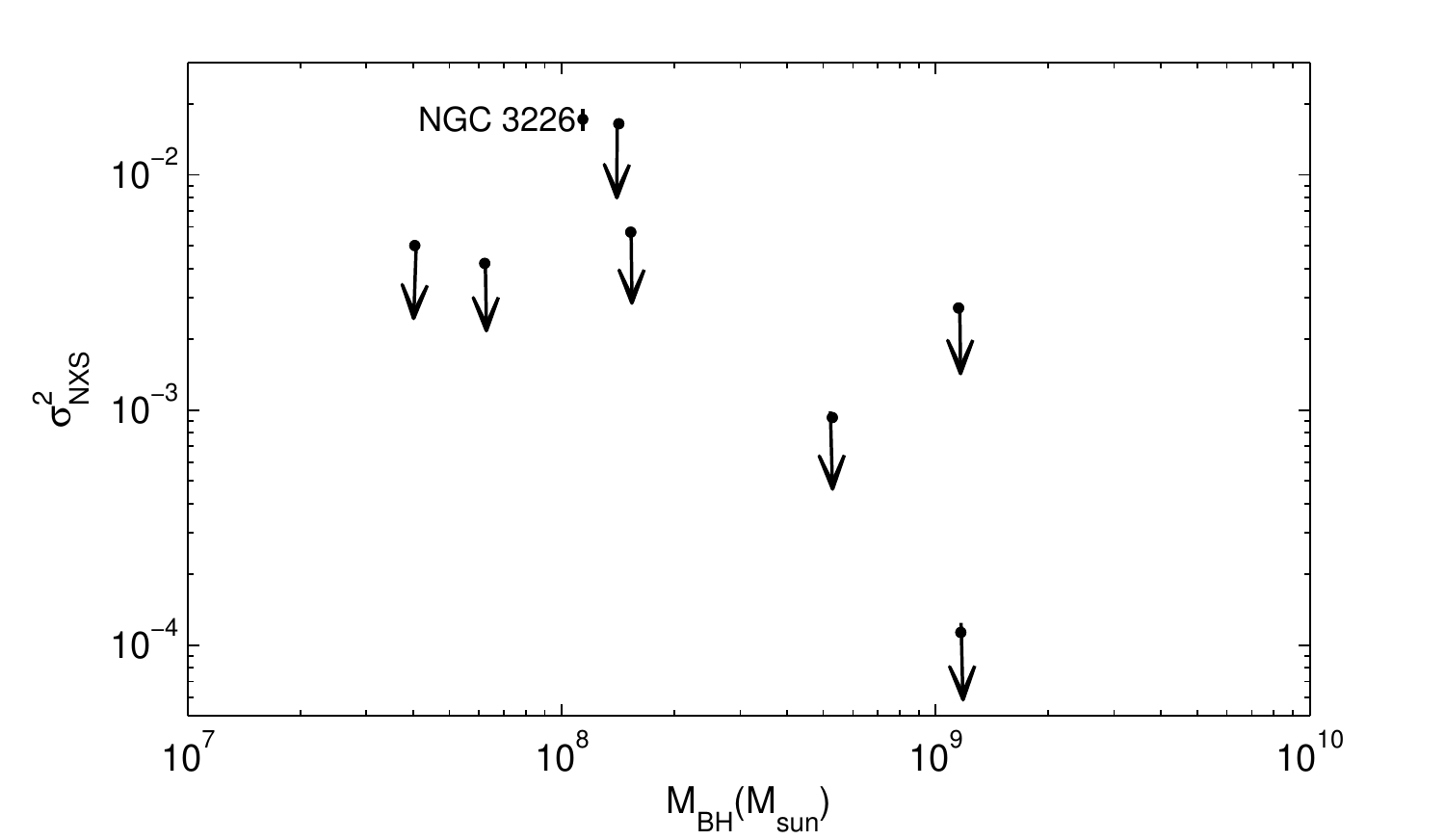}}
\caption{$\sigma_{NXS}^2$ derived from the \xmm\ observations as a function of the BH mass for our sample of LINER~1s. Arrows represent upper limits. NGC~3226 is the only source showing clear short time-scale ($\sim$1~day) variability and thus a non-upper limit value on $\sigma_{NXS}^2$.}
\label{NXSvsMbh}
\end{figure}

The uncertainty owing  to the red noise process  has been presented by
\citet{vaughan03mnras:varagn} to depend on the power-spectrum shape of
the   source   which   we   do   not  know   a   priori.    Therefore,
\citet{oneill05mnras:varagn}  estimated  the error  on  the red  noise
process directly from the data.  Our observations are not well sampled
to use \citet{oneill05mnras:varagn} method  to determine any error due
to the stochastic  nature of the AGN X-ray  variability.  The error on
$\sigma_{NXS}^2$  shown  in  equation~\ref{errsig}  was also  used  to
estimate  upper  limits  to  the   excess  variance  in  the  case  of
non-variability  detection whenever  $\sigma_{NXS}^2$  is negative  or
consistent with zero. Only one object, NGC~3226, in our sample shows a
clear   short  time-scale  variability   during  the   longest  100~ks
observation  with  $\sigma_{NXS}^2=0.02\pm0.002$  (comparable  to  the
value of 0.014  found by \citealt{binder09apj:ngc3226}).  Upper limits
were obtained  for the rest  of the sample.   Table~\ref{varsigma} and
Fig.~\ref{NXSvsMbh}  summarizes  the  results  that are  discussed  in
\S~\ref{xrayvar}.

\subsection{X-ray spectral results}
\label{specresulsec}

The spectral  analysis was performed  using XSPEC \citep{arnaud96conf}
version  12.6.0.  The  photo-electric  cross sections  and  the  solar
abundances of  \citet{wilms00ApJ} are  used throughout to  account for
absorption  by  neutral  gas.   An absorbed  Galactic  column  density
derived for every single source from \citet{kalberla05aa:nh} (obtained
with                              the                             W3NH
tool\footnote{http://heasarc.gsfc.nasa.gov/cgi-bin/Tools/w3nh/w3nh.pl})
was applied to the  different spectral models.  Spectral uncertainties
are  given  using  $\Delta$\chisq\  of  2.71,  corresponding  to  90\%
confidence for  one interesting parameter, and to  95\% confidence for
upper limits.

\subsubsection{snapshot observations}

We   began   our   spectral   analysis   with   the   study   of   the
\chandra\  snapshot  observations.  Table~\ref{specfit-param-snapshot}
gives  the best  fit parameters  to the  snapshot observations  of our
sample of LINER~1s.  Complicated  models, like partial covering and/or
two power-law components could not be  tested due to the low number of
counts. A power-law modified by Galactic absorption gave a good fit in
the  case of NGC~266  and NGC~4143.   An additional  intrinsic neutral
absorption  was needed  in the  remainder  of the  cases.  The  photon
indices  vary   between  $1.3\pm0.2$   for  the  hardest   spectra  to
$2.1\pm0.7$ for the softest ones with  a mean value of about 1.7.  The
hydrogen column density  of the intrinsic absorber had  an upper limit
of  $\sim4\times10^{21}$~cm$^{-2}$   in  the  case   of  NGC~4750  and
NGC~5005.   A value consistent  with $10^{21}<N_{H}<10^{22}$~cm$^{-2}$
was derived for  the rest of the snapshot  observations.  In one case,
NGC~3718, we find a somewhat larger column density with $N_{H}$ in the
order of  $\sim10^{22}$~cm$^{-2}$.  For NGC~5005,  a thermal component
\citep[{\sl    mekal}   model,][using    the   abundance    table   of
  \citealt{wilms00ApJ}]{mewe85aaps:mekal}            with            a
$0.8_{-0.2}^{+0.3}$~keV temperature, was included in the model to take
account  for some  low energy  features,  most likely  due to  diffuse
emission from hot gas.  In order to rigorously confirm the validity of
our best fit spectral-parameter values  derived using the C-stat and a
modeled  background, we compared  our results  to the  results derived
from fits  applying the  $\chi^2$ statistics to  all of  the snapshots
observations.  Spectral  parameters derived using the  C-stat were all
in agreement,  within the error  bars, with the results  derived using
the $\chi^2$ statistics; with smaller deviation from the central value
of one  interesting parameter.  We  decided, as a consequence,  to use
the C-stat fits to calculate  model fluxes in the soft, 0.5-2~keV, and
in  the hard, 2-10~keV,  bands.  Table~\ref{specfit-fluxes}  gives the
corresponding  0.5-2~keV and  2-10~keV observed  fluxes  and corrected
luminosities.

%-----------
% Table  4
%-----------
\begin{table*}[!th]
\caption{Best fit parameters to the \chandra\ snapshot observations of our sample of LINER~1s.}
\label{specfit-param-snapshot}
\newcommand\T{\rule{0pt}{2.6ex}}
\newcommand\B{\rule[-1.2ex]{0pt}{0pt}}
\begin{center}{
\resizebox{0.8\textwidth}{!}{
\begin{tabular}{l c c c c c c}
\hline
\hline
Galaxy Name\T\B & Obs. ID & N$_{h}$ & $\Gamma$ &  Pl Norm. at 1~keV & kT & EM$^{(a)}$ \\
   \T\B & & ($10^{20}$~cm$^{-2}$) & & ($10^{-5}$~Photons~keV$^{-1}$~cm$^{-2}$~s$^{-1}$) & (keV) & (10$^{62}$~cm$^{-3}$) \\
\hline
NGC~266\T  & 1610       & ($\ldots$)          & 1.4~[0.9-1.9]   & 2~[1-3]       & ($\ldots$) & ($\ldots$)  \\
NGC~315\T  & 855        & 20~[10-30]   & 1.3~[1.1-1.5]   & 13~[10-16]    & ($\ldots$) & ($\ldots$)  \\
NGC~3226\T & 1616       & 74~[47-105]  & 1.7~[1.3-2.1]   & 22~[15-34]    & ($\ldots$) & ($\ldots$)  \\
NGC~3718\T & 3993       & 114~[97-132] & 1.5~[1.4-1.7]   & 66~[53-81]    & ($\ldots$) & ($\ldots$)  \\
NGC~4143\T & 1617       & ($\ldots$)          & 1.9~[1.6-2.1]   & 7.2~[6.6-8.3] & ($\ldots$) & ($\ldots$)  \\
NGC~4750\T & 4020       & $<$31        & 1.8~[1.4-2.3]   & 5~[3-8]       & ($\ldots$) & ($\ldots$)  \\
NGC~4772\T & 3999       & 46~[24-53]   & 1.69~[1.29-1.74]& 7~[4-11]      & ($\ldots$) & ($\ldots$)  \\
NGC~5005\T & 4021       & $<$39        & 2.1~[1.4-2.8]   & 5~[3-11]      & 0.8~[0.6-1.1] & 3~[1-4]  \\
\hline
\end{tabular}}}
\end{center}
\begin{list}{}{}
\item[{\bf Notes.}]$^{(a)}$The emission measure (EM) of the {\sl mekal} model, EM=$\int$n$_{e}$n$_{H}$dV.
\end{list}
\end{table*}
%-----------
% Table  4
%-----------

\subsubsection{Long-exposure observations}

%-----------
% Table  5
%-----------
\begin{table*}[!th]
\caption{Best fit parameters to the LINER~1s in our sample observed with a relatively long \chandra\ and \xmm\ exposure time.}
\label{specfit-param}
\newcommand\T{\rule{0pt}{2.6ex}}
\newcommand\B{\rule[-1.2ex]{0pt}{0pt}}
\begin{center}{
\resizebox{0.8\textwidth}{!}{
\begin{tabular}{l c c c c c c c c}
\hline
\hline
Galaxy Name\T\B & Obs. ID & N$_{h}$ & $\Gamma$ &  Pl Norm. at 1~keV & kT & EM$^{(a)}$ & $\chi^2_\nu$ & d.o.f. \\
 \T\B & & ($10^{20}$~cm$^{-2}$) & & ($10^{-5}$~Photons~keV$^{-1}$~cm$^{-2}$~s$^{-1}$) & (keV) & (10$^{62}$~cm$^{-3}$) & &  \\
\hline
NGC~315 \T & 4156 & 10~[9-13] & 1.5~[1.4-1.6] & 18~[16-21] & 0.55~[0.47-0.59] & 15~[13-17] & \multirow{2}{*}{1.13} & \multirow{2}{*}{465} \\
             \T & 0305290201 & L.           & 2.1~[1.9-2.2]   & 20~[17-25]    & L.      & L.      &   &      \\
\hline
NGC~2681     \T & 2060       & $<$29      & 1.5~[1.2-1.8] & 0.6~[0.5-0.8]& 0.67~[0.63-0.70] & 0.4~[0.3-0.5] & \multirow{2}{*}{0.90} & \multirow{2}{*}{74} \\
             \T & 2061       & L.           & L.              & L.            & L.      & L.      &  &          \\
\hline
NGC~2787     \T & 4689       & 16~[8-24]    & 2.4~[2.1-2.6]   & 3~[2-4]       & ($\ldots$) &  ($\ldots$)  &  \multirow{2}{*}{1.12} & \multirow{2}{*}{118}       \\
             \T & 0200250101 & L.           & L.              & 4~[3-5]       & ($\ldots$) &  ($\ldots$)  & &       \\
\hline
NGC~3226   \T & 860        & 25~[$<$60]   & 1.7~[1.5-2.0]   & 13~[9-19]     & ($\ldots$) & ($\ldots$)   &  \multirow{3}{*}{0.98} & \multirow{3}{*}{467}       \\
             \T & 0101040301 & 89~[82-96]   & 1.8~[1.7-1.9]   & 25~[23-27]    & ($\ldots$) & ($\ldots$)   &   &       \\
             \T & 0400270101 & 42~[39-44]   & 2.05~[2.0-2.1]  & 27~[26-28]    & ($\ldots$) & ($\ldots$)   &   &       \\
\hline
NGC~3718     \T & 0200430501 & 138~[121-155]&1.8~[1.7-1.9]    & 57~[48-66]    & ($\ldots$) & ($\ldots$)   &  \multirow{2}{*}{0.88} & \multirow{2}{*}{122}       \\
             \T & 0200431301 & L.           &L.               & 47~[40-55]    & ($\ldots$) & ($\ldots$)   &   &       \\
\hline
NGC~3998     \T & 6781       & 3~[2-4]      &2.1~[2.0-2.2]    & 282~[267-298] & ($\ldots$) & ($\ldots$)   &  \multirow{2}{*}{1.10} & \multirow{2}{*}{590}      \\
             \T & 0090020101 & L.           & 1.84~[1.82-1.85]& 323~[318-328] & ($\ldots$) & ($\ldots$)   &  &  \\
\hline
NGC~4143     \T & 0150010601 & 6~[3-9]      & 2.2~[2.1-2.3]   & 17~[15-19]    & ($\ldots$) & ($\ldots$)   &  0.98 & 118       \\
\hline
NGC~4203     \T & 10535      & ($\ldots$)      & 2.3~[2.2-2.4]   & 83~[78-89]    & ($\ldots$) & ($\ldots$)   &  0.83 & 51        \\
\hline
NGC~4278     \T & 4741       & $<$6.78  & 2.1~[2.0-2.3] & 43~[39-47] & 0.62~[0.58-0.66] & 2.6~[2.3-3.0] & \multirow{6}{*}{0.93} & \multirow{6}{*}{310} \\
             \T & 7077       & L.           & 2.3~[2.2-2.4]   & 18~[17-20]    &   L. & L. & & \\
             \T & 7078       & L.           & 2.3~[2.2-2.5]   & 42~[39-46]    &   L. & L. & & \\
             \T & 7079       & L.           & 2.4~[2.3-2.5]   & 38~[35-41]    &   L. & L. & & \\
             \T & 7080       & L.           & 2.0~[1.8-2.2]   & 11~[10-13]    &   L. & L. & & \\
             \T & 7081       & L.           & 2.1~[2.0-2.3]   & 12.5~[11.4-12.9]& L. & L. & & \\
             \T & 0205010101 & 3.8[3.1 4.6] & 2.05~[2.03-2.07]& 81~[79-82]    & ($\ldots$) &  ($\ldots$)  &  1.01 & 487       \\
\hline
NGC~5005     \T & 0110930501 & 9~[2-18] & 1.7~[1.5-1.8]   & 8~[6-9]      & 0.64~[0.61-0.67] & 5.2~[4.7-5.7] & 1.2 & 107 \\
\hline
\end{tabular}}}
\end{center}
\begin{list}{}{}
\item[{\bf Notes.}](L.) represents a linked paramter in the fit. $^{(a)}$The emission measure (EM) of the {\sl mekal} model, EM=$\int$n$_{e}$n$_{H}$dV.
\end{list}
\end{table*}
%-----------
% Table  5
%-----------

%-----------
% Table  6
%-----------
\begin{table*}[!th]
\caption{Absorbed fluxes and corrected luminosities derived from the best fit model to our sample of LINER~1s and the corresponding \eddratio.}
\label{specfit-fluxes}
\newcommand\T{\rule{0pt}{2.6ex}}
\newcommand\B{\rule[-1.2ex]{0pt}{0pt}}
\begin{center}{
\resizebox{0.9\textwidth}{!}{
\begin{tabular}{l c c c c c c c}
\hline
\hline
Galaxy Name\T\B & Obs. ID   & 0.5-2~keV Flux & 2-10~keV Flux & Corr. 0.5-2~keV Lum. & Corr. 2-10~keV Lum. & Pl$^{a}$ & Log($L_{2-10~keV}$/$L_{Edd}$)\\
  \T\B &       & \multicolumn{2}{c}{(Logarithmic scale; erg~s$^{-1}$~cm$^{-2}$)} & \multicolumn{2}{c}{(10$^{41}$~erg~s$^{-1}$)} & \%\ & \\
\hline
NGC~266\T&1610& -13.40~[-13.50 -13.30] & -12.90~[-13.10 -12.80] & 0.23~[0.14 0.29] & 0.59~[0.37 0.74] & 100 & -5.77\\
NGC~315\T&855 & -12.74~[-12.78 -12.71] & -11.99~[-12.03 -11.95] & 1.56~[1.42 1.71] & 5.42~[4.94 5.94] & 100 & -5.42\\
       \T&4156& -12.80~[-12.82 -12.78] & -12.04~[-12.05 -12.02] & 2.42~[2.36 2.53] & 5.06~[4.83 5.18] &  96 & -5.45\\
 \T&0305290201& -12.80~[-12.82 -12.78] & -12.37~[-12.39 -12.35] & 2.42~[2.36 2.53] & 2.37~[2.26 2.48] &  94 & -5.78\\
NGC~2681\T&2060&-13.56~[-13.58 -13.55] & -13.47~[-13.49 -13.45] & 0.011~[0.010 0.012] & 0.012~[0.011 0.013] & 73 & -5.80\\
        \T&2061&-13.56~[-13.58 -13.55] & -13.47~[-13.49 -13.45] & 0.011~[0.010 0.012] & 0.012~[0.011 0.013] & 73 & -5.80\\
NGC~2787\T&4689& -13.30~[-13.33 -13.26]& -13.28~[-13.32 -13.25] & 0.0054~[0.0049 0.0058] & 0.004~[0.002 0.007]&100&-7.70\\
 \T&0200250101&-13.34~[-13.38 -13.31] & -13.33~[-13.37 -13.29] & 0.0048~[0.0044 0.0052]&0.0032~[0.0029 0.0034]&100&-7.74\\
NGC~3226\T& 1616 &-12.76~[-12.81 -12.71] & -12.07~[-12.12 -12.02] & 0.32~[0.28 0.36] &0.59~[0.51 0.66] &100 & -5.38\\
        \T&860& -12.76~[-12.80 -12.73] & -12.32~[-12.36 -12.29] & 0.19~[0.17 0.20] & 0.33~[0.30 0.35] & 100 & -5.64\\
    \T&0101040301&-12.77~[-12.77 -12.76] & -12.07~[-12.07 -12.06] & 0.37~[0.36 0.38] & 0.61~[0.60 0.62]&100 & -5.37\\
    \T&0400270101&-12.56~[-12.57 -12.56] & -12.21~[-12.22 -12.21] & 0.39~[0.38 0.40] & 0.42~[0.41 0.43]&100 & -5.53\\
NGC~3718\T&3993  &-12.40~[-12.41 -12.37] & -11.50~[-11.52 -11.48] & 0.50~[0.48 0.54] & 1.15~[1.10 1.23]&100 & -4.64\\
        \T&0200430501&-12.58~[-12.59 -12.56]&-11.76~[-11.78 -11.74]&0.44~[0.42 0.45] & 0.66~[0.63 0.68]&100 & -4.88\\
        \T&0200431301&-12.65~[-12.68 -12.63]&-11.84~[-11.86 -11.82]&0.36~[0.34 0.38] & 0.55~[0.51 0.58]&100 & -4.96\\
NGC~3998\T&6781 & -11.25~[-11.26 -11.23] & -11.19~[-11.21 -11.17] & 1.47~[1.44 1.54] & 1.54~[1.47 1.61]&100 & -5.98\\
  \T&0090020101&-11.183~[-11.186 -11.180]&-10.975~[-10.978 -10.973]&1.70~[1.69 1.71] & 2.53~[2.51 2.54]&100 & -5.76\\
NGC~4143\T& 1617 &-12.81~[-12.87 -12.75] & -12.63~[-12.70 -12.57] & 0.05~[0.04 0.06] & 0.07~[0.06 0.08]&100 & -6.43\\
  \T&0150010601&-12.53~[-12.54 -12.52]&-12.51~[-12.52 -12.49]&0.112~[0.109 0.115] & 0.096~[0.092 0.098]&100 & -6.30\\
NGC~4203\T&10535&-11.75~[-11.77 -11.72]& -11.86~[-11.89 -11.84] & 0.51~[0.48 0.53] & 0.38~[0.35 0.40] & 100 & -5.25\\
NGC~4278\T&4741&-11.99~[-12.02 -11.97] & -12.05~[-12.08 -12.02] & 0.33~[0.31 0.35] & 0.28~[0.26 0.30] &  94 & -6.37\\
        \T&7077&-12.30~[-12.32 -12.28] & -12.50~[-12.52 -12.48] & 0.16~[0.15 0.17] & 0.10~[0.09 0.11] &  83 & -6.83\\
        \T&7078&-12.00~[-12.02 -11.97] & -12.18~[-12.21 -12.16] & 0.33~[0.31 0.35] & 0.20~[0.19 0.21] &  93 & -6.51\\
        \T&7079&-12.04~[-12.06 -12.02] & -12.25~[-12.27 -12.23] & 0.30~[0.29 0.32] & 0.17~[0.16 0.18] &  90 & -6.58\\
        \T&7080&-12.46~[-12.49 -12.44] & -12.56~[-12.59 -12.54] & 0.11~[0.10 0.12] & 0.08~[0.07 0.09] &  80 & -6.89\\
        \T&7081&-12.42~[-12.44 -12.40] & -12.57~[-12.59 -12.55] & 0.12~[0.11 0.13] & 0.08~[0.07 0.09] &  81 & -6.90\\
 \T&0205010101&-11.808~[-11.812 -11.804]&-11.718~[-11.722 -11.714]&0.55~[0.54 0.56]&0.60~[0.59 0.61]&  100&-6.04\\
NGC~4750\T& 4020 &-13.14~[-13.21 -13.07] & -12.84~[-12.92 -12.78] & 0.08~[0.07 0.09] & 0.12~[0.10 0.14]&100 & -5.30\\
NGC~4772\T& 3999 &-13.25~[-13.33 -13.17] & -12.67~[-12.76 -12.59] & 0.04~[0.03 0.05] & 0.07~[0.06 0.08]&100 & -5.72\\
NGC~5005\T& 4021 &-12.86~[-12.92 -12.81] & -12.95~[-13.01 -12.90] & 0.08~[0.06 0.11] & 0.06~[0.05 0.07]& 72 & -6.11\\
        \T&0110930501&-12.55~[-12.57 -12.54]&-12.49~[-12.51 -12.48]&0.17~[0.16 0.18] & 0.18~[0.17 0.19]  & 76 & -5.64\\
\hline
\hline
\end{tabular}}}
\end{center}
\begin{list}{}{}
\item[{\bf Notes.}]$^{(a)}$The power-law component fraction to the 0.5-10~keV corrected luminosity.
\end{list}
\end{table*}
%-----------
% Table  6
%-----------

\begin{figure}[]
\includegraphics[angle=0,width=0.47\textwidth]{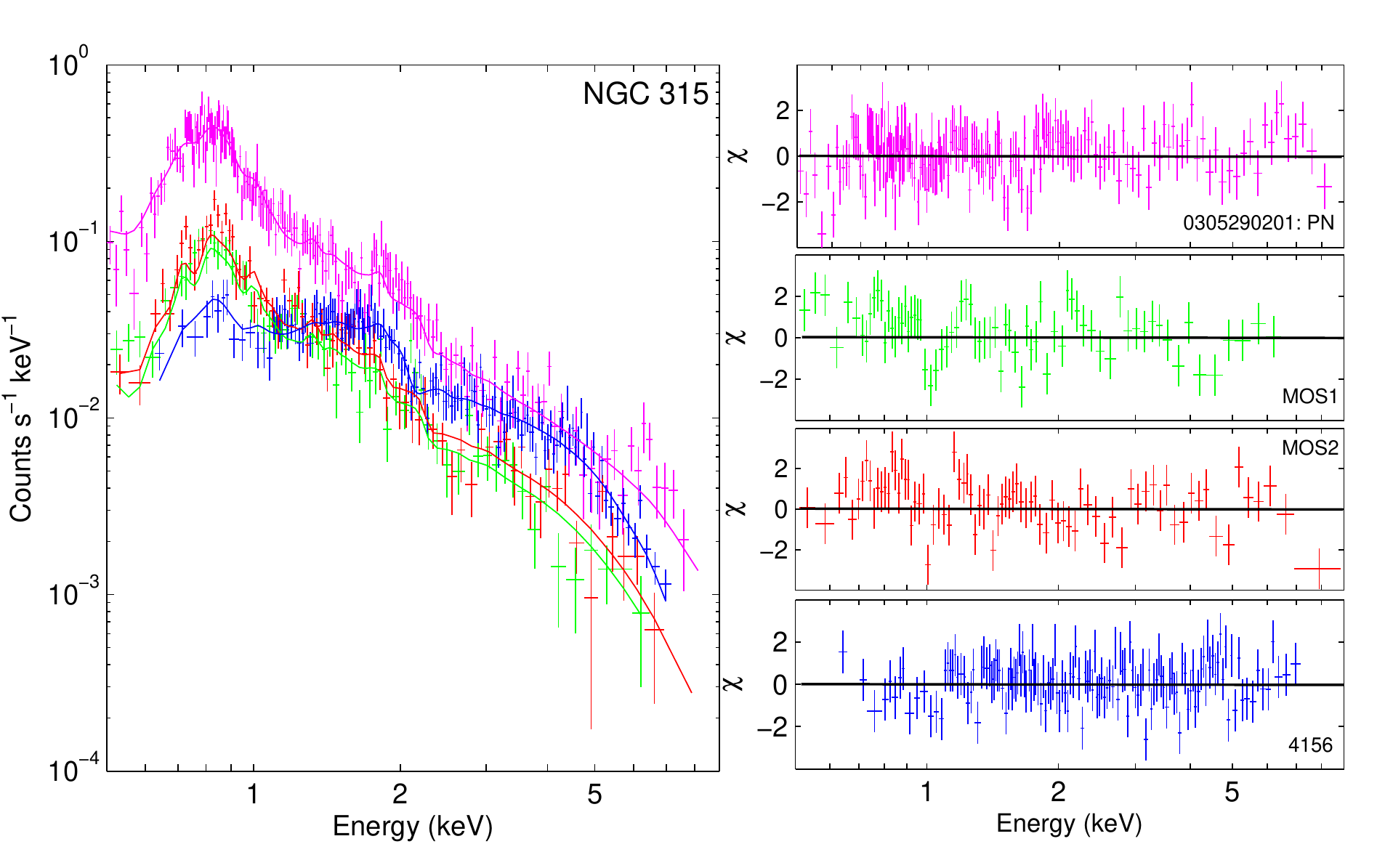}\\
\includegraphics[angle=0,width=0.47\textwidth]{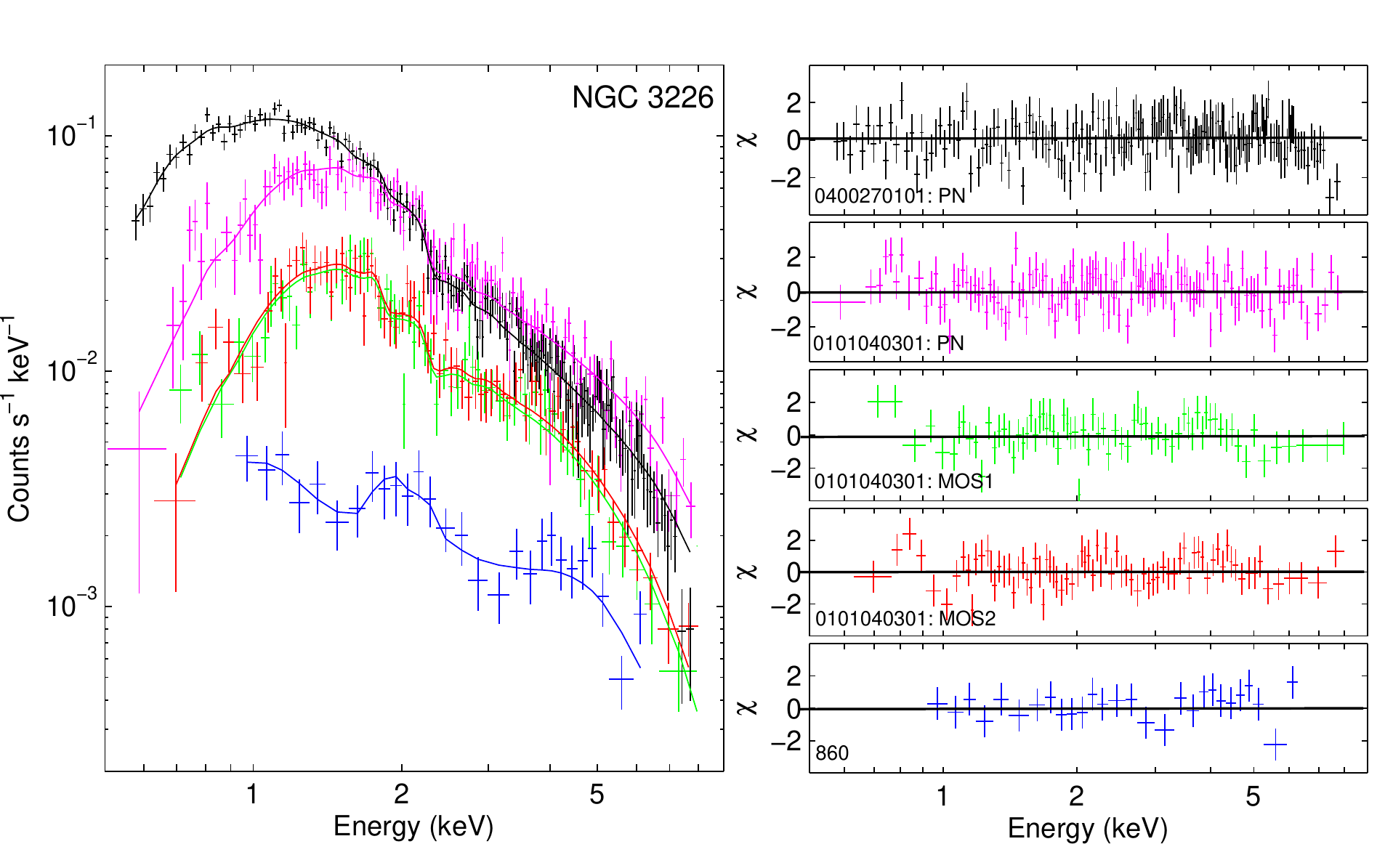}
\caption{{\sl Upper left panel}. Data and best fit model of the different spectra of NGC~315. A hardening in the \chandra\ ACIS spectrum (in blue) above $\sim$1.5~keV is seen, relative to the \xmm\ spectra (burgundy, green, and orange). {\sl Upper right panel}. Residuals of the best fit model in terms of sigma. {\sl Lower left panel}. Data and best fit model of the different spectra of NGC~3226. It is clear that more absorption below 2~keV from cold material is taking place between the two  \xmm\ observations (black representing the long $\sim$100~ks observation). {\sl Lower right panel}. Residuals of the best fit model in terms of sigma.}
\label{bestfitmod}
\end{figure}

\onlfig{10}{
\begin{figure*}[]
\begin{center}
\includegraphics[height=.19\textheight,angle=0]{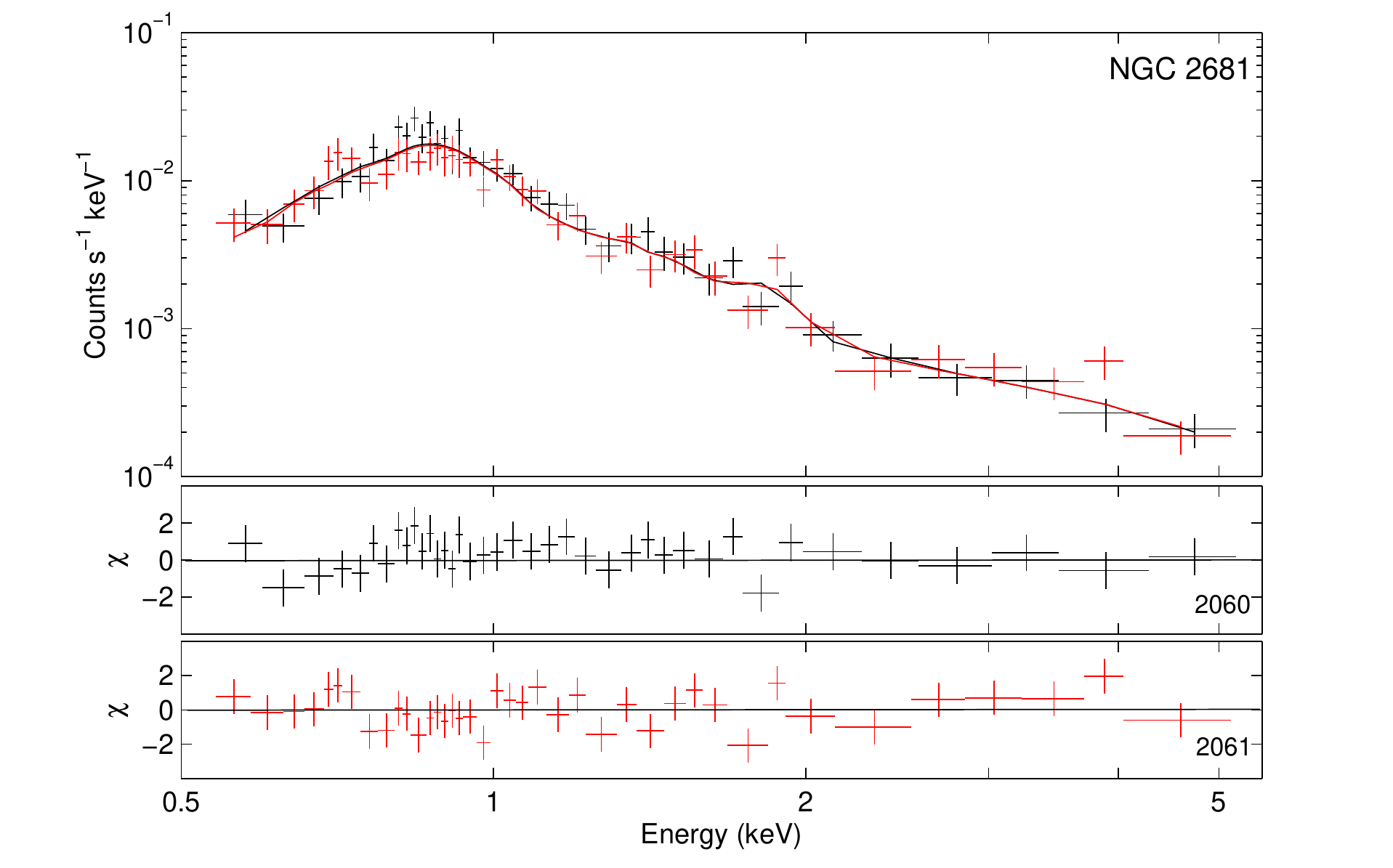}
\includegraphics[height=.19\textheight,angle=0]{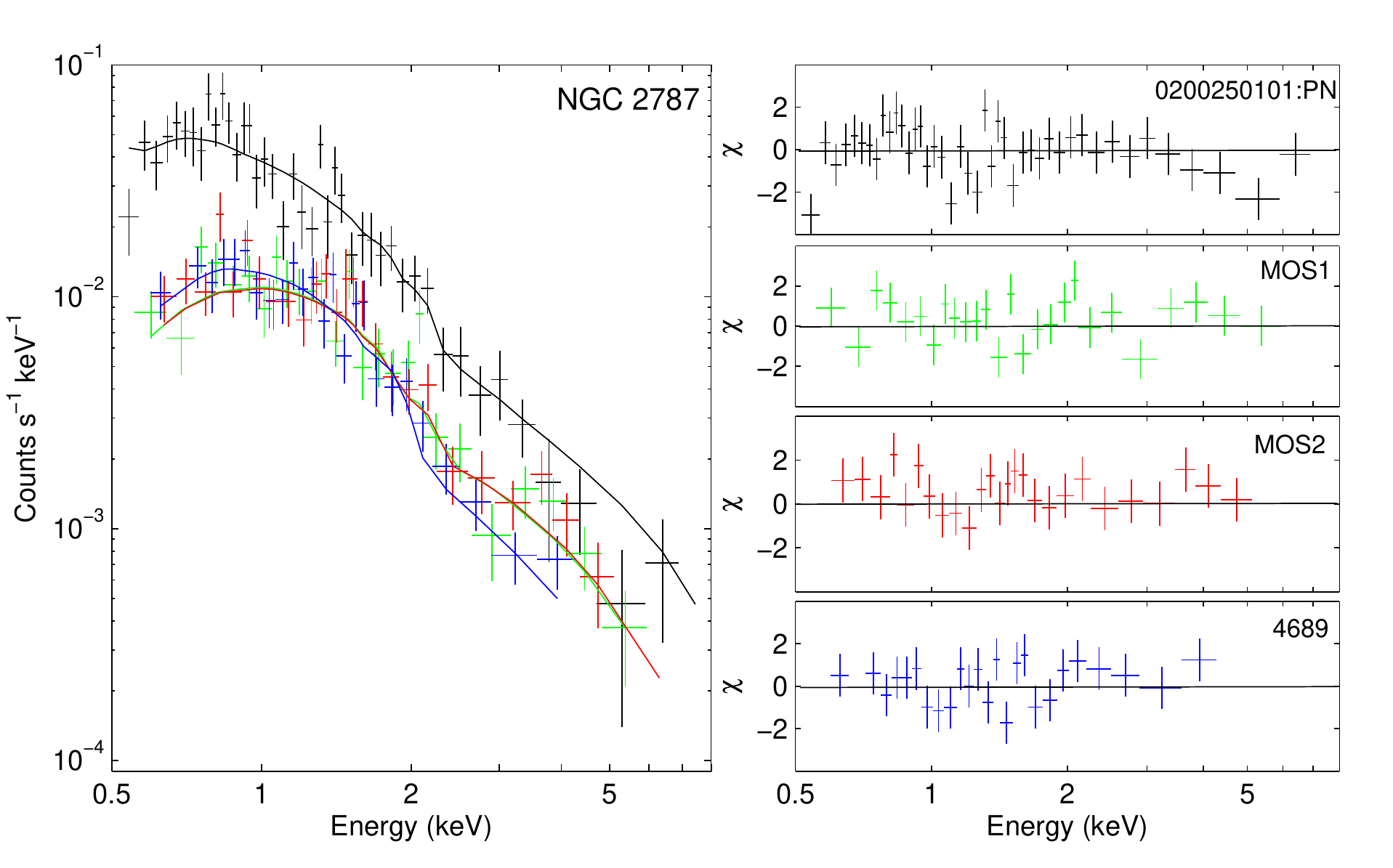}\\
\includegraphics[height=.19\textheight,angle=0]{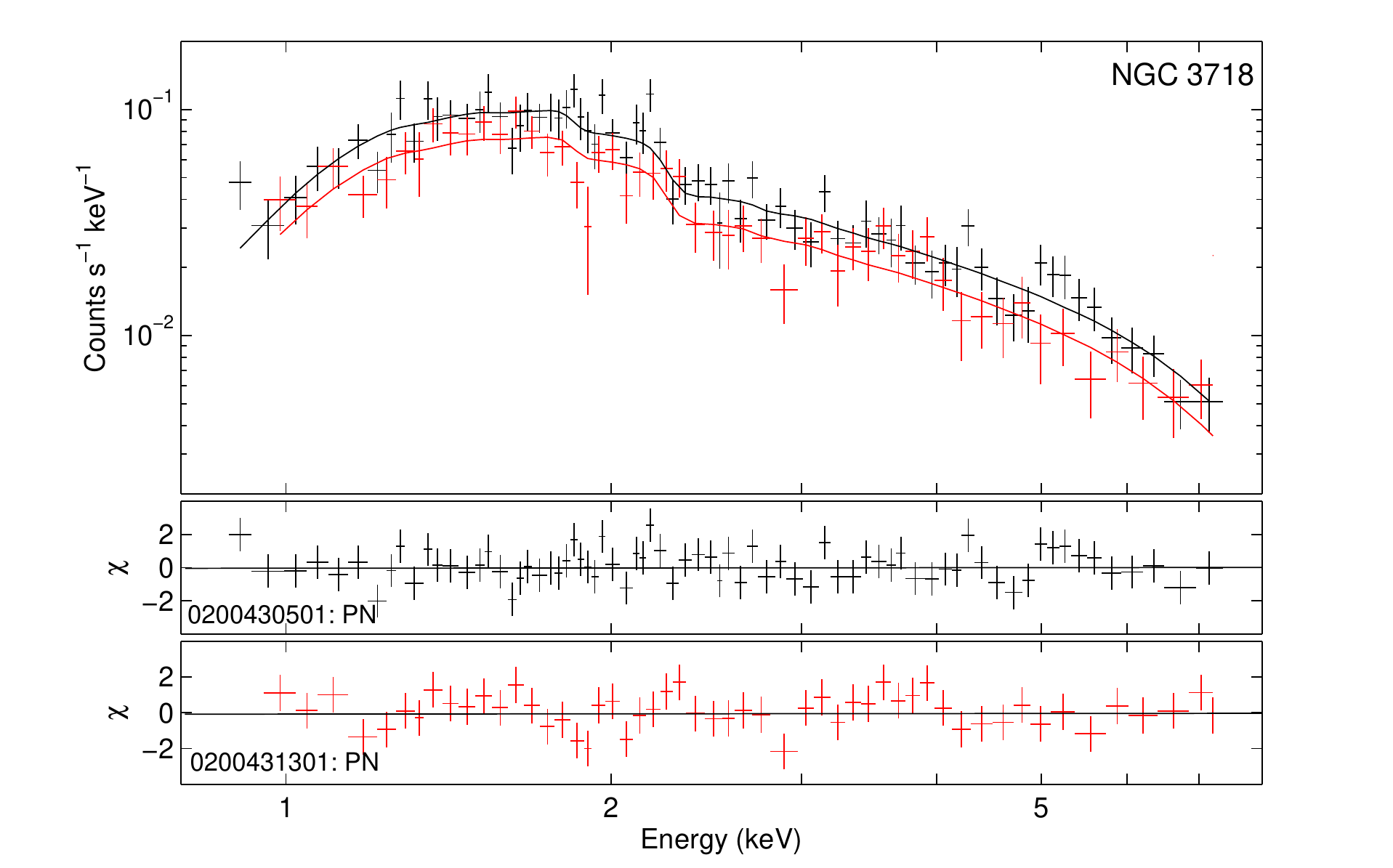}
\includegraphics[height=.19\textheight,angle=0]{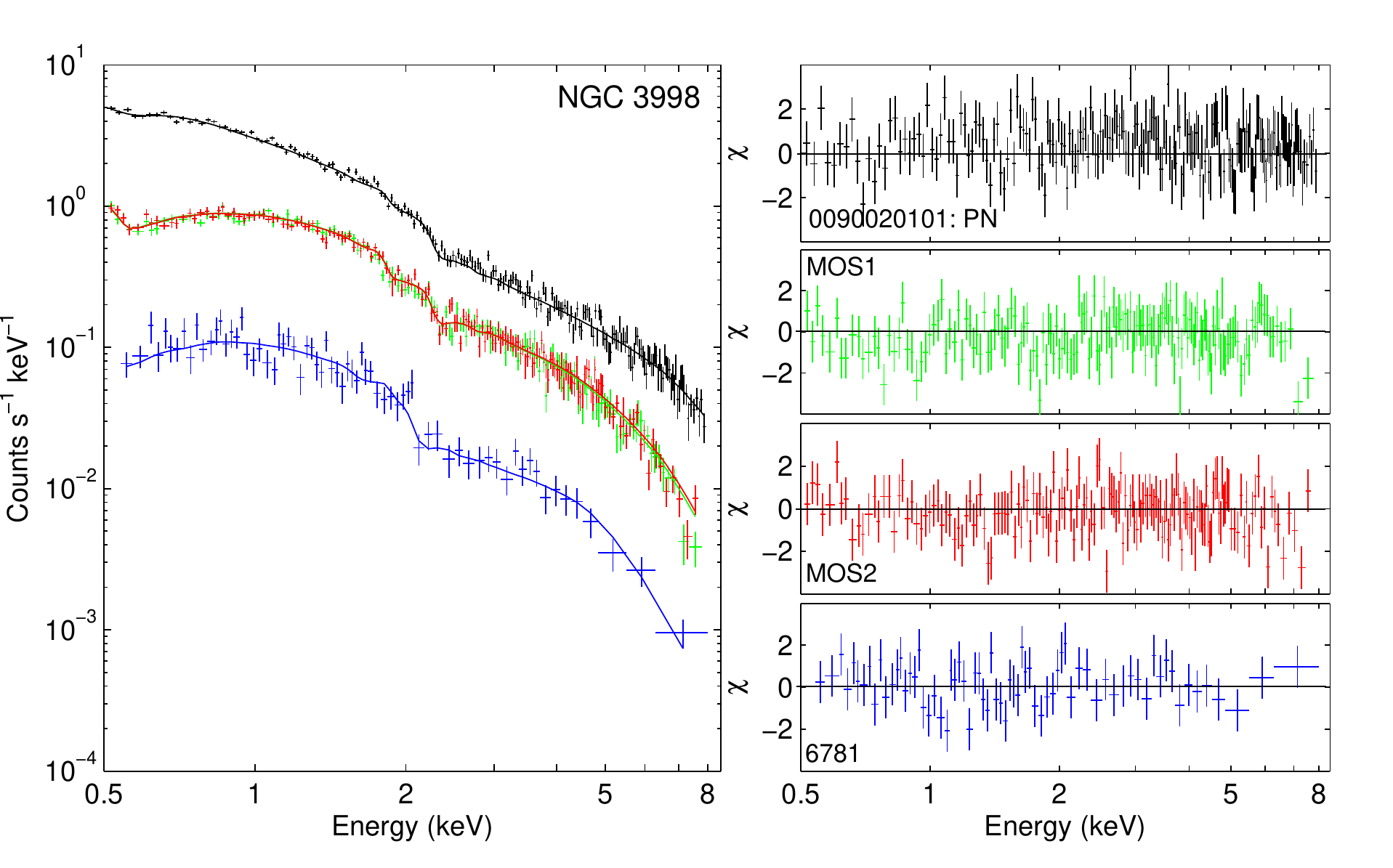}\\
\includegraphics[height=.19\textheight,angle=0]{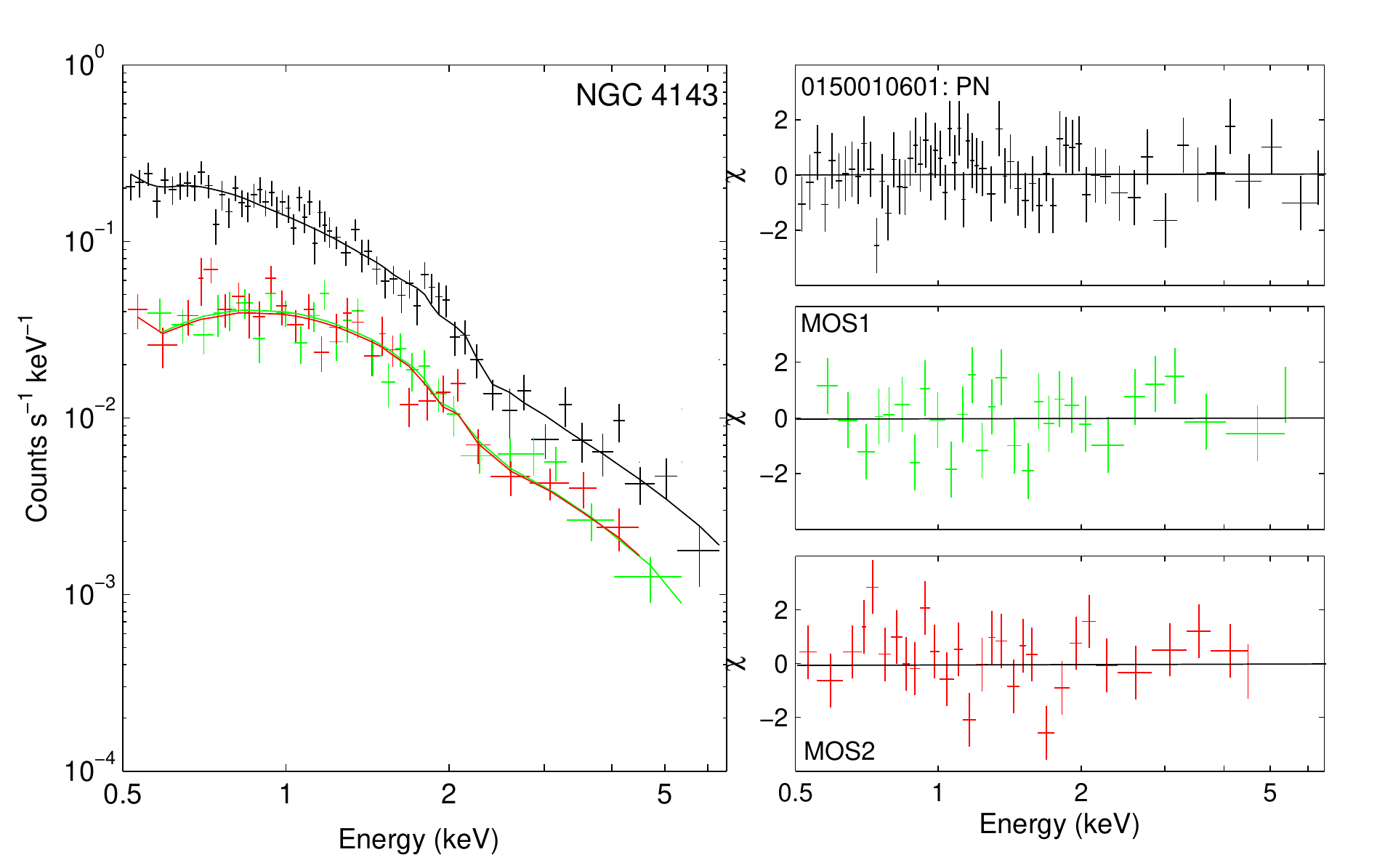}
\includegraphics[height=.19\textheight,angle=0]{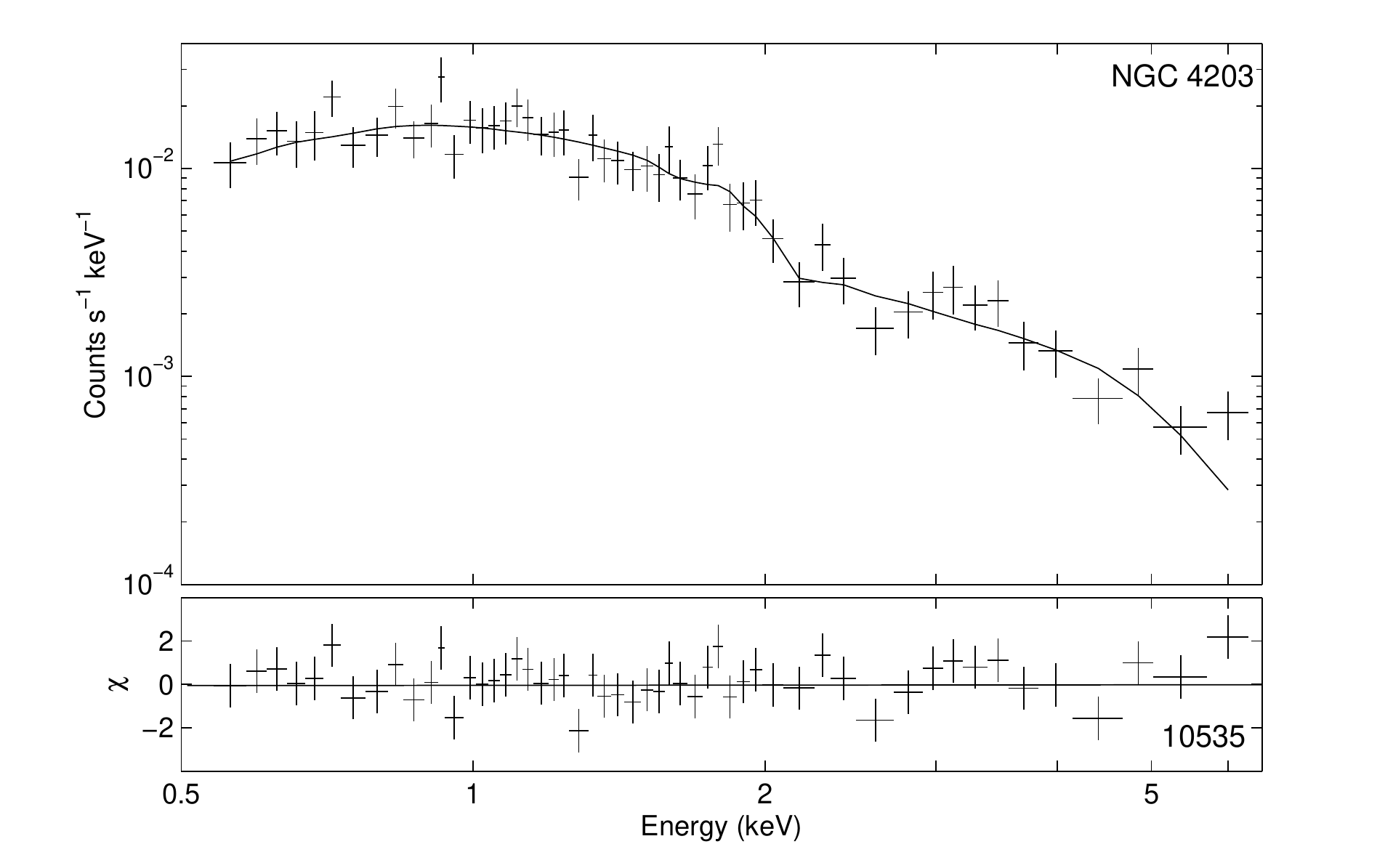}\\
\includegraphics[height=.19\textheight,angle=0]{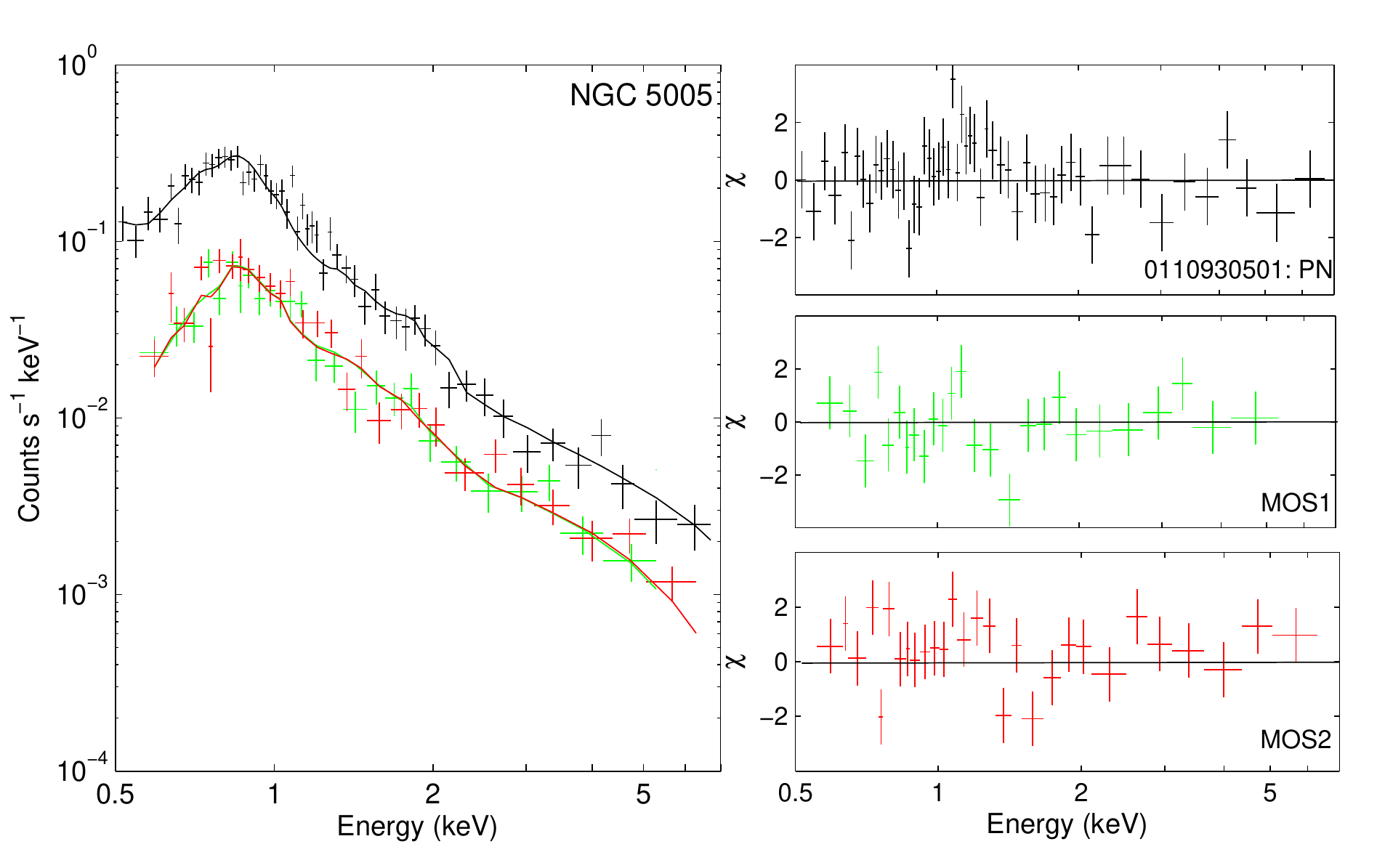}
\caption{Data and best fit model of the spectra of the LINER~1s in our sample with a relatively long exposure time. Residuals of every fit are given in terms of sigma.}
\label{bestfitallLINERs}
\end{center}
\end{figure*}
}

We then  turned to the  analysis of observations with  relatively long
exposure times.   We started with  a simple absorbed power-law  fit to
each of  the spectra of  the different sources, separately.   The fits
were acceptable for  6/10 sources but residuals at  energies less than
2~keV persisted  in the other 4 sources  (NGC~315, NGC~2681, NGC~4278,
and  NGC~5005),  suggesting the  presence  of  diffuse  hot gas.

In order to have a better signal to noise ratio and photon statistics,
we decided to fit the different \chandra\ and/or \xmm\ spectra of each
source simultaneously (the normalizations of the different models in a
fit were left free between the different EPIC instruments to take care
of  any  potential  cross  calibration  uncertainties).   Simultaneous
fitting  routines  of different  observations  performed at  different
times is done for the first time for any sample of LINER sources.  For
this purpose,  whenever a source  is observed with both  \chandra\ and
\xmm,   we    carried   out   careful   imaging    analysis   of   the
\chandra\  observations   to  disentangle  the   different  components
(diffuse emission, LMXBs, and/or jet emission) that are blended in one
point-like source  in the \xmm\ extraction  region of 25\arcsec-radius
circle (see Appendix~B and  online Fig.~1--4).  These components, that
we   assume   non-variable,   are   included   in   the   simultaneous
\chandra/\xmm\ fit.   We do not  expect diffuse hot  gas to vary  on a
time-scale  of  a   few  years  \citep{fabbiano89aa:difgaz},  although
off-nuclear point-like  sources and  jet X-ray emission  could exhibit
variation  on such  time-scales  \citep[e.g.,][]{harris03nar:m87}.  We
present  in Appendix~B  the  surrounding medium  around the  different
LINER~1s that  are observed with both  \chandra\ and \xmm,  and we give
spectral results to the different components.

We  tested multiple  models  on the  data  in order  to determine  the
mechanism responsible for the  observed X-ray spectra, noticeably: (1)
a simple absorbed  power-law, (2) same as (1)  but including a thermal
{\sl mekal} component to take  account of any diffuse hot gas features
in the soft  band, (3) two power-law components  with different photon
index values with one representing  the hard 2-10~keV emission and the
other representing a possible  0.5-2~keV soft excess emission commonly
seen    in   the    nuclei   of    luminous   galaxies\citep[e.g.,][]{
  porquet04aa:pgquasar}.  We investigated  spectral  variability in  a
source by  permitting one  spectral parameter of  a given fit  to vary
independently between different observations.  We then used the F-test
to evaluate the  improvement in the fit where  a 99\%\ probability for
an improvement to  occur by chance is considered  valid.  The case for
NGC~4278 is already  analyzed in Y10 and best  fit spectral parameters
and fluxes are taken from Y10.

We  find  that 6/10  sources  are  best fit  with  model  (1) with  no
additional need for any thermal or two power-law components (NGC~2787,
NGC~3226, NGC~3718,  NGC~3998, NGC~4143,  and NGC~4203).  The  rest of
the  sources were  best fit  with model  (2) showing  features  at low
energies  below  2~keV,  indicating  emission  from  diffuse  hot  gas
(NGC~315, NGC~2681, NGC~4278, and NGC~5005). Model (3) did not improve
the  quality of the  fit in  any of  the cases,  giving worse  fits at
times.   We found that  the intrinsic  hydrogen column  density varies
significantly        in         NGC~3226        decreasing        from
$(8.9\pm0.7)\times10^{21}$~cm$^{-2}$                                 to
$4.2_{-0.3}^{+0.2}\times10^{21}$~cm$^{-2}$.   This is clearly  seen in
the lower panels  of Fig.~\ref{bestfitmod} where the soft  part of the
spectrum  during the  short \xmm\  observation is  much  more absorbed
compared to the long  observation.  Additionally, the power-law photon
index varies in seven  sources (NGC~315, NGC~3226, NGC~3718, NGC~3998,
NGC~4143, NGC~4278,  and NGC~5005) with the most  drastic change being
the one observed in  NGC~315 where $\Gamma$ increased from $1.5\pm0.1$
during  the \chandra\  observation to  $2.1_{-0.2}^{+0.1}$  during the
\xmm\ one  (upper panels of Fig.~\ref{bestfitmod}).   This increase is
accompanied   by    a   decrease    in   the   2-10~keV    flux   from
$9.8\times10^{-13}$    to    $4.6\times10^{-13}$~erg~s$^{-1}$.    This
behavior is typical of X-ray emission originating in a RIAF structure,
which is the accretion flow believed to exist at the center of NGC~315
\citep{wu07apj:riaf}, where the Eddington ratio, which is proportional
to  L$_{2-10~keV}$,  is inveresly  proportional  to  the photon  index
$\Gamma$   \citep{gu09mnras:gamVSeddllagn}.   Best   fit   models  and
residuals  of the other  LINER~1s in  our sample  are shown  in online
Fig.~\ref{bestfitallLINERs}.

The photon  indicies we derived for  all of the sources  in our sample
observed  with   a  relatively  long  exposure   time  varied  between
$1.5\pm0.3$  and  $2.4_{-0.3}^{+0.2}$  with   a  mean  value  of  2.0.
Intrinsic column density covered two orders of magnitude, with $N_{H}$
varying between $\sim10^{20}$~cm$^{-2}$ for unabsorbed sources, and up
to  $\sim10^{22}$~cm$^{-2}$  for   the  only  mildly  absorbed  source
NGC~3718.   The thermal  component  had a  temperature  mean value  of
0.63~keV, consistent  with all LINER-type sources  embedded in diffuse
emission  \citep{flohic06apj}.   Table~\ref{specfit-param}  gives  the
best fit model parameters for our sample of LINER~1s with a relatively
long exposure time, and Table~\ref{specfit-fluxes} gives the 0.5-2~keV
and 2-10~keV  observed fluxes and  corrected luminosities, as  well as
the   corresponding  ``Eddington   ratios'',  \eddratio.    The  hard,
2-10~keV,   luminosity   spans   three   orders  of   magnitude   from
3.2$\times10^{38}$~erg~s$^{-1}$   to   5.4$\times10^{41}$~erg~s$^{-1}$
which  resulted  in  a  \eddratio\ range  from  $2.0\times10^{-8}$  to
$2.3\times10^{-5}$ which  is at least  one to two orders  of magnitude
smaller   than   \eddratio\   seen   in  luminous   AGN   \citep[e.g.:
][]{porquet04aa:pgquasar,nandra07mnras:felinesey}.

\begin{figure}[]
\centerline{\includegraphics[angle=0,width=0.5\textwidth]{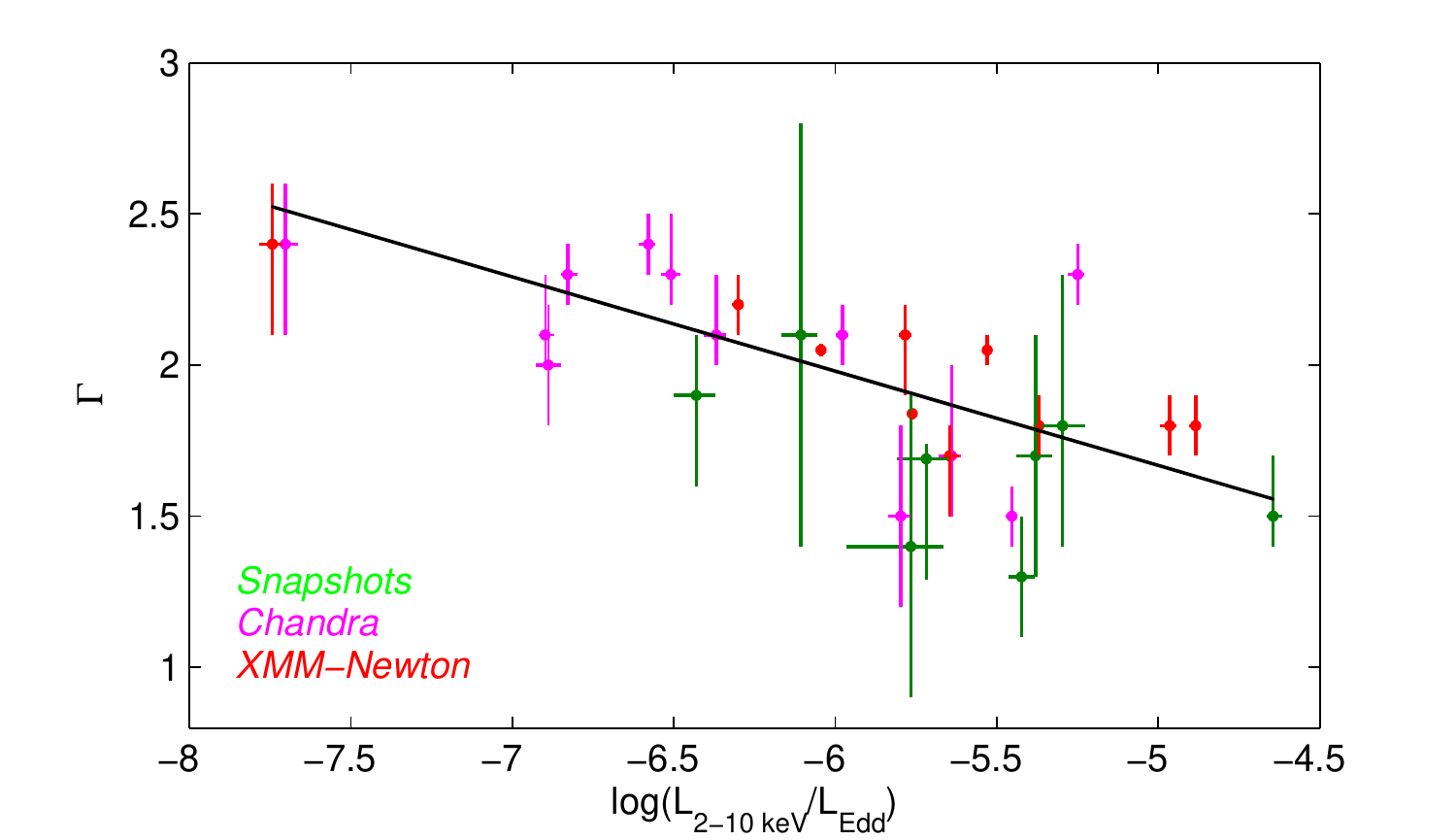}}
\caption{Photon index, $\Gamma$, as a function of the \eddratio\ for our sample of LINER~1s. It is clear these two quantities are strongly anticorrelated. The solid black line designates the least square best fit to a straight line. Snapshots, \chandra, and \xmm\ observations are shown in different colours.}
\label{gmaVSedd}
\end{figure}

\begin{figure*}[!t]
\includegraphics[angle=0,width=0.47\textwidth]{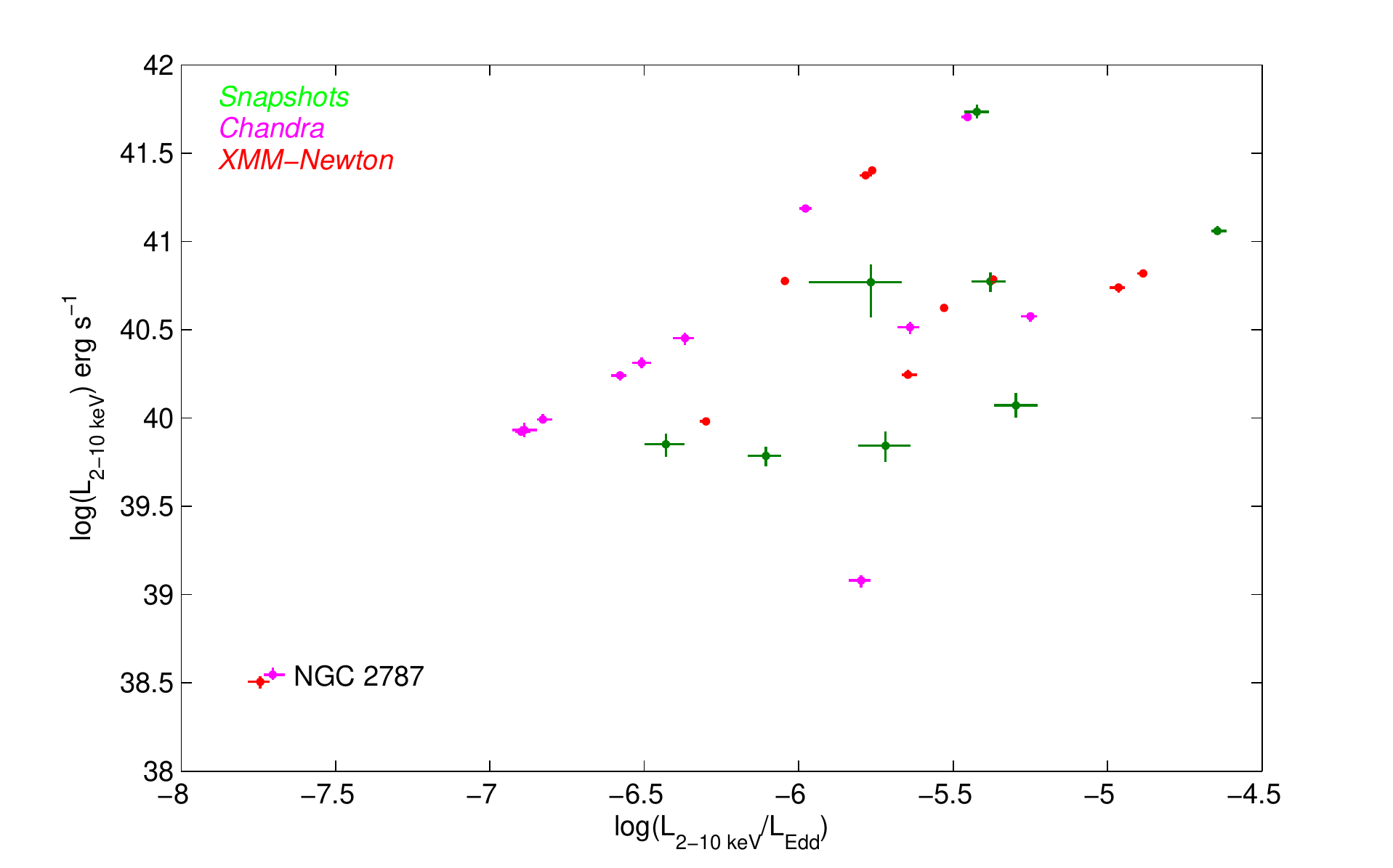}
\includegraphics[angle=0,width=0.47\textwidth]{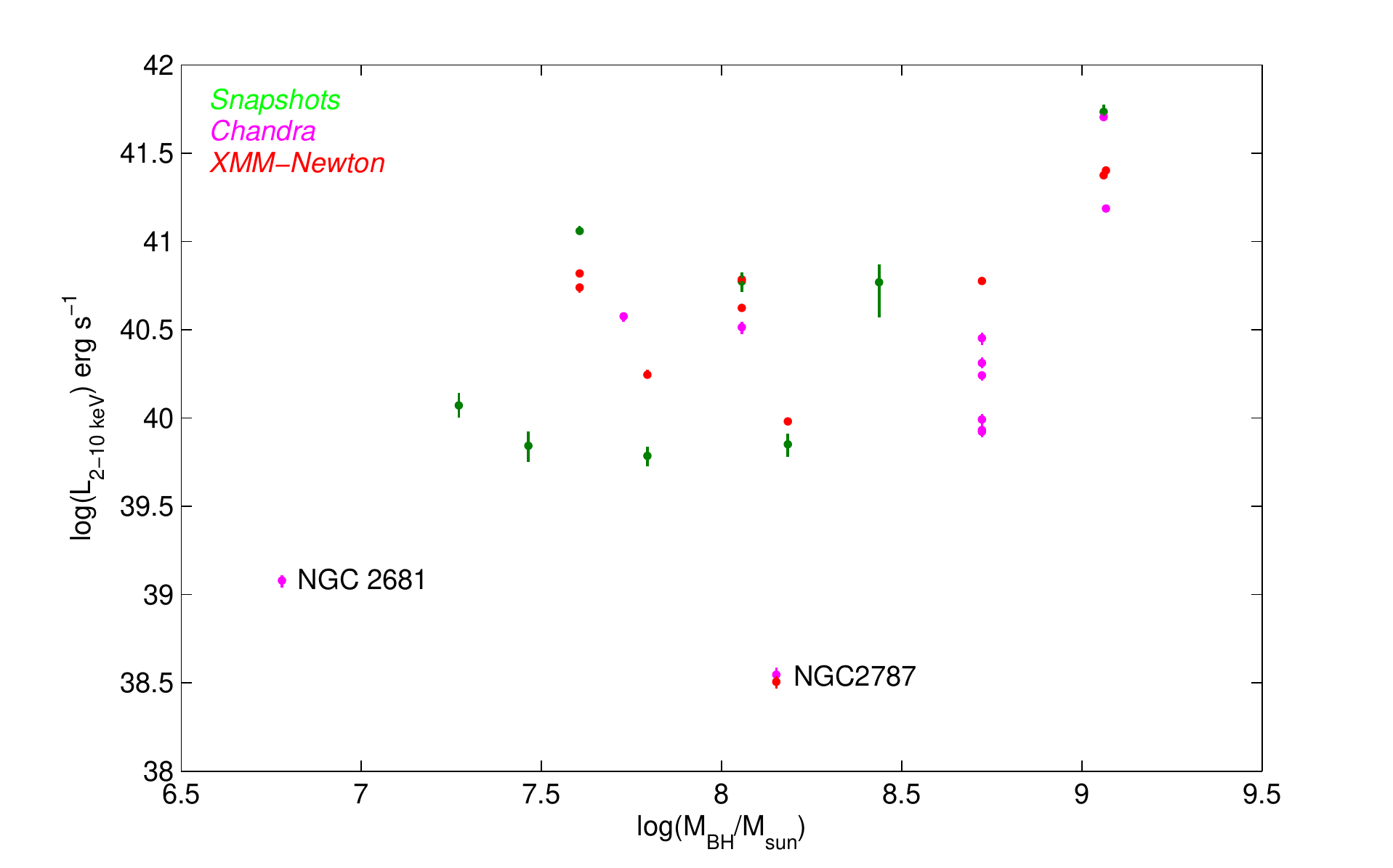}\\
\includegraphics[angle=0,width=0.47\textwidth]{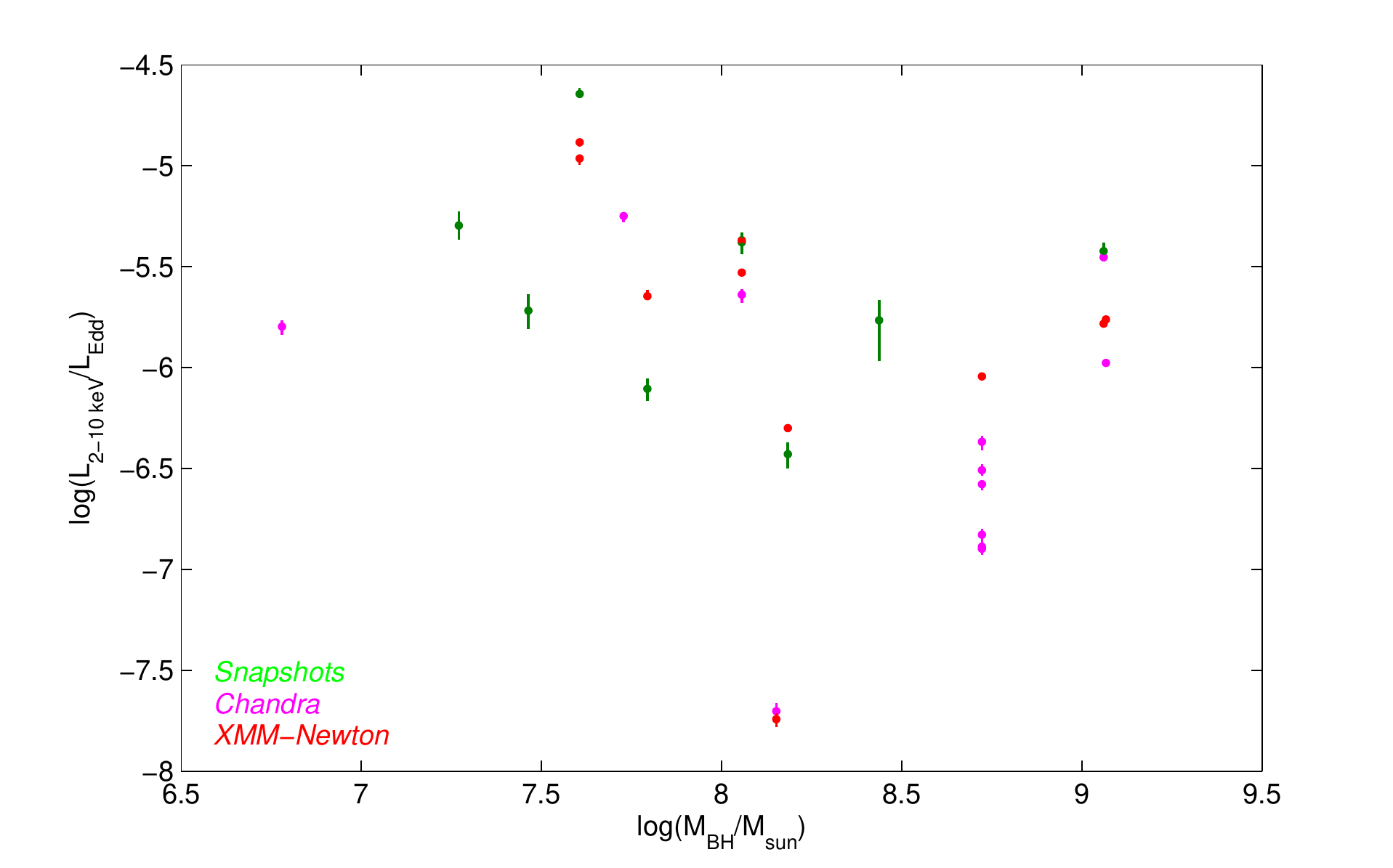}
\includegraphics[angle=0,width=0.47\textwidth]{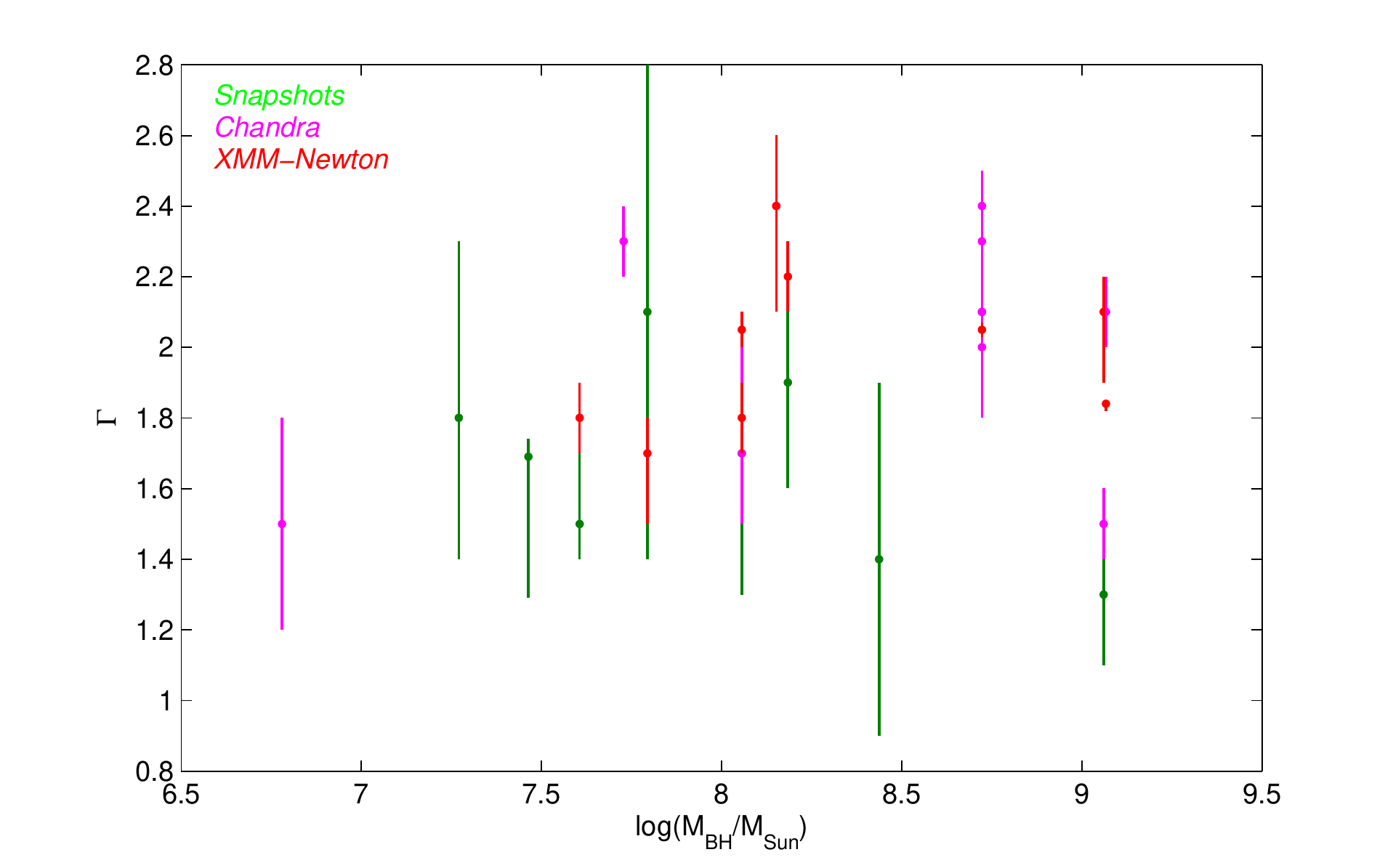}
\caption{{\sl Top left panel.} The positive correlation between the 2-10~keV luminosity and the Eddington ratio. {\sl Top right panel.} The positive correlation between the 2-10~keV luminosity and the BH mass. {\sl bottom panels.} No correlation is found between the BH mass and neither the Eddington ratio ({\sl bottom-left}) nor $\Gamma$ ({\sl bottom-right}). See text for more details.}
\label{other-corr-fig}
\end{figure*}

%-----------
% Table  7
%-----------
\begin{table*}[!th]
\newcommand\T{\rule{0pt}{2.6ex}}
\newcommand\B{\rule[-1.2ex]{0pt}{0pt}}
\begin{center}{
\caption{ Spearman-rank correlation and the \citet{bevington03book} method to check for the dependence between the fit parameters and the intrinsic parameters of our sample of LINER~1s. See text for details.}
\label{spear-rank}
\resizebox{0.8\textwidth}{!}{
\begin{tabular}{l c c c c}
\hline
\hline
Correlations \T \B & Coefficient $r_s$ & Probability (\%) &  Coefficient $r$ & Probability (\%)\\
 \T \B & \multicolumn{2}{c}{Spearman-rank} & \multicolumn{2}{c}{\citet{bevington03book}} \\
\hline
$\Gamma-$\eddratio       \T  & -0.62  & $>$99.9  & -0.65 & $>$99.9 \\
$L_{2-10~keV}-$\eddratio  \T  & 0.61   & $>$99.9  &  0.65 & $>$99.9  \\
$L_{2-10~keV}-M_{BH}$     \T  & 0.43   & 99.0     &  0.70  & $>$99.9 \\
\eddratio$-M_{BH}$       \T  & -0.41  & 97.9     &  -0.11 & 72 \\
$\Gamma-M_{BH}$          \T  &  0.33  & 90       & 0.26 & 90 \\
\hline
\end{tabular}}}
\end{center}
\end{table*}
%-----------
% Table  7
%-----------

We looked for any sign of narrow Fe~K$\alpha$ emission line at 6.4 keV
in the EPIC-pn  \xmm\ observations of our sample  of LINER~1s. None of
the observations  have clear  evidence for the  line.  Good  signal to
noise ratio  around 6~keV is  only found in  the spectra of  4 sources
that  enable the  estimate  of  an upper  limit:  112~eV for  NGC~3718
(obs.ID: 0200430501),  38~eV for NGC~3226  (obs.ID: 0400270101), 33~eV
for  NGC~3998, and  22~eV for  NGC~4278.  Even  though a  hint  for an
Fe~K$\alpha$ emission line around 6.4  keV seems to be apparent in the
EPIC-pn spectrum  of NGC~315, the addition  of a gaussian  line to the
best fit model does not improve  the quality of the fit with an F-test
probability of 60\%\ for an improvement to occur by chance.

\subsection{X-ray correlations}
\label{correlations}

We looked  for any correlations  between the X-ray  properties, mainly
the  photon  index  $\Gamma$  and  the 2-10~keV  luminosity,  and  the
LINER~1s  intrinsic   parameters,  black  hole  mass   and  the  ratio
\eddratio,  which  could be  directly  linked  to  the Eddington  ratio
$L_{bol}/L_{Edd}$ considering that $L_{bol}=const.\times L_{2-10~keV}$
\citep[$const.\approx16$,][]{ho09apj:riaf}.    We   investigated   the
validity of a correlation by fitting  the data with a least square fit
to a  straight line  using the equations  of \citet{york66cjp:linefit}
and by  minimizing the  weighted residuals in  both parameters  $x$ and
$y$.   We assessed  the goodness  of  the fit  following the  criteria
explained in \citet{bevington03book}.  If  a dependent variable $y$ is
correlated  to  a  variable  $x$  with  a line  slope  $b$,  then  the
reciprocity in fitting $x$ as a  function of $y$ should lead to a line
with a slope $b'$.  Therefore, the linear correlation coefficient $r$,
$r\equiv\sqrt{bb'}$, varies  between 0, when there  is no correlation,
to  $\pm1$,  when there  is  complete  correlation.  This  correlation
coefficient  $r$ cannot  be used  directly to  indicate the  degree of
correlation, instead a probability  should be calculated that a random
sample   of   N   uncorrelated   data   points  would   yield   to   a
linear-correlation coefficient as large as or larger than $r$.

We found  a strong anticorrelation between the  photon index $\Gamma$,
and \eddratio, for our sample.  The fit of these two quantities with a
least  square  fit  to  a  straight line  resulted  in  the  following
equation:

\begin{equation}
\Gamma=(-0.31\pm0.06)log(L_{2-10~keV}/L_{Edd})+(0.11\pm0.40)
\end{equation}

with   a   linear-correlation   coefficient  $r\approx-0.65$   and   a
probability greater than 99.99\%  that \eddratio\ and $\Gamma$ yield a
linear-correlation coefficient $\ge  r$.  Additionally, we performed a
spearman-rank  correlation between $\Gamma$  and \eddratio\  and found
that these  two quantities are  correlated with a  probability greater
than 99.99\%\  and a  spearman-rank correlation coefficient  of -0.62.
Figure~\ref{gmaVSedd} shows  the anticorrelation between  $\Gamma$ and
\eddratio,  and  the  least  square  best  fit  to  a  straight  line.
Snapshots,  \chandra, and  \xmm\ observations  are shown  in different
colours.

Using  the same  criteria as  above, we  found a  positive correlation
between the hard X-ray luminosity, $L_{2-10~keV}$, and \eddratio, with
a  linear  correlation coefficient  $r\approx0.65$  and a  probability
$P\ge99\%$   that   these   two    quantities   would   yield   to   a
linear-correlation coefficient as large as or larger than $r$. Another
strong positive correlation  we found is the increase  of the 2-10~keV
luminosity  with   increasing  BH   mass  with  $r\approx0.7$   and  a
probability  greater  than  99.9$\%$.   We  did not  find  any  strong
dependence of the spectral slope $\Gamma$ or \eddratio\ on the BH mass
with  correlation coefficients of  0.26 and  -0.11 respectively  and a
probability  $\le90\%$  that  these  quantities  are  correlated.   To
strengthen our  conclusions we performed  a spearman-rank test  on the
four   different   correlations  and   the   results   are  shown   in
Table~\ref{spear-rank}. Our previous results are in agreement with the
spearman-rank test, except for a weak anticorrelation emerging between
\eddratio\ and  the BH mass  not seen using our  principal correlation
test.   Fig.~\ref{other-corr-fig}  shows  the  dependence of  the  fit
parameters to  the LINER~1s parameters.   We discuss these  results in
\S~\ref{accmode} and \S~\ref{other-corr}.

\section{Discussion}

In  the  following  discussion,  we  are comparing  X-ray  timing  and
spectral results of our sample of LINER~1s to other results derived on
broad  samples of  LINERs  (including type~1  and  type~2 LINERs,  and
transition  nucleus)  and/or  low  luminosity  AGN.   We  compare  our
findings, X-ray  variability and  absence of an  Fe~K$\alpha$ emission
line,  to type~1  luminous  AGN (Seyfert  galaxies  and quasars).   We
discuss  correlations   between  the  fit   parameters,  $\Gamma$  and
$L_{2-10~keV}$, and  the intrinsic  parameters of our  LINER~1 sample,
\eddratio\ and $M_{BH}$,  and compare our results to  luminous AGN and
XRBs in order to find out the accretion mechanism in LINER~1s.

\subsection{X-ray variability}
\label{xrayvar}

One  of  the  most  important  characteristics of  AGN  is  the  X-ray
flux-variability   on  different  time-scales.    An  anti-correlation
between  the variability  amplitude, characterized  by  the normalized
excess  variance $\sigma_{NXS}^2$, with  both the  2-10~keV luminosity
and the black hole mass has been established for a considerable number
of      AGN      \citep{nandra97apj:variance,      turner99apj:seyvar,
  papadakis04mnras:rmsVSmbh,  oneill05mnras:varagn}.  Such variability
on time-scales of less than a day was never detected for LINER sources
in  the past,  with  observations taken  with  low spatial  resolution
telescopes,      e.g.:{\sl      ASCA},      \citep{komossa99aa:liners,
  terashima02apjs:LLAGNASCA}.   \citet{ptak98apj:variance} showed that
LINER  and  low  luminosity  AGN   sources  do  not  follow  the  same
anticorrelation  of  luminous AGN  showing  stronger variability  with
decreasing  luminosity.  The authors  attributed the  non-detection of
short time-scale  variability in  LINERs and low  luminosity AGN  to a
bigger X-ray emitting region, e.g.  RIAF, compared to luminous AGN.

Three  sources in  our  sample of  LINER~1s  (NGC~2787, NGC~4143,  and
NGC~4203)  show  hint  of   inter-day  variability  with  a  K-S  test
probability between  4\%\ and 2\%\ that the  X-ray emission originates
from a constant source.  Two sources in our sample exhibit significant
short time-scale variability, both  already reported in the literature
(NGC~4278: Y10 and NGC~3226: \citealt{binder09apj:ngc3226}).

NGC~4278  exhibited  a  short  time-scale  variability  ($t\sim1.5$~h)
during the  \xmm\ observation where  the X-ray flux level  was highest
(compared  to  the other  6  \chandra\  observations,  where no  short
time-scale variability  was detected).  During  this 1.5~h time-scale,
the  flux of  NGC~4278  increased  by 10\%.   Y10  proposed that  this
variability  could be  the result  of  a more  compact X-ray  emission
region, e.g.  an accretion disk truncated at a lower radius during the
\xmm\ observation compared to the \chandra\ observations.

On the other hand,  NGC~3226 \citep{binder09apj:ngc3226}, which is the
only  source observed  for $\sim$100~ks  with \xmm,  shows significant
flux  variability  during the  entire  observation, increasing  almost
continuously.        \citet{binder09apj:ngc3226}       reported      a
$\sigma_{NXS}^2\approx0.014$, comparable to  0.02 reported here, which
is  a  variability amplitude  similar  to  the  one observed  in  more
luminous AGN, but on shorter time-scales \citep{oneill05mnras:varagn}.
\citet{binder09apj:ngc3226},     assuming     a     BH     mass     of
$1.4\times10^8$~M$_{\odot}$     and    an    Eddington     ratio    of
$2\times10^{-5}$, predicted,  using \citet{mchardy06nat:var} relation,
a  variability  amplitude  of  $\sim2-3\times10^{-4}$  on  a  one  day
time-scale.  The  discrepancy between  the observed and  the predicted
value     of    $\sigma_{NXS}^2$    is     the    fact     that    the
\citet{mchardy06nat:var}  relation   is  derived  for   objects  in  a
high/soft  state.   As   we  show  in  paragraph~\ref{accmode},  LINER
sources,  in contrast  to luminous  Seyfert  galaxies, could  be in  a
low/hard  state  similar  to   XRBs  in  their  low/hard  state.   The
\eddratio\  derived for  the NGC~3226  longest observation,  where the
variability was observed, did not increase compared to the other three
observations,  in  fact   it  decreased  from  4.27$\times10^{-6}$  to
2.95$\times10^{-6}$.   Therefore,  the  significant  variability  seen
during  the  $\sim100$~ks  observation  cannot  be  attributed  to  an
increase      in      $\dot{m}$      (as      also      denied      by
\citealt{binder09apj:ngc3226}). \citet{markowitz05apj:psdllagn} showed
that the  PSD break time-scale of  the LLAGN NGC~4258  is greater than
4.5~days  at $>90$\%  confidence  level.  The  authors suggested  that
LLAGN,  like XRBs  in their  low/hard state,  might have  longer break
time-scales  compared  to  luminous  AGN  and XRBs  in  the  high/soft
state. The X-ray variability detected in the case of NGC~3226 could be
the result of a  break time-scale of $\sim1$~day ($\sim10^{-5}$~Hz) in
the NGC~3226  PSD, however, this assumption is  purely speculative and
PSD measurement  of NGC~3226, which  is not possible with  the present
100~ks observation  due to  a low  number of counts  for this  kind of
study, could help confirm or  discard this idea.  Similar to NGC~3226,
variability  on short  time-scales (half  a day  to several  days) was
observed in the LINER~1 source NGC~3998 \citep{pianmnras10} and in the
low luminosity AGN M~81 \citep{ishisaki96PASJ:m81, iyomoto01MNRAS:m81,
  pianmnras10}.

\begin{figure*}[!t]
\centerline{\includegraphics[angle=90,width=0.99\textwidth]{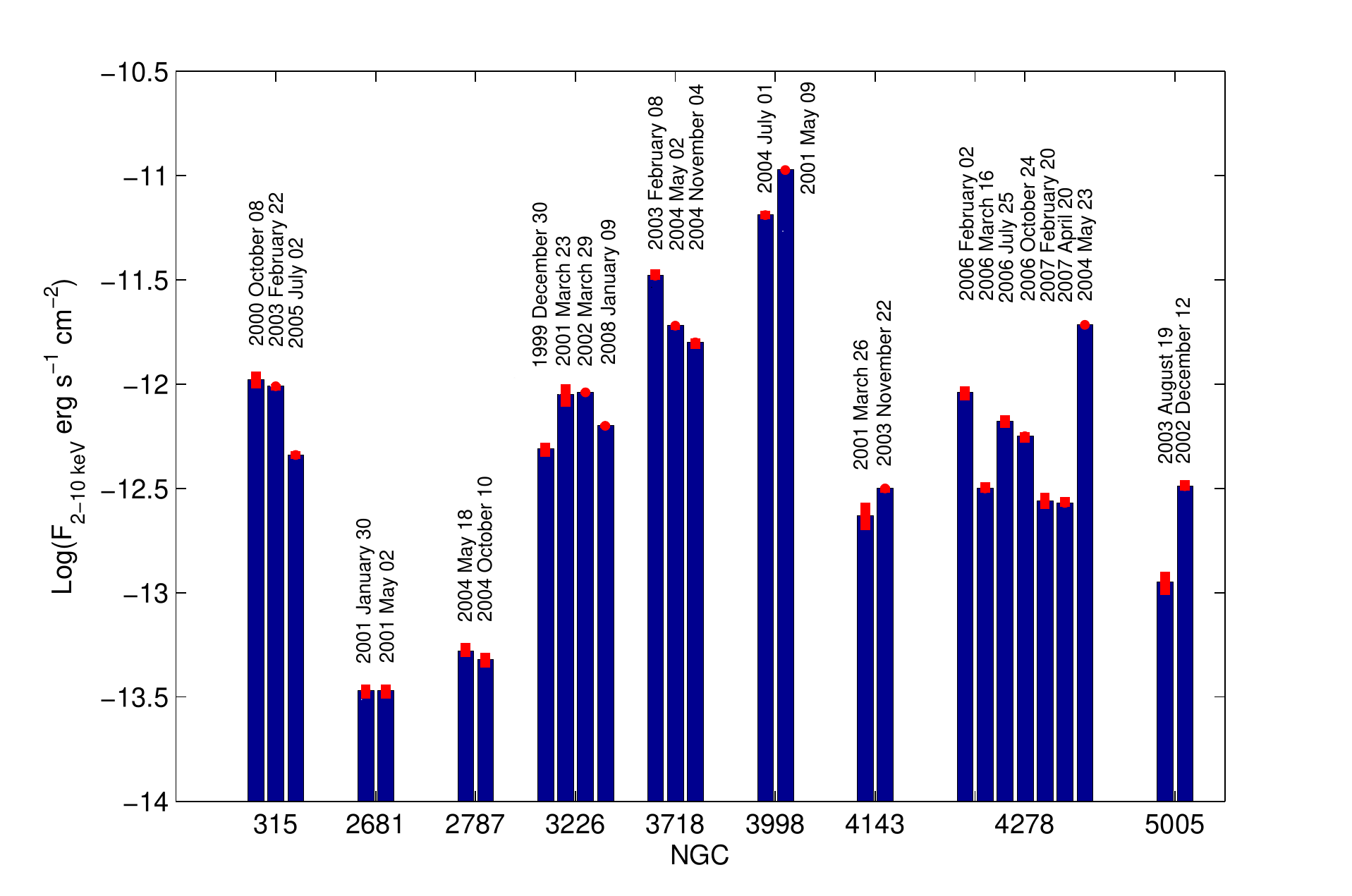}}
\caption{Long term X--ray variability of the LINER~1s observed more than once. Red ticks represent the error on the 2-10 keV corrected flux.}
\label{longtermflux}
\end{figure*}

Although all of the normalized  excess variance derived for our sample
consist   of  upper   limits,  except   for  NGC~3226,   we   show  in
Fig.~\ref{NXSvsMbh}  $\sigma_{NXS}^2$  as   a  function  of  BH  mass.
Fig.~\ref{NXSvsMbh} rules  out any variability  on time-scales shorter
than  50~ks in  our  sample  of LINER~1s,  commonly  observed in  more
luminous  AGN \citep{oneill05mnras:varagn},  except for  NGC~4278 that
shows a 10\%\ flux increase in a $\sim1.5$~h time-scale.

Variability  on  time-scales of  months  to  years  is common  in  our
sample. Seven  out of  nine sources observed  for more than  once show
variability  on  months (NGC~5005  and  NGC~4278)  to years  (NGC~315,
NGC~3226,     NGC~3718,    NGC~3998,    NGC~4143,     and    NGC~4278)
time-scales\footnote{All of  the sources reported  to show variability
  on  years time-scale and  not on  months time-scale  lack inter-year
  observations and  therefore, monthly variability  cannot be tested.}
with a  5 times flux increase  in the most  variable source, NGC~4278,
and a 1.4 times flux  increase in the least variable source, NGC~4143.
The only two sources that  do not show variability on long time-scales
are  the  ones  with  a  very low  \eddratio:  NGC~2681  and  NGC~2787
(Table~\ref{specfit-fluxes}).   Both sources  were  observed twice  in
less   than  a  5   months  period   and  therefore,   variability  on
years-timescale  cannot  be  tested.   For illustration  purposes,  we
report in Figure~\ref{longtermflux} the fluxes derived for the sources
observed more than once. All  but NGC~2681 and NGC~2787 exhibit months
and/or years time-scale variability.

With  the increasing  number of  X-ray observations  of LINER  and low
luminosity AGN  sources, more  sources are revealing  short time-scale
variability ($\le1$~day).  This variability  is detected with the help
of  the current generation  of X-ray  telescopes (\chandra,  \xmm, and
{\sl  Swift}) which  have a  good  spatial resolution  to isolate  the
nucleus X-ray  emission from contaminating  sources. Nevertheless, and
before  any  firm conclusions  are  made  about  variability from  low
accretion  rate  sources, a  homogeneous  well  sampled population  of
LINERs  and/or  low  luminosity  AGN,  with  long  X-ray  observations
($\sim100$~ks), should be available.

\subsection{X-ray spectral shape}

Most recently, \citet{zhang09apj:llagnxray}  studied the X-ray nuclear
activity of  a distance limited  sample, D$<$15~Mpc, of  187 galaxies.
The  authors found  that the  X-ray emission  from $\sim$60\%\  of the
elliptical and  early type spiral galaxies is  consistent with nuclear
accreting supermassive BH.   They fit the spectra of  each source with
an absorbed  power-law and  found a photon  index $\Gamma\approx1.5-2$
and an intrinsic column density  covering almost 4 orders of magnitude
from $10^{20}$~cm$^{-2}$ to  $10^{24}$~cm$^{-2}$.  The authors found a
luminosity     ranging      from     $\sim10^{38}$~erg~s$^{-1}$     to
$\sim10^{42}$~erg~s$^{-1}$          corresponding         to         a
L$_{0.3-8~keV}/$L$_{Edd}$      ratio     between      $10^{-4}$     to
$10^{-8}$. Similar results  were derived in previous work  such as the
work of  \citet{gonzalezmartin06aa,gonzalezmartin09aa} on a  sample of
82 nearby  LINER sources. The authors  found that 60\%\  of the sample
shows  a hard unresolved  X-ray point-like  source in  the 4.5-8.0~keV
band consistent with  an AGN assumption.  The data  were fit mainly by
an absorbed power-law, with a mean photon index $\Gamma=2.1\pm0.5$ and
an  intrinsic  column  density  ranging  from  $10^{20}$~cm$^{-2}$  to
$10^{24}$~cm$^{-2}$, and  a thermal component with  a mean temperature
$kT=0.5\pm0.3$~keV.

The  results we  derive  for our  homogeneous  and optically  selected
sample  of  LINER~1s showing  definite  detection  of broad  H$\alpha$
emission are  in agreement  with all of  the results derived  on broad
samples of LINERs and low luminosity AGN in the past. X-ray spectra in
the  whole 0.5-10~keV  band of  the majority  of our  sample  are well
fitted with  an absorbed  power-law (9/13). This  could mean  that the
X-ray emission  in the soft and  in the hard band  is originating from
the  same  region. In  the  remaining  4  sources (NGC~315,  NGC~2681,
NGC~4278, and NGC~5005), we included  a thermal component to take into
account some low  energy residuals, most likely from  diffuse hot gas.
We  found a  photon index  ranging  from $1.3\pm0.2$  for the  hardest
source to $2.4^{+0.2}_{-0.3}$ for the softest one with a mean value of
$1.9\pm0.2$ and a dispersion  $\sigma=0.3$, similar to values reported
in  \citet{komossa99aa:liners}  and \citet{terashima02apjs:LLAGNASCA}.
The absorption column density observed  in our sample spans two orders
of  magnitude from  $10^{20}$~cm$^{-2}$ for  unabsorbed  sources (e.g.
NGC~3998,  NGC~4143) to  $10^{22}$~cm$^{-2}$ for  the  mildly absorbed
source  NGC~3718.   This  is  consistent  with the  fact  that  strong
absorption is  not expected in objects showing  broad optical emission
line, e.g.   LINER~1s, as found  in luminous type~1 AGN.   This result
was already confirmed by \citet{terashima02apjs:LLAGNASCA} who studied
the {\sl  ASCA} X-ray observations of  21 LINER and  17 low luminosity
Seyfert  galaxies. The authors  found a  discrepancy in  the intrinsic
absorbing     column     density     between    type     1     LINERs,
$N_{H}<10^{22}$~cm$^{-2}$ (except  for NGC~1052 which  is not included
in our  sample, see \S~\ref{sec:sample}),  and type 2  LINERs absorbed
with   a    column   density    consistent   with   $N_{H}$    a   few
$10^{22}$~cm$^{-2}$.  The  origin of  the slight excess  absorption in
some of the  sources in our sample of LINER~1s could  be due to either
the host  galaxy and/or  some compton thin  material intrinsic  to the
central engine. We find a mean value to the temperature of the thermal
component  of $0.63\pm0.06$~keV typical  of diffuse  hot gas  in early
type galaxies \citep[$kT\approx0.5-2.0~keV$,] []{fabbiano89aa:difgaz}.
Similar $kT$ values were  reported in \citet{flohic06apj} for a sample
of 19 LINER sources observed with \chandra.  The 2-10~keV luminosities
of    the   LINER~1s    in   our    sample   span    a    range   from
3.2$\times10^{38}$~erg~s$^{-1}$   to   5.4$\times10^{41}$~erg~s$^{-1}$
which  resulted   in  \eddratio\  range   from  $2.0\times10^{-8}$  to
$2.3\times10^{-5}$, in agreement with results derived on broad samples
of  LINERs and  low luminosity  AGN.  These  corresponding ``Eddington
ratios''  are  at least  an  order  of  magnitude smaller  than  those
reported                for                luminous                AGN
\citep{porquet04aa:pgquasar,nandra07mnras:felinesey}.

Although the  photon indices measured  for our sample of  LINER~1s are
similar      to     those      of     type~1      Seyfert     galaxies
\citep{nandra97apj:SEYfekline},  this does  not necessarily  mean that
the X-ray emission  in LINER~1s is originating from  an accretion flow
similar    to   that   of    more   luminous    galaxies   \citep[e.g.
  NGC3998,][]{ptak04apj:ngc3998}.   Other  X-ray  timing and  spectral
aspects could help shed light on the accretion mechanism in LINER~1s.

\subsection{The absence of an Fe K$\alpha$ line}

It is now believed that a narrow emission Fe K$\alpha$ line at 6.4~keV
is  a common feature  in the  X-ray spectra  of Seyfert  galaxies. The
origin of this neutral narrow emission line is generally attributed to
fluorescence  originating  from   parsec-scale  distances  (torus)  to
distances      closer     than      the     broad      line     region
\citep{shu10apjs:iteffect}.  An X-ray Baldwin effect was discovered by
\citet{iwasawa97apj:iteffect} where  the EW  of the Fe  line decreases
with  increasing  luminosity.  \citet{page04mnras:iteffect}  suggested
that this Baldwin  effect observed in X-rays could be  the result of a
luminosity-dependent  covering fraction  of the  putative  torus.  The
increase in radiation pressure flattens  the torus leading to a bigger
opening  angle  for the  torus  and,  hence,  smaller covering  factor
\citep{konigl94apj:torus}.  This effect  was later confirmed on bigger
samples       of       radio-quiet       and      radio-loud       AGN
(\citealt{nandra97apj:iteffect};        \citealt{bianchiaa07:iteffect};
\citealt{chaudharyaa10:iteffect},         but         also         see
\citealt{jim05aa:iteffect}).   LINER~1s  individually  studied in  the
literature with  a high  signal to noise  ratio around 6.4~keV  do not
show  any sign  of  Fe~K$\alpha$ emission  line  with stringent  upper
limits on the EW:  25~eV for NGC~3998 \citep{ptak04apj:ngc3998}, 35~eV
for  NGC~3226  \citep{binder09apj:ngc3226},   and  22~eV  for  NGC4278
(Y10). Therefore, it appears that the X-ray Baldwin effect do not hold
down to very low luminosity AGN.

We do  not detect  any significant Fe~K$\alpha$  emission line  in our
sample with  upper limits  obtained for the  sources with  the highest
signal to noise ratio around 6~keV (38~eV: NGC~3226, 112~eV: NGC~3718,
33~eV: NGC~3998,  and 22~eV: NGC~4278).   If the broad line  region is
responsible for the emission of an Fe~K$\alpha$ line, such as the case
for     the     intermediate     Seyfert-LINER     source     NGC~7213
\citep{lobban10mnras:ngc7213,bianchi08mnras:ngc7213},  we would expect
to detect it  in at least the sources with high  signal to noise ratio
around  6.4~keV,  since our  sample  consists  of  LINER~1s showing  a
definite detection  of {\sl  broad} H$\alpha$ emission.   Instead, the
disappearance  of  the torus  structure  at  low  Eddington ratios  as
suggested  by  \citet{ho08aa:review} and  \citet{zhang09apj:llagnxray}
could explain  the lack of Fe  emission lines in  our sample, assuming
that the torus is responsible for the formation of the fluorescence Fe
line. Interestingly, we find that the highest upper limit on the EW of
the  Fe~K$\alpha$ line  is found  for NGC~3718  which has  the highest
hydrogen column density  ($N_H\approx10^{22}$~cm$^{-2}$) in our sample
of LINER~1s.

\subsection{Accretion mode in LINER~1s?}
\label{accmode}

The accretion mechanism responsible for  the bulk of energy from radio
to   X-rays   in   LINER   sources   is   still   poorly   understood.
\citet{ho09apj:riaf}  recently demonstrated  that  through local  mass
loss from evolved stars and  Bondi accretion of hot gas, the accretion
rate supply needed  for the luminosities observed in  LINERs and other
LLAGN  is easily  attained.  The  author  argued that  the gas  supply
present at the  center of nearby galaxies should  generate more active
nuclei and the  luminosity deficit seen in the  nearby universe is the
result of a low  radiative efficiency. Indeed, radiatively inefficient
accretion         flow          \citep[RIAF,         see         ][for
  reviews]{narayan08:riafreview,quataert01aspc:riaf}  models have been
applied  to a  growing number  of LINERs  and LLAGN  to  explain their
energy  budget  and  their  spectral  energy  distribution:  M~81  and
NGC~4579           \citep{quataert99apj:m81ngc4579},          NGC~4258
\citep{gammie99apj:ngc4258},  NGC~3998  \citep{ptak04apj:ngc3998}, and
NGC~1097  \citep{nemmen06apj:ngc1097}.   \citet{wu07apj:riaf} fit  the
SED  of a  small sample  of eight  FR~I sources  with RIAF  and/or jet
models  and confirmed  the  prediction of  \citet{yuan05apj:jetvsriaf}
that    below   a    critical   value    of   the    Eddington   ratio
(\eddratio$\approx10^{-6}$)  the X-ray  emission becomes  dominated by
the jet  rather than the RIAF. \citet{ho09apj:riaf}  suggested an even
lower value to distinguish between objects in the ``low'' state, where
an  outer thin  disk persists,  and those  in the  ``quiescent'' state
containing  a pure  RIAF.  Since  we  are only  considering the  X-ray
properties  of this sample  in our  study, we  decided to  examine the
$\Gamma-$\eddratio\ relation which was shown to be a good indicator of
accretion rate in luminous AGN \citep{shemmer06apj:rqagn}.

In  luminous AGN,  a positive  correlation is  found between  the hard
X-ray power-law slope and the ratio  of the 2-10 keV luminosity to the
Eddington   luminosity   \citep{wang04apjl:rqagn,  shemmer06apj:rqagn,
  sobolewska09mnras:gamvsedd}.   \citet{greene07apj:imagn} showed that
the relation  holds for intermediate-mass  ($10^5-10^6$~M$_\odot$) BHs
in  active galaxies and  \citet{porquet04aa:pgquasar} showed  that the
relation extends up to more luminous objects when studying a sample of
21 low-redshift  quasars.  A viable  explanation is that  whenever the
disk  emission increases,  the corona,  the origin  of the  hard X-ray
emission, cools  more efficiently exhibiting a steepening  of the hard
X-ray spectrum.  This relation  was examined for LLAGN (local Seyferts
and  LINERs) by  \citet{gu09mnras:gamVSeddllagn}  (see also  \citealt{
  constantin09ApJ:liners})  who  found  a significant  anticorrelation
between the hard X-ray photon  index $\Gamma$ and the Eddington ratio,
L$_{bol}$/L$_{Edd}$=30$\times$\eddratio,   for   the   local   Seyfert
galaxies in  their sample.  However,  no strong correlation  was found
when considering  only the  LINER sources in  their sample  owing most
likely to  heterogeneous fitting models  as they have  collected their
data points  from different studies.  The authors  suggested that this
anticorrelation found  in their  sample, which is  in contrast  to the
positive correlation  found for more luminous AGN,  could signify that
LLAGN   resemble  XRBs  in   the  low/hard   state  where   a  similar
anticorrelation  is found  \citep{yamaoka05cjaa:XRBhs, yuan07ApJ:xrbs,
  wu08apj:XRBhs}.     \citet{wu08apj:XRBhs}    suggested   that    the
anticorrelation could  mean an accretion  mode consistent with  a RIAF
whereas  a  positive  correlation  could  mean the  existence  of  the
classical  thin  accretion  disk.   A plausible  explanation  for  the
hardening of  the spectrum as the  accretion rate increases  in a RIAF
context (as  seen in  the $\Gamma$-\eddratio\ anticorrelation)  is the
increase of the  optical depth of the RIAF which  leads to an increase
in the Compton $y$-parameter resulting in a harder X-ray spectrum.

Our  well  defined  optically-selected  sample  of  LINER~1s  and  our
homogeneous data analysis techniques  allowed us to establish a strong
anticorrelation  between $\Gamma$  and  \eddratio\ for  our sample  of
LINER~1s (see \S~\ref{correlations}), not  seen in the LINER sample of
\citet{gu09mnras:gamVSeddllagn}.  This  strong anticorrelation support
the idea suggested by \citet{gu09mnras:gamVSeddllagn} that LLAGN might
be  similar  to XRBs  in  the low/hard  state  where  the emission  is
presumably  generated  in  a RIAF  structure.   \citet{qiao10pasj:LHS}
predicted  such an  anticorrelation in  the low/hard  state  for their
accretion flow model consisting  of an outer-cool optically-thick disk
and  inner-hot optically-thin RIAF  within the  framework of  disk and
corona     with     mass     evaporation     \citep{liu02apj:LHSriaf}.
\citet{qiao10pasj:LHS}  found  that  their  model  can  reproduce  the
anticorrelation between the X-ray photon index and the Eddington ratio
observed for  the X-ray  binary \object{XTE~J1118+480}.  

\begin{figure}[!t]
\centerline{\includegraphics[angle=0,width=0.5\textwidth]{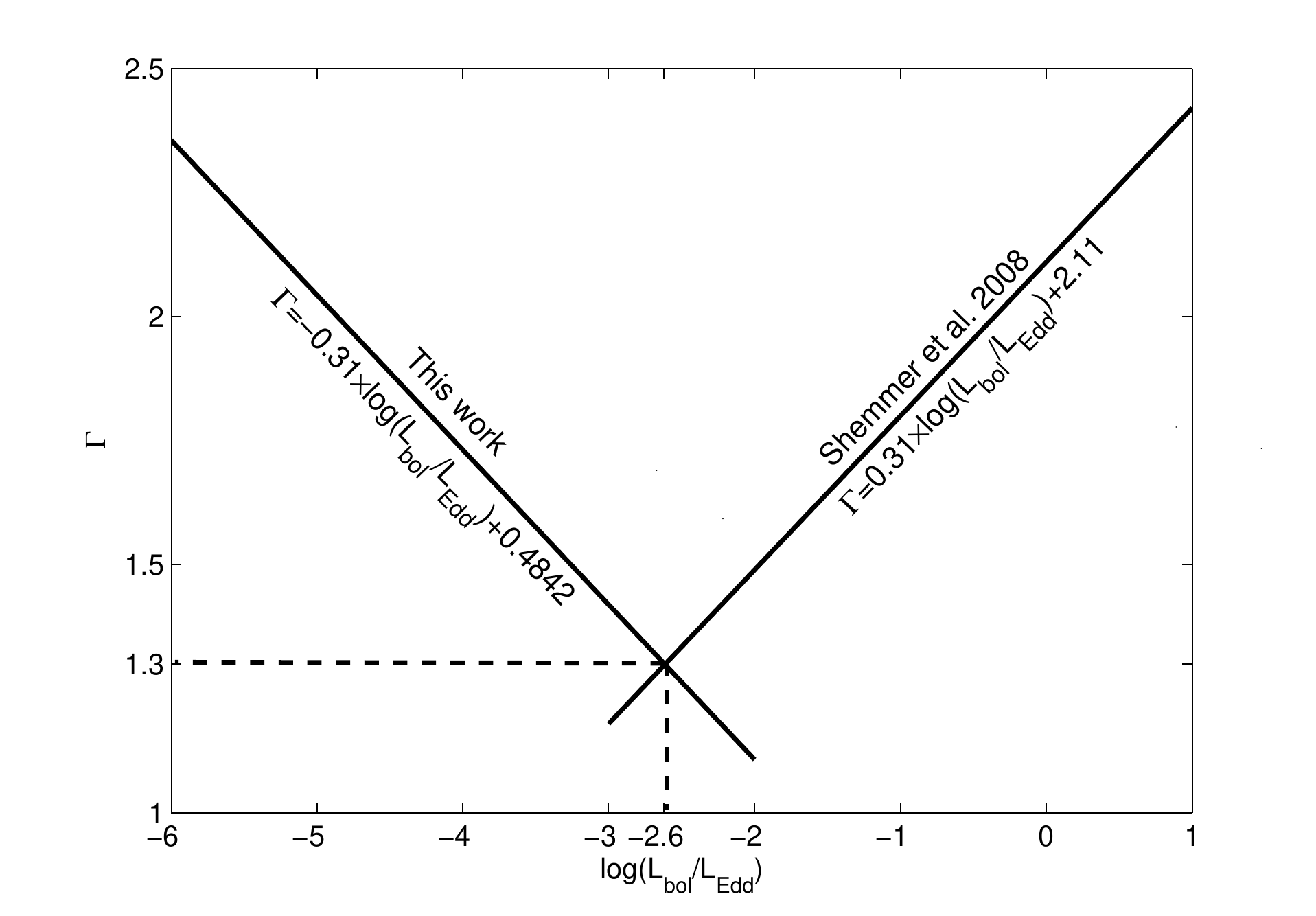}}
\caption{Dependence of $\Gamma$ as a function of the Eddington ratio. The positive and negative correlations represent the ones found for a sample of luminous AGNs \citep{shemmer08apj:gamvsedd} and for our sample of LINER~1s respectively. Note here that the intercept of our anticorrelation has changed since now we are considering $L_{bol}$ instead of $L_{2-10~keV}$. The crosspoint between the two lines represents a probable transitional point from a standard thin accretion disk to a RIAF in AGNs.}
\label{diff-state}
\end{figure}

\citet{wu08apj:XRBhs}  noted  that  the  transition  between  the  two
different accretion  modes in  their sample of  XRBs is  different for
different sources but roughly converge to the transitional point where
$\Gamma=1.5\pm0.1$   and  $log(L_X(0.5-25$~keV)$/L_{Edd})=-2.1\pm0.2$.
Assuming that the accretion mode  in the high luminosity AGN sample of
\citet{  shemmer08apj:gamvsedd}  is the  standard  thin  disk and  the
accretion  mode  in  our  sample  of  LINER~1s is  a  RIAF,  then  the
transition point between  the two accretion modes in  AGNs would be at
$\Gamma\approx1.3_{-0.3}^{+0.2}$   and  $log(L_{bol}/L_{Edd})  \approx
-2.6_{-0.8}^{+0.7}$ (Fig.~\ref{diff-state})  in good agreement, within
the error  bars, with  the values  reported for the  XRBs and  for the
sample  of \citet{constantin09ApJ:liners}.   One should  keep  in mind
that these two  values are affected by several  parameters, mainly the
BH  mass  and  the  bolometric  luminosity,  that  are  calculated  by
different means  between \citet{shemmer08apj:gamvsedd} and  this work.
Indeed, \citet{shemmer08apj:gamvsedd} calculated the BH masses and the
Eddington ratios, $L_{bol}/L_{Edd}$, of  their sample of luminous AGNs
using  the  $\nu  L_{\nu}$(5100\AA)  and FWHM(H$\beta$),  whereas  our
calculation of the BH masses of our sample are based on the $M-\sigma$
relation  and  $L_{bol}/L_{Edd}$ is  calculated  using the  bolometric
correction $L_{bol}=16~L_X$ of \citet{ho08aa:review}.

We  do  not  detect  any   sharp  cut  below  the  critical  value  of
\eddratio\  of 10$^{-6}$ indicating  a change  in the  X-ray emission,
becoming    dominated   by   synchrotron    emission   from    a   jet
\citep{yuan05apj:jetvsriaf}.  This  could be due  to selection effects
where none of our sources,  except for NGC~315, comprises strong jets.
Moreover,  a transition between  a ``low''  and a  ``quiescent'' state
whenever   the   accretion  rate   drops   below   a  critical   value
($\dot{m}\approx10^{-6}$)    as    suggested    most    recently    by
\citet{ho09apj:riaf}  could not be  tested since,  as stressed  by the
author, this kind of analysis should be conducted on large samples.

\subsection{Other correlations and implications}
\label{other-corr}

We  found a positive  correlation between  the hard  X-ray luminosity,
$L_{2-10~keV}$,  and  \eddratio.   An   increase  in  the  hard  X-ray
luminosity with increasing \eddratio\  is independent of the accretion
model and seen  in different types of galaxies  from transition nuclei
to     high     luminosity      Seyfert     nuclei     and     quasars
\citep{ho09apj:riaf,greene07apj:seyquas}.    Another  strong  positive
correlation we found for our sample of LINER~1s is the increase of the
2-10~keV luminosity with  increasing BH mass. A similar  trend is seen
in a  sample of  112 early-type galaxies  within a distance  of 67~Mpc
\citep{pellegriniapj10:xemiearlygal}.  Therefore, both the BH mass and
the \eddratio\ (equivalent to the Eddington ratio L$_{bol}$/L$_{Edd}$)
are driven factors  of the hard $2-10~$keV luminosity.   This could be
seen  in  the  right   upper  panel  of  Fig.~\ref{other-corr-fig}  in
\S~\ref{correlations} where NGC~2787 with a relatively high BH mass of
$1.4\times10^8$~M$_{\odot}$  has  a   low  2-10~keV  luminosity  as  a
consequence of  a low \eddratio\ ($2\times10^{-8}$).  We  did not find
any strong dependence of the  spectral slope $\Gamma$ or the Eddington
ratio on the BH mass (although a spearman-rank correlation showed that
\eddratio\  is weakly  anticorrelated to  the  BH mass).   This is  in
contrast with  the anti-correlation  between the X-ray  spectral slope
and  the  BH  mass  seen  in  a sample  of  low  redshift  PG  quasars
\citep{porquet04aa:pgquasar}.   It  appears  that  in  our  sample  of
LINER~1s, which  have a very low  accretion rate, the BH  mass -- here
spanning 2  orders of magnitude in  log($M_{BH}$) space --  is not the
main driver of the X-ray photon index.

\section{Summary and conclusion}

In the  present work, we studied  the X-ray properties of  a sample of
nearby LINER galaxies showing  definite detection of broad H${\alpha}$
emission  \citep{ho97apjs:broadHal}.    Such  a  sample   insures  the
responsibility of accretion into a SMBH for the detection of the broad
emission lines, guarantees the non-existence of large obscuration, and
enables X-ray comparison of this class with both XRBs and type 1 AGN.

Only  two sources  in our  sample  exhibit significant  hours to  days
time-scale  variability. The NGC~4278  flux increased  by a  factor of
10\%\ on a  $\sim1.5$ hour period, to remain constant  for the rest of
the  observation  (see  Y10).   On  the  other  hand,  NGC~3226  shows
variability  for  the whole  observation  of $\sim100$~ks,  increasing
continuously.  Three other sources  show hint of inter-day variability
with  a K-S  test probability  between 2\%\  and 4\%\  that  the X-ray
emission  originates   from  a  constant   source.   Short  time-scale
variability study from a homogeneous well sampled population of LINERs
and/or low luminosity AGN, with long X-ray observations ($\sim100$~ks)
should  be  conducted  before  any  firm conclusions  are  made  about
variability  from low  accretion  rate sources.   On  the other  hand,
variability on longer  (months to years) time-scales is  common in our
sample where 7  out of 9 sources exhibit  long time-scale variability.
The two  sources not exhibiting variability  are the ones  with a very
low \eddratio.

The X-ray spectra  of our sample of LINER~1s are  typical of all types
of LINER sources, fit with  an absorbed power-law, or a combination of
a  thermal component  and an  absorbed power-law.   We found  a photon
index for  our sample between  $1.3\pm0.2$ for the hardest  source and
$2.4^{+0.2}_{-0.3}$  for  the  softest   one  with  a  mean  value  of
$1.9\pm0.2$ and a dispersion $\sigma=0.3$.  None of the sources in our
sample is heavily absorbed  with NGC~3718 having the highest intrinsic
hydrogen  column  density  of  $\sim10^{22}$~cm$^{-2}$.   The  thermal
component had a mean  temperature kT$\approx0.6$~keV, typical of other
LINER sources  embedded in diffuse  emission \citep{flohic06apj}.  Our
sample  spans  three  orders  of  magnitude  in  both  luminosity  and
\eddratio\ space, ranging from 10$^{38}$ to 10$^{41}$~erg~s$^{-1}$ and
from 10$^{-8}$ to  10$^{-5}$ respectively, which is at  least an order
of magnitude  smaller than the  Eddington ratios observed  in luminous
AGN (Seyferts and quasars).

We do not detect any significant Fe~K$\alpha$ emission line at 6.4~keV
in the spectra of our sample  of LINER~1s. We obtained upper limits on
the Fe  line for  the four  sources with the  highest signal  to noise
ratio  around  6.4~keV   (38~eV:  NGC3226,  112~eV:  NGC~3718,  33~eV:
NGC3998, and 22~eV: NGC~4278).  The lack  of a narrow Fe line could be
due  to the  disappearance of  a torus  structure in  LLAGN  and LINER
sources \citep{ho08aa:review,zhang09apj:llagnxray}.  This implies that
the  X-ray  Baldwin  effect  or the  ``Iwasawa-Tanigushi''  effect  of
decreasing  Fe~K$\alpha$ EW with  increasing 2-10~keV  luminosity does
not extend down to LINER~1s.

Finally, we  looked for correlations between the  X-ray properties and
the  AGN  properties  of  our  LINER~1  sample.   We  found  a  strong
anticorrelation between  the power-law photon index  and the Eddington
ratio suggesting  that LINER~1s differ from more  luminous Seyfert and
quasar galaxies  that show a  positive correlation between  the photon
index and the Eddington  ratio.  This anticorrelation, established for
the first time  for LINER~1s, suggest that LINER~1s  mode of accretion
could be similar  to that of XRBs in their low-hard  state, and a RIAF
could be  responsible for  the emitting energy  from the  nucleus.  We
found  that  the  2-10~keV  luminosity  in our  sample  is  positively
correlated to  two parameters,  the BH mass  and the  Eddington ratio,
\eddratio, confirming  the results found  for broad samples  of LINERs
and low  luminosity AGN.  On  the other hand,  it appears that  in our
sample of LINER~1s, which have a  very low accretion rate, the BH mass
is  not  the  main driver  of  the  X-ray  photon  index, as  the  two
quantities do not show any strong correlation.

\acknowledgements This research has made use of the data obtained from
the {\sl Chandra  Data Archive} and the {\sl  Chandra Source Catalog},
and software provided  by the {\sl Chandra X-ray  Center} (CXC) in the
application  packages   CIAO  and  Chips.   This  work   is  based  on
observations with  \xmm, an ESA  science mission with  instruments and
contributions directly funded by ESA Member States and the USA (NASA).
This research  has made use of  the SIMBAD database,  operated at CDS,
Strasbourg,  France.  This  research  has made  use  of the  NASA/IPAC
Extragalactic Database  (NED) which is operated by  the Jet Propulsion
Laboratory,  California Institute of  Technology, under  contract with
the National Aeronautics and Space Administration. G.Y.  would like to
thank  I.  Papadakis  for helpful  and enlightening  discussions.  The
authors would  also like  to thank the  referee for  fruitful comments
that improved the quality of the manuscript.

%
% Appendix
% 

\begin{appendix}
\section{Notes on individual sources}
\label{appa}

{\sl NGC~266.}  \citet{terashima03apj:rloud} studied the 2~ks snapshot
observation made  with \chandra\ and  discussed here.  They  model the
X-ray spectrum with an absorbed power-law and derive a hydrogen column
density and a photon  index of $<0.82\times10^{22}$~cm$^{-2}$ and 1.4,
respectively.      They     find     a     2-10    keV     flux     of
$1.6\times10^{-13}$~erg~s$^{-1}$~cm$^{-2}$.  The  values of the photon
index and the  2-10~keV flux are with a good  agreement with the value
we report here  (within the error bars), but  no additional absorption
is added to our model.

{\sl   NGC~315.}   \citet{worrall94apj:ngc315}  first   suggested  the
presence of an active galactic  nucleus at the center of NGC~315 using
\rosat\    data.    This   assumption    was   later    confirmed   by
\citet{matsumoto01pasj:ngc315} when studying  a 37~ks ASCA observation
(see  also \citet{terashima02apjs:LLAGNASCA}).   The  authors fit  the
hard 2-10~keV  spectrum with  a power-law and  find a photon  index of
$\sim2$  and  a   luminosity  of  $3.1\times10^{41}$~erg~s$^{-1}$.   A
resolved    X-ray    jet    emission    was    first    reported    by
\citet{worrall03mnras:ngc315}  when  studying  the \chandra\  snapshot
observation  (obs. ID:  855).   The fit  to  the jet  emission with  a
power-law    gave   a    photon   index    of   $2.5\pm0.7$    and   a
$3.5\times10^{40}$~erg~s$^{-1}$  luminosity.    The  authors  fit  the
unresolved core emission with  a moderately absorbed power-law with an
intrinsic  hydrogen column  density  of $\sim5\times10^{21}$~cm$^{-2}$
and a photon index of $1.4\pm0.4$. \citet{worrall07mnras:ngc315} found
similar  results  when   studying  the  longer  \chandra\  observation
reporting a harder core spectrum  than the jet with photon indicies of
$\sim1.6$       and      $\sim2.2$       respectively.       Moreover,
\citet{croston08mnras:radgal}, after solar flare cleaning, studied the
only \xmm\  observation and fit  the $60\arcsec$ core spectrum  with a
combination  of a  mekal and  a power-law.   Our analysis  of  the two
\chandra\ observations and the \xmm\  one gives similar results to all
of  the  above  studies  with  a simultaneous  fit  to  the  different
extracted spectra where a  combination of a mekal ($kT\approx0.5$~keV)
and  a  mildly  absorbed  ($N_{h}\approx10^{22}$~cm$^{-2}$)  power-law
($\Gamma$ between 1.5 and 2) were used.  The increase in the power-law
slope   from   $1.5\pm0.1$  during   the   \chandra\  observation   to
$2.1_{-0.2}^{+0.1}$ during the \xmm\  one is accompanied by a decrease
in     the    2-10~keV     flux     from    $9.8\times10^{-13}$     to
$4.6\times10^{-13}$~erg~s$^{-1}$.        This       behavior      (see
\S~\ref{accmode}) is  typical of X-ray emission originating  in a RIAF
structure which is believed  to be the accretion mechanism responsible
for   the  bulk   of  energy   from   radio  to   X-rays  in   NGC~315
\citep{wu07apj:riaf}.

{\sl NGC 2681.}  One of  the two observations performed with \chandra\
has  been already  reported  in \citet{satyapalapj05:sfVSliner}.   The
authors fit  the 0.5-8~keV  spectrum with a  combination of  a thermal
component with $kT\approx0.7$~keV and  a power-law with a photon index
$\Gamma\approx1.6$.  No  intrinsic absorption was  required.  The same
observation was treated in \citet{gonzalezmartin09aa} and same results
were  derived after  fitting the  spectrum with  a {\sl  mekal}  and a
power-law.   \citet{gonzalezmartin09aa}  derived  a 2-10~keV  flux  of
$\sim2\times10^{-13}$~erg~s$^{-1}$~cm$^{-2}$.  We  fit the spectrum of
the  two  \chandra\ observations  of  NGC~2681  simultaneously with  a
combination of a thermal component and an absorbed power-law. We found
similar   results   to   that    derived   in   the   previous   works
($\Gamma\approx1.5$  and $kT\approx0.6$~keV) with  a 2-10~keV  flux of
$3\times10^{-13}$~erg~s$^{-1}$~cm$^{-2}$.

{\sl NGC 2787.} \citet{ho01apjl}, after studying a \chandra\ snapshot,
gave  this source  a class  III X-ray  morphology, showing  hard X-ray
nucleus  embedded in  diffuse  emission.  \citet{terashima03apj:rloud}
derived         a        2-10~keV         flux         of        about
$3\times10^{-14}$~erg~s$^{-1}$~cm$^{-2}$,  after   assuming  a  photon
index  of  2 and  a  Galactic absorption.   \citet{gonzalezmartin09aa}
analyzed  both  \chandra\ and  \xmm\  long  observations  and fit  the
\chandra\ spectrum with a power-law with a rather soft $\Gamma$ of 2.3
and fit the \xmm\ spectrum with  a combination of two power-laws and a
thermal     component    with     a     power-law    absorption     of
$\sim10^{22}$~cm$^{-2}$.  We  found a good fit for  both \chandra\ and
\xmm\  spectra simultaneously  with  a single  absorbed power-law  and
found little  absorption of $\sim2\times10^{21}$~cm$^{-2}$  and a soft
power-law  photon   index,  $\Gamma=2.4$,  and  a   2-10~keV  flux  of
$4\times10^{-14}$~erg~s$^{-1}$~cm$^{-2}$.

{\sl   NGC~3226.}   \citet{georgeapj01:ngc3226}  fit   the  0.5-10~keV
spectrum  extracted  from  the  long  \chandra\  observation  with  an
absorbed       ($N_{H}\approx5\times10^{21}$~cm$^{-2}$)      power-law
($\Gamma\approx1.9$).   \citet{terashima03apj:rloud} fit  the spectrum
of  the  2.5~ks  snapshot  \chandra\  observation  with  a  moderately
absorbed    power-law    with   $N_{H}\approx10^{22}$~cm$^{-2}$    and
$\Gamma\approx2.2$.  \citet{gondoin0411:ngc3226}  fit the data  of the
35~ks  \xmm\  observation  with  a  partial  covering  absorber  to  a
bremsstrahlung  and  found that  the  X-ray  emitting  region, with  a
temperature  $kT\approx0.9$~keV is  90\% covered  by an  absorber with
$N_{H}\approx5\times10^{21}$~cm$^{-2}$.     \citet{binder09apj:ngc3226}
studied the $\sim$100~ks \xmm\ observation and fit the spectrum with a
partially  covered  power-law   with  $\Gamma\approx1.9$,  a  covering
fraction  of  90\%,  and  an  intrinsic hydrogen  column  density  of
$10^{21}$~cm$^{-2}$. We  fit the  spectra of both  \xmm\ and  the long
\chandra\ observations  simultaneously with an  absorbed power-law and
found  an intrinsic  column density  varying between  the observations
from  $\sim3\times10^{21}$~cm$^{-2}$ to $\sim9\times10^{21}$~cm$^{-2}$
and a mean photon index $\Gamma=1.9$.

{\sl  NGC~3718.} \citet{satyapalapj05:sfVSliner} studied  the snapshot
\chandra\  observation   and  fit   the  spectrum  with   an  absorbed
($N_{H}\approx10^{22}$~cm$^{-2}$)  power-law  ($\Gamma\approx1.5$)  in
excellent agreement with  our fit results to the  same observation. We
studied two \xmm\ observations of NGC~3718, being in the field of view
of  the   observations  of  the  heavily   absorbed  Seyfert~2  galaxy
\object{UGC~6527}. We  fit the spectra with an  absorbed power-law and
found an intrinsic hydrogen column  density similar to the one derived
for  the  \chandra\ observation  of  $\approx10^{22}$~cm$^{-2}$ but  a
somewhat softer power-law  with $\Gamma\approx1.8$.  This softening is
accompanied     with     a     2-10~keV     flux     decrease     from
$3.3\times10^{-12}$~erg~s$^{-1}$~cm$^{-2}$                           to
$1.6\times10^{-12}$~erg~s$^{-1}$~cm$^{-2}$.

{\sl   NGC~3998.}    \citet{ptak04apj:ngc3998}   studied   the   10~ks
\xmm\  observation  and fit  the  spectrum  with  a slightly  absorbed
($N_{H}\approx10^{20}$~cm$^{-2}$) power-law ($\Gamma\approx1.9$). Same
results  were   found  for  observations  made   with  {\sl  BeppoSAX}
\citep{pellegriniaa00:mgc3998} and  {\sl ASCA} \citep{ptak99apjs:TCM}.
\citet{gonzalezmartin09aa}   fit  the   \chandra\   spectrum  with   a
combination of two absorbed power-laws and a thermal component. We fit
the  \xmm\ and  the  \chandra\ spectra  simultaneously  with a  mildly
absorbed  ($N_{H}\approx10^{20}$~cm$^{-2}$)   power-law  and  found  a
varying $\Gamma$ from  1.8 to 2.1 respectively, occurring  with a flux
decrease     from     $1.1\times10^{-11}$~erg~s$^{-1}$~cm$^{-2}$    to
$6.5\times10^{-12}$~erg~s$^{-1}$~cm$^{-2}$.

{\sl     NGC~4143.}       \citet{terashima03apj:rloud}     fit     the
\chandra\  snapshot  observation   with  an  absorbed  power-law  with
$N_{H}<10^{21}$~cm$^{-2}$  and $\Gamma\approx1.7$.   We  fit the  same
snapshot  observation with  a power-law  without a  requirement  of an
intrinsic absorption and found a similar power-law photon index within
the error bars, $\Gamma\approx1.9$.  We fit the \xmm\ observation with
an absorbed power-law and found a $N_{H}=6\times10^{20}$~cm$^{-2}$ and
a $\Gamma\approx2.2$.

{\sl NGC~4203.} A power-law fit to the {\sl ASCA} spectrum resulted in
a  $\Gamma\approx1.8$  \citep{iyomotoapj98:ngc4203}.  \citet{ho01apjl}
gave  NGC~4203  a class  I  X-ray  morphology  showing dominant  X-ray
nucleus. We find that the 40~ks \chandra\ spectrum is well fitted with
a   simple   power-law    affected   by   Galactic   absorption   with
$\Gamma\approx2.3$, softer  than the  result reported for  {\sl ASCA},
most likely due to contamination  from X-ray sources in the {\sl ASCA}
extraction region of 1\arcmin.

{\sl NGC~4278.} See Y10.

{\sl NGC~4750.} \citet{dudik05apj:hardcoreliner}, and according to the
only \chandra\ snapshot observation,  gave this source a morphological
X-ray type II, exhibiting  multiple, hard off-nuclear point sources of
comparable brightness  to the nuclear  source. We fit the  spectrum of
this    same   observation   with    an   absorbed    power-law   with
$N_{H}<3\times10^{21}$~cm$^{-2}$ and $\Gamma=1.8$.

{\sl NGC~4772.}  We fit the  spectrum of the only, \chandra\ snapshot,
observation with  an absorbed ($N_{H}\approx5\times10^{21}$~cm$^{-2}$)
power-law ($\Gamma\approx1.7$).

{\sl NGC~5005.}  \citet{terashima02apjs:LLAGNASCA}  fit the {\sl ASCA}
spectrum      with     a      combination      of     an      absorbed
($N_{H}<9\times10^{21}$~cm$^{-2}$) power-law  ($\Gamma\approx1$) and a
thermal  component  ($kT\approx0.8$). \citet{dudik05apj:hardcoreliner}
gave NGC~5005 a  morphological X-ray type III, showing  a hard nuclear
point source embedded  in diffuse emission. \citet{gonzalezmartin09aa}
fit the \xmm\ spectrum with  a combination of a thermal component with
$kT\approx0.3$~keV             and             an             absorbed
($N_{H}\approx6\times10^{21}$~cm$^{-2}$)                      power-law
($\Gamma\approx1.5$). We find a  hotter thermal component when fitting
the  same  data set  with  $kT\approx0.6$~keV  and  a mildly  absorbed
power-law with $N_{H}\approx10^{21}$~cm$^{-2}$ and $\Gamma\approx1.7$.

\section{Surrounding sources of the centers of galaxies observed with \chandra}
\label{appb}

In this Appendix,  we report the spectral analysis  of resolved and/or
unresolved off-nucleus sources  detected with the \chandra\ telescope,
but blended  within the central  LINER in the \xmm\  extraction region
(\S~\ref{xmmobs}).   Fig.~1-4 show  the surrounding  sources  of these
LINER~1s  detected with  \chandra\ within  a  25\arcsec-radius circle.
The surrounding medium of NGC~4278 is already reported in Y10.

{\sl NGC~315.} This source is the only source in our sample that shows
a   resolved    X-ray   jet.     We   extracted   from    the   longer
\chandra\ observation the  spectrum of the jet from  an ellipse with a
semi-major  axis   of  about  11.3\arcsec\  and   semi-minor  axis  of
5.6\arcsec.    The  base   of  the   ellipse  extends   down   to  the
$1.1\times99\%$ PSF of the central source.  We fit the spectrum with a
combination  of  an  absorbed  power-law  and a  thermal  {\sl  mekal}
component and found  a good fit with a reduced $\chi^2$  of 0.9 for 42
d.o.f.    We  find   a  hydrogen   column  density   upper   limit  of
$2\times10^{21}$~cm$^{-2}$       and        a       photon       index
$\Gamma=2.0^{+0.4}_{-0.2}$.  The  thermal component had  a temperature
of $0.6^{+0.5}_{-0.9}$~keV.   We find a corrected  0.5-10~keV flux for
the                  jet                  emission                  of
$(1.2\pm0.1)\times10^{-13}$~erg~s$^{-1}$~cm$^{-2}$,  which corresponds
to   a   0.5-10~keV   luminosity   of   $6\times10^{40}$~erg~s$^{-1}$,
corresponding  to 8\%\ of  the nuclear  flux.  The  power-law emission
contributes  to 95\%\  to  the total  emission  of the  jet.  For  the
diffuse emission, we extracted the spectrum from an annulus with inner
circle delimited by  $1.1\times99\%$ PSF of the central  source and an
outer radius  of 25\arcsec\ excluding the jet  extraction region.  The
same model fit to  the jet gave a good fit with  a reduced $\chi^2$ of
1.3 for  61 d.o.f.   We find an  intrinsic absorption to  the powerlaw
component  $N_{H}=6^{+12}_{-10}\times 10^{21}$~cm$^{-2}$ and  a photon
index   $\Gamma=1.7\pm0.8$,   possibly   representing  emission   from
unresolved X-ray  binaries.  The  thermal component has  a temperature
similar   to   the   one   derived   for   the   jet   emission   with
$kT=0.6\pm0.2$~keV.    We  found  a   corrected  0.5-10~keV   flux  of
$2.9^{+0.2}_{-0.3}\times10^{-13}$~erg~s$^{-1}$~cm$^{-2}$    with   the
power-law  contributing only  to 30\%\  of the  total  emission.  This
corresponds  to a  luminosity of  $\sim10^{41}$~erg~s$^{-1}$  which is
$\sim14$\%\ of the total core luminosity.

{\sl NGC~2787.} An X-ray source  south-east of the nucleus of NGC~2787
at a distance less than 10\arcsec\  is present. We fit the spectrum of
this source with an absorbed power-law.  We used the same redshift and
Galactic absorption  as for the  NGC~2787 nucleus, assuming  that this
X-ray source is located in  NGC~2787 and not a background quasar.  The
fit is  acceptable with  a reduced $\chi^2$  of 1.0 for  19~d.o.f.  We
found a power-law  photon index $\Gamma=1.5^{+0.5}_{-0.4}$, typical of
X-ray          binaries           in          nearby          galaxies
\citep{irwin03apj:lmxbpop,fabbiano06aa:lmxbpop}, and an upper limit on
the       intrinsic       hydrogen       column       density       of
$2\times10^{21}$~cm$^{-2}$. We derived  a 0.5-10~keV corrected flux of
$7\pm1\times10^{-14}$~erg~s$^{-1}$~cm$^{-2}$,  which   resulted  in  a
luminosity   of,   adapting  the   NGC~2787   distance  of   7.48~Mpc,
$5\pm1\times10^{38}$~erg~s$^{-1}$.   This luminosity  is close  to the
NGC~2787 core luminosity  of $\sim9\times10^{38}$~erg~s$^{-1}$. Such a
source could  be a  luminous low mass  X-ray binary (LMXB)  similar to
some  seen in  early type  galaxies  \citep{fabbiano06aa:lmxbpop}. The
rest of  the medium  in an annulus  of outer radius  25\arcsec\ around
NGC~2787 is formed  by six other X-ray sources,  much fainter than the
closest one to the center.   We could not perform spectral analysis on
the different sources aside, so we fit the spectrum of the six sources
simultaneously with a power-law  only affected by Galactic absorption.
We found  a photon index  of $\sim$2 for  the six sources and  a total
corrected               0.5-10~keV               flux               of
$8^{+1}_{-2}\times10^{-15}$~erg~s$^{-1}$~cm$^{-2}$,   resulting  in  a
0.5-10~keV   luminosity   of  $5^{+1}_{-2}\times10^{37}$~erg~s$^{-1}$,
corresponding  to 5\%\ of  the nucleus  luminosity, when  adapting the
NGC~2787 distance of 7.48~Mpc.

{\sl    NGC~3226.}      Two    X-ray    sources     are    within    a
$\sim$12\arcsec\  distance from  the  nucleus of  NGC~3226 (source  1:
CXOU~J102334.1+195347, source 2: CXOU~J102326.7+195407).  Both sources
were reported  in \citet{georgeapj01:ngc3226}.  Based  on the hardness
ratio between the  counts in the 0.3-2~keV band and  the counts in the
2-10~keV band, the  authors estimated the sources to have  a flux of a
few  $10^{-14}$~erg~s$^{-1}$~cm$^{-2}$,  and  therefore  a  luminosity
between   a  few   times   $10^{38}$~erg~s$^{-1}$  to   a  few   times
$10^{39}$~erg~s$^{-1}$.  We  fit the spectrum of both  sources with an
absorbed power-law,  using the cash statistic  due to a  low number of
counts.   We  find  that source  1  and  source  2 have  an  intrinsic
absorption    upper    limit    of   $6\times10^{22}$~cm$^{-2}$    and
$5\times10^{21}$~cm$^{-2}$,  respectively. The  photon index  we found
for source 2  is typical, within the error  bars, of accreting objects
with  a $\Gamma=1.2^{+0.9}_{-0.6}$.   On the  other hand,  we  found a
harder spectrum for source 1 with a $\Gamma=0.2\pm1.3$.  The corrected
0.5-10~keV   flux  we   derive  for   source  1   and  source   2  are
$\sim4\times10^{-14}$~erg~s$^{-1}$~cm$^{-2}$                        and
$\sim2\times10^{-14}$~erg~s$^{-1}$~cm$^{-2}$,    respectively.    This
implies,  assuming  the  distance  of  NGC~3226  to  both  sources,  a
corrected  0.5-10~keV luminosity of  $3\times10^{39}$~erg~s$^{-1}$ and
$2\times10^{39}$~erg~s$^{-1}$     for     source~1    and     source~2
respectively. Both  luminosities are well  beyond the luminosity  of a
typical  neutron star  LMXB of  $\sim3\times10^{38}$~erg~s$^{-1}$, and
hence could be  BHs greater than or equal to a  few solar masses, very
young supernovae, or microquasars \citep{georgeapj01:ngc3226}.

{\sl NGC~3998.} Only one source is detected within a 25\arcsec\ circle
around NGC~3998  in the  \chandra\ image.  We  fit the source  with an
absorbed power-law, using the cash  statistic. We found an upper limit
on the intrinsic column density of $10^{21}$~cm$^{-1}$ and a power-law
photon index of $1.4^{+0.9}_{-0.5}$. We derived a corrected 0.5-10~keV
flux     of     $3\pm1\times10^{-14}$~erg~s$^{-1}$~cm$^{-2}$,    which
corresponds  to  a  luminosity  of  $7\pm1\times10^{38}$~erg~s$^{-1}$,
adapting the  NGC~3998 distance of 14.1~Mpc. This  corresponds to only
0.2\%\  of  the  total  core  luminosity of  NGC~3998  and  match  the
luminosity of XRBs in nearby galaxies.

\end{appendix}

%------------------------------------
%       References
%------------------------------------

%\bibliographystyle{aa} \bibliography{mybib}

\end{document}